\documentclass[fleqn,usenatbib]{mnras}

\usepackage{newtxtext,newtxmath}
\usepackage{xspace}

\usepackage[T1]{fontenc}
\usepackage{ae,aecompl}
\usepackage{url}
\usepackage{appendix}

\usepackage{graphicx}	
\usepackage{amsmath}	
\usepackage{pifont}
\usepackage{float}
\usepackage{xcolor}
\usepackage[utf8]{inputenc}

\newcommand{\nup}{$\nu_{\rm peak}^S$}
\newcommand {\magg}{\hphantom{>}}  
\newcommand{\xmark}{\ding{55}}%

\title[The spectra of IceCube neutrinos - III]{The spectra of IceCube Neutrino (SIN)  candidate sources - III. Optical spectroscopy and source characterization of the full sample}

\author[S. Paiano et al.]{Simona Paiano$^{1,2}$\thanks{E-mail:
simona.paiano@inaf.it}, Renato Falomo$^{3}$, Aldo Treves$^{4,5}$, Paolo Padovani$^{6,7}$, \newauthor Paolo Giommi$^{8,9,10}$, Riccardo Scarpa$^{11,12}$,  Susanna Bisogni$^{2}$, Ester Marini$^{13}$ \\
$^{1}$INAF - IASF Palermo, via Ugo La Malfa, 153, I-90146, Palermo, Italy \\ 
$^{2}$INAF - IASF Milano, via Corti 12, I-20133, Milano, Italy \\
$^{3}$INAF - Osservatorio Astronomico di Padova, vicolo dell'Osservatorio 5, I-35122, Padova, Italy\\
$^{4}$Universit\`a dell'Insubria, via Valeggio, 22100, Como, Italy\\
$^{5}$INAF - Osservatorio Astronomico di Brera, via Bianchi 46, I-23807, Merate (Lecco), Italy\\
$^{6}$European Southern Observatory, Karl-Schwarzschild-Str. 2, D-85748 Garching bei M\"unchen, Germany\\
$^{7}$Associated to INAF - 
Osservatorio di Astrofisica e Scienza dello Spazio, Via Piero Gobetti 93/3, I-40129 Bologna, Italy\\
$^{8}$Institute for Advanced Study, Technische Universit{\"a}t M{\"u}nchen, Lichtenbergstrasse 2a, D-85748 Garching bei M\"unchen, Germany\\
$^{9}$Center for Astro, Particle and Planetary Physics (CAP3), New York University Abu Dhabi, PO Box 129188 Abu Dhabi, United Arab Emirates\\
$^{10}$Associated to Agenzia Spaziale Italiana, ASI, via del Politecnico s.n.c., I-00133 Roma Italy \\
$^{11}$Instituto de Astrofisica de Canarias, C/O Via Lactea, s/n E38205 - La Laguna (Tenerife) - Spain\\
$^{12}$Universidad de La Laguna, Dpto. Astrofsica, s/n E-38206 La Laguna (Tenerife) - Spain\\
$^{13}$INAF - Osservatorio Astronomico di Roma, via Frascati 33, I-00077 Monte Porzio Catone (RM), Italy\\
}
\date{Received:~; Accepted:~}

\pubyear{2022}

\begin{document}
\label{firstpage}
\pagerange{\pageref{firstpage}--\pageref{lastpage}}
\maketitle

\begin{abstract}
A correlation between astrophysical high-energy neutrinos and blazars has been suggested by various authors.
In particular, a likely association between IceCube events and intermediate and high-energy peaked BL Lac objects has led to a sample of 47 objects having a high probability of being neutrino sources. 
In the first paper of this series we reported optical spectroscopy of 17 objects, which together with data taken from the literature covered 80 per cent of the sample. 
Here we present spectroscopy obtained at large aperture telescopes 
of a further 17 objects (plus four additional targets), 
which completes the sample coverage. 
For twelve objects we are able to determine the redshift 
($0.07 < z <1.6$), while for the others we set a lower limit on it, 
based on either the robust detection of intervening absorption systems 
or on an estimation derived from the absence of spectral signatures of 
the host galaxy. With these new data we expand and reinforce the main 
results of our previous papers, namely the fact that in terms of their broad-band properties our sources appear to be indistinguishable from the rest 
of the blazar population and the relatively large ($>$~34 per cent and 
possibly as high as 80 per cent) fraction of  
\textit{masquerading} BL Lac objects, for which the low equivalent width of the 
emission lines is due to the brightness of the boosted continuum, rather than 
being an intrinsic property, in our sample.
\end{abstract}


\section{Introduction}  
\label{sec:introduction} 
High-energy (TeV-PeV) neutrino astrophysics was born recently thanks to IceCube \citep{icecube2013}. The quasi-isotropy of the distribution of neutrino events in the sky indicates a dominant component of extragalactic objects, which are also most probably linked to the sources of high energy cosmic rays 
\citep[e.g.][]{resconi2017, kurahashi2022}.

Many authors \citep[e.g.][and references therein]{righi2019, giommi2020, plavin2021, hovatta2021, buson2022} 
have suggested a correlation between IceCube neutrinos and blazars, i.e. active galactic nuclei whose emission is dominated by a relativistic jet pointing towards the observer. In particular, \citet{giommi2020} (hereafter G20) considered all the 70 IceCube neutrino tracks known in 2020 and found a 3.2$\sigma$ post-trial correlation with intermediate- and high-energy (IBL and HBL)\footnote{Based on the rest-frame frequency of the low-energy synchrotron peak (\nup), blazars are divided into
low- (LBL \nup~$<10^{14}$~Hz [$<$ 0.41 eV]), intermediate- (IBL:
$10^{14}$~Hz$ ~<$ \nup~$< 10^{15}$~Hz [0.41 -- 4.1 eV)], and high-energy
(HBL: \nup~$> 10^{15}$~Hz [$>$ 4.1 eV]) peaked objects respectively
\citep{padovani95,abdo2010}.} blazars, i.e. objects where the synchrotron peak frequency is $>10^{14}$ Hz. Specifically, 30 neutrino events with well defined positions were matched with a set of 47 blazars (see Table~5 of G20), with $16\pm4$ of those possibly being neutrino sources.

In order to test this scenario out, optical spectra are required, since the 
majority of the G20 sample did not have redshift determinations. 
These are obviously needed to derive the luminosity of the source, vital for any modelling, determine the properties of the spectral lines, and also possibly estimate the mass of the central black hole, $M_{\rm BH}$.
Our group has therefore started ``The spectra of IceCube Neutrino (SIN) candidate sources'' project, which has so far led to two papers. 
In the first paper (\citealt{Paiano_2021}: hereafter Paper~I) we obtained optical spectroscopy of 17 of the G20 sources, which together with 
spectra gathered from the literature, covered $\sim$80 per cent of the G20 sources. 
All objects but one (a quasar), were shown to be BL Lac objects (BLLs): the redshift was measured for half of them, and for the rest a lower limit was given. 

In the second paper (\citealt{Padovani_2022}: hereafter Paper II) it was shown that the BLLs under scrutiny were in many cases (40 -- 80 per cent) masquerading BLLs, i.e. objects where the weakness of the emission lines 
and their low values of the Equivalent Width (EW) is due to a very bright Doppler boosted continuum, which is washing out the lines. Note that 
TXS~0506+056, which is the blazar with the most significant association with neutrino events \citep{icecube2018a,icfermi,baikal2022} is also a masquerading BLL \citep{Padovani_2019}. Paper II discussed also in some details possible jet models for high-energy neutrino production.

We report here on optical spectroscopy of 17 targets, plus other four sources (referred to as ``extra'') associated to neutrino events posterior to G20. This completes the spectroscopy coverage for the G20 sample. 
The spectra were taken at large diameter telescopes, i.e. the Gran Telescopio Canarias (GTC), the Very Large Telescope (VLT), and the Large Binocular Telescope (LBT).

This paper is organized as follows: in Section~\ref{sec:sample} we present the sample, in Section~\ref{sec:obsdata} we describe the data and their reduction. In Section~\ref{sec:results} we present the observational results, the main properties of the features found in the spectra and discuss the redshift measurements. In Section~\ref{sec:notes} we give details on individual objects. In Section~\ref{sec:characterisation} we characterize the G20  
sources, while Section~\ref{sec:extra} does that for the extra objects. 
Finally, Section \ref{sec:discussion} discusses our results and Section \ref{sec:conclusions} summarises our conclusions. Appendix 
\ref{sec:appendix} shows flux calibrated and de-reddened spectra for our sources, some examples of close-ups around the detected spectral lines, and spectral decompositions of the observed optical spectra into a power-law and an elliptical template galaxy.

We use a $\Lambda$CDM cosmology with Hubble constant $H_0 = 70$ km s$^{-1}$ Mpc$^{-1}$, matter density $\Omega_{\rm m,0} = 0.3$, and dark energy density $\Omega_{\Lambda,0} = 0.7$.

\setcounter{table}{0}
\begin{table*}
\begin{center}
\caption{The sample of neutrino candidate blazars. } 
\begin{tabular}{llllllll}
\hline 
Object  Name  &    Counterpart  Name   &  IC event & Sep. & RA    &  DEC  &  Mag.  &  \nup   \\   
           &                    &           & (degrees) & (J2000)  &(J2000) &  (g)         & (Hz)  \\
\hline
4FGL~J0258.1+2030  & 5BZB J0258+2030 & IC-191231A/IC-211125A &  2.11/2.08 & 02:58:07.3 &  +20:30:02.0 & 20.8 & 14.1  \\ 
\hline
4FGL~J0339.2$-$1736  & 3HSP J033913.7$-$173600 & IC-141109A & 1.24 & 03:39:14.7 & $-$17:36:00.8 & 16.8 & 15.6 \\ 
\hline
4FGL~J0545.0+0613c  & NVSS J054341+062553   & IC-200421A &  2.88 & 05:45:29.1 &  +06:19:58.0 & 21.3 & ...  \\ 
\hline
4FGL~J0658.6+0636  & 3HSP J065845.0+063711 & IC-201114A &  0.85 & 06:58:44.9 &  +06:37:11.5 & 18.9 & 15.5 \\       
\hline
4FGL~J0946.2+0104   & 3HSP~J094620.2+010451 & IC-190819A & 2.24 & 09:46:20.2 & +01:04:51.6 & 20.0 & $>$18.0 \\ 
\hline
4FGL~J1003.4+0205    & 3HSP~J100326.6+020455 & IC-190819A & 2.35 & 10:03:26.6 & +02:04:55.5 & 19.7 & 15.8 \\
\hline
4FGL~J1055.7$-$1807    & VOU~J105603$-$180929  & IC-171015A  & 2.6  & 10:56:03.5  & $-$18:09:30.2  & 20.5 & 14.1 \\ 
\hline 
4FGL~J1124.0+2045 & 3HSP~J112405.3+204553 & IC-130408A & 3.61 & 11:24:05.3 & +20:45:52.9 & 18.8 & 15.3 \\ 
\hline
4FGL~J1124.9+2143   & 3HSP~J112503.6+214300 & IC-130408A & 3.48 & 11:25:03.5 & +21:43:04.0 & 18.5 & 15.8  \\ 
\hline
4FGL~J1314.7+2348     & 5BZB~J1314+2348      & IC-151017A & 3.68 & 13:14:43.8 & +23:48:26.6 & 17.1 & $\geq$14 \\
\hline
4FGL~J1321.9+3219 & 5BZB~J1322+3216      & IC-120515A  & 1.54 & 13:22:47.4 & +32:16:08.7  & 19.3 & 14.5  \\
\hline
4FGL~J1439.5$-$2525 & VOU~J143934$-$252458 & IC-170506A & 1.81 & 14:39:34.6 & $-$25:24:58.3 & 19.0 &  14.0 \\
\hline
4FGL~J1507.3$-$3710 & VOU~J150720$-$370902 & IC-181014A & 2.0 & 15:07:20.8 & $-$37:09:02.8 & 17.5 & 14.5  \\
\hline
4FGL~J1528.4+2004 & 3HSPJ152835.8+200420     & IC-110521A & 3.23  & 15:28:35.7 &  +20:04:20.2 & 20.2 & 16.2   \\
\hline
4FGL~J1702.2+2642 & 5BZB J1702+2643  &  IC-200530A & 0.62  & 17:02:09.6 & +26:43:15.0 &  18.5 & 13.9\\
\hline
4FGL~J1808.2+3500 & CRATES~J180812+350104    & IC-110610A &  0.55 & 18:08:11.5 & +35:01:18.8  &  15.4 &  14.5   \\ 
\hline
4FGL~J2030.9+1935 & 3HSP~J203057.1+193612 & IC-100710A & 1.82 & 20:30:57.1 &  +19:36:12.8 & 18.6 & 15.8 \\
\hline
4FGL~J2030.5+2235 & 3HSP~J203031.6+223439 & IC-100710A & 1.31 & 20:30:31.6 & +22:34:39.3 & 20.3 & 16.2 \\ 
\hline
4FGL~J2133.1+2529c & 3HSP~J213314.3+252859 & IC-150714A & 2.18 & 21:33:14.3 & +25:28:59.1 & 18.8 & 15.2  \\
\hline
4FGL~J2350.6$-$3005 & 3HSP~J235034.3$-$300604 & IC-190104A & 3.32 & 23:50:34.3 & $-$30:06:03.2 & 18.1 & 15.7  \\
\hline
4FGL~J2358.1$-$2853  & CRATES~J235815$-$285341 & IC-190104A & 2.47 & 23:58:16.9 & $-$28:53:34.0 & 19.5 & 14.0 \\
\hline
\end{tabular}
\label{tab:targets}
\end{center}
\raggedright
\footnotesize \textit{Notes.} Column~1: {\it Fermi} name of the target in the 4FGL catalogue \citep{4FGLDR3}; Column~~2: Counterpart name of the $\gamma$-ray source; Column~~3: IceCube tracks; Column~~4: Angular separation (degrees) between the target and the centroid of the IceCube track; Column~~5~-~6: Right ascension and declination of the optical counterpart; Column~7: g-magnitude from SDSS survey or PANSTARRs; Column~8:  Frequency (in logarithmic scale) of the synchrotron peak. 
\\
\end{table*}

\section{The sample} 
\label{sec:sample} 

In Paper~I we considered a sample of 47 $\gamma$-ray sources matched to 30 
neutrino events with well-defined IceCube positions (area of the error ellipse 
smaller than that of a circle with radius r~$= 3^{\circ}$) and we presented
spectroscopic data for 17 objects of them with previously unknown or uncertain redshift in the literature. 

In this work we report on optical spectroscopy of further 21 sources (see 
Table~\ref{tab:targets}) with unknown or uncertain redshift or with low quality 
optical spectrum with the aim of determining the distance and the main spectral 
properties. Among them, 17 sources belong to the original G20 sample while 4 objects (4FGL~J0258.1+2030, 4FGL~J0545.0+0613, 4FGL~J0658.6+0636 and 4FGL~J1702.2+2642) are newly added because they were proposed as potential counterparts of Icecube detections recorded after the publication of G20. 
Most of these sources satisfy all the G20 
selection criteria ($|b_{\rm II}|>10^{\circ}$, \nup~$>10^{14}$~Hz, and area of the error ellipse), apart from 4FGL~J0258.1+2030/IceCube-191231A, which violates the area constrain but represents a rare case of an object associated with more than one neutrino track, and 4FGL~J0545.0+0613c whose spectral energy distribution (SED) is dominated by thermal emission and for which is therefore impossible to determine \nup. More details are reported in the notes of the single sources in Section~\ref{sec:notes}.  
A journal of the observations of the 21 sources is given in Table~\ref{tab:observations}. 

\setcounter{table}{1}
\begin{table*}
\begin{center}
\caption{Journal of the observations.}
\begin{tabular}{llllllll}
\hline 
Object Name  & E(B-V) & Telescope & Instrument & Date & t$_{\rm exp}$ & seeing  & Air Mass \\   
 &      &      &      &     & (s)       &   (")  &  \\
\hline
4FGL~J0258.1+2030 &  1.10      &   GTC         &  OSIRIS    & 08 October 2020  & 9000     & 1.3      &   1.1 \\
4FGL~J0339.2$-$1736 &  0.07      &   VLT         &  FORS2     & 10 October 2020  & 2700     & 0.5      &   1.0 \\
4FGL~J0545.0+0613c &  0.99     &   GTC         &  OSIRIS    & 10 October 2020  & 9000     & 0.8      &   1.2 \\
4FGL~J0658.6+0636 &  0.22      &   GTC         &  OSIRIS    & 17 November 2020 & 4599     & 1.2      &   1.1 \\
4FGL~J0946.2+0104 &  0.13      &   VLT         &  FORS2    & 07 December 2020 & 3600     & 1.1      &   1.1 \\
4FGL~J1003.4+0205 &  0.02      &   VLT         &  FORS2    & 18/19 December 2020 & 5400     &   0.7 &   1.2 \\
                  &            &   GTC          &  OSIRIS     & 20  December 2020 & 4500     &   1.3  &   1.3  \\
4FGL~J1055.7$-$1807 &  0.03      &   VLT         &  FORS2     & 23 December 2020 & 5400     & 0.4      &   1.3 \\ 
4FGL~J1124.0+2045 &  0.02      &   GTC         &  OSIRIS    & 12 December 2020 & 3600     & 1.4      &   1.2 \\
4FGL~J1124.9+2143 &  0.02      &   GTC         &  OSIRIS    & 12 December 2020 & 3600     & 1.0      &   1.1 \\
4FGL~J1314.7+2348 &  0.01      &   GTC         &  OSIRIS    & 11 February 2021 & 3600     & 1.5      &   1.2 \\
4FGL~J1321.9+3219 &  0.02      &   GTC         &  OSIRIS    & 10 February 2021 & 4500     & 1.5      &   1.3 \\
4FGL~J1439.5$-$2525 &  0.09      &   VLT         &  FORS2     & 08 February 2021 & 2700     & 1.1      &   1.5 \\
4FGL~J1507.3$-$3710 &  0.06      &   VLT         &  FORS2     & 07 February 2021 & 2400     & 0.9      &   1.2 \\
4FGL~J1528.4+2004 &  0.06      &   LBT         &  MODS      & 08 June 2021     & 10800    & 0.7      &   1.0 \\
4FGL~J1702.2+2642 &  0.05      &   GTC         &  OSIRIS    & 13 June  2020    & 3600     & 0.7      &   1.2 \\
4FGL~J1808.2+3500 &  0.04      &   LBT         &  MODS      & 10 May   2021    & 10800    & 1.5      &   1.0 \\
4FGL~J2030.9+1935 &  0.08      &   GTC         &  OSIRIS    & 30 April 2020    & 1500     & 1.0       &   1.0 \\ 
                  &            &               &    & 10 October 2020    & 1500     & 1.1       &   1.0 \\ 
4FGL J2030.5+2235  &  0.21      &   LBT         &  MODS      & 01 June 2022     & 7200  & 0.7      &   1.0 \\
4FGL~J2133.1+2529c & 0.10      &   GTC         &  OSIRIS    & 07 October 2020  & 2700     & 1.4      &   1.3 \\
4FGL~J2350.6$-$3005 &  0.01      &   VLT         &  FORS2     & 12 October 2020  & 2400     & 0.8      &   1.2 \\
4FGL~J2358.1$-$2853 &  0.02      &   VLT         &  FORS2     & 10 October 2020  & 5400     & 0.8      &   1.1 \\
\hline
\end{tabular}
\label{tab:observations}
\end{center}
\raggedright
\footnotesize \textit{Notes}.Column~2: E(B-V) taken from the NASA/IPAC Infrared Science Archive (\url{https://irsa.ipac.caltech.edu/applications/DUST/}); Column~3: Telescope used for the observation; Column~4: Instrument; Column~5: Date of the observation; Column~6: Total integration time (sec); Column~7: Average seeing during the observation (arcsec); Column~8: Air mass during the observation.\\
\end{table*}

\section{Observations and data reduction} 
\label{sec:obsdata} 

The optical spectra were obtained using three different telescopes. 
Twelve sources located in the northern hemisphere were observed with the 
10.4m GTC at the Roque de Los Muchachos (La Palma) equipped with the OSIRIS spectrograph \citep{cepa2003}, configured with the grism R1000B (slit width of 1.2") that covers the spectral range 4100~-~7750~$\textrm{\AA}$ with a resolution R~$\sim$~600. 
For 7 targets, the 8~m VLT-UT1 at Paranal using FORS2 \citep{appenzeller1998} was utilized. 
In this case, the grism GRIS\_300V+10 with slit width of 1.3" was used, sampling the range 4700~-~8500~$\textrm{\AA}$ with R~$\sim$~200.
Spectra of three sources were taken with the LBT, two twin 8.4m telescopes located in southeastern Arizona (Mt. Graham) using the MultiObject Double Spectrographs MODS-1 and MODS-2 \citep{pogge2010} in dual grating mode with the grisms G400L and G670L and a slit width of 1.2" and covering the spectral ranges 4000-9000~$\textrm{\AA}$ (R~$\sim$~1000).

Data reduction of the GTC observations was performed adopting standard IRAF procedures \citep{tody1986, tody1993} for long slit spectroscopy following the same scheme given in \citet{paiano2017tev}. 
For FORS2 data we adopted the reduction pipeline provided by the EsoReflex environment \citep{freudling2013} including extraction of 1D spectra.
Spectroscopy data reduction for LBT data was carried out at the Italian LBT Spectroscopic Reduction Center using the pipeline SIPGI \citep{gargiulo2022} and using the standard procedure for long-slit spectroscopy with bias subtraction, flat-fielding, and bad-pixel correction. 
In all cases the accuracy of the wavelength calibration based on the scatter of the polynomial fit (pixel vs wavelength) is $\sim$0.1~$\textrm{\AA}$ over the whole observed spectral range. 
In order to perform an optimal correction of cosmic rays and other artifacts, the observation of each source was divided in at least three individual exposures. 
This procedure allows us also to check for possible spurious features. 
The final spectrum results from the combination of all individual exposures.

Spectro-photometric standard stars were secured for each night in order to perform the relative flux calibration.  
In addition, we assessed the absolute flux calibration from the measured magnitude of the source (see Table~\ref{tab:results}) as derived from the acquisition images obtained during the observations.  
For each image we measured the observed magnitude of a number of stars in the SDSS and/or Pan-STARRS catalogues and derived the calibration constant of the frame, with  
an accuracy $\sim$ 0.2 magnitudes.  
A comparison with literature magnitudes (see Tab.~\ref{tab:targets}) shows 
some variability in 4FGL~J0658.6+0636 and 4FGL~J1439.5$-$2525.

Finally all spectra were de-reddened applying the extinction law of \citet{cardelli1989} and assuming the value of Galactic extinction E(B-V) derived from the NASA/IPAC Infrared Science Archive\footnote{\url{http://irsa.ipac.caltech.edu/applications/DUST/}}
\citep{schlafly2011}. 


\section{Results} 
\label{sec:results}

The flux calibrated and de-reddened spectra of the neutrino source candidates 
are displayed in Fig.~\ref{fig:spectra} and are also available in the online 
database ZBLLAC\footnote{\url{https://web.oapd.inaf.it/zbllac/}} \citep{landoni2020}. 
For each spectrum, we evaluated the signal-to-noise ratio (S/N) in a number 
of spectral regions (the average value is given in Table~\ref{tab:results}), 
and performed a search for emission/absorption lines. 

For 12 out of 21 targets we provide a firm redshift (range: $0.0655-1.645$, median $\sim$ 0.3). 
Except for 4FGL~J0545.0+0613c, that displays a quasar spectrum with very broad emission lines (for details see the relevant note in Section~\ref{sec:notes}), all targets show a spectrum consistent with that of a BLL, i.e., characterized by a combination of a non-thermal continuum, described by a power law, and host galaxy features, when relevant. 
For 10 objects the spectral continuum is well dominated by the non-thermal continuum with spectral index $\alpha$ (F$_\lambda\sim\lambda^{-\alpha}$) in the 0.2~-~1.4 interval.
In the other 10 objects, the spectrum shows the contribution of the elliptical host galaxy and a firm determination of the redshift (see Table~\ref{tab:results}) was obtained from typical absorption lines (Ca~II~3934,3968 doublet, the G-band~4305, Mg~I~5157, Ca+Fe~5269, and Na~I~5892).
For 4FGLJ~1321.9+3219 the redshift (z~=~0.8126) was derived from a single emission line corroborated by a Mg~II intervening absorption system (see note in Section~\ref{sec:notes}).
For 4 sources (4FGL~J1003.4+0205, 4FGL~J1055.7$-$1807, 4FGL~J1124.0+2045, and 4FGL~J2358.1$-$2853) a spectroscopic redshift lower limit is provided by the detection of intervening absorption systems due to Mg~II and Fe~II.
For the 5 lineless objects, we can set a redshift lower limit based on the absence of the host galaxy absorptions (assuming a standard average luminosity of $\langle M(R) \rangle =-22.9$ \citep{sbarufatti2005imaging} and on the miniminum detectable EW (see details in \citealt{paiano2017tev}). All redshift 
measurements are reported in Table~\ref{tab:results}.

\setcounter{table}{2}
\begin{table*}
\begin{center}
\caption{Properties of the optical spectra of the neutrino candidate blazars studied in this work.}
\begin{tabular}{lclcclll}
\hline
Object Name            &  g   & r     & S/N & EW$_{\rm min}$   & z &  Line & $\alpha$  \\  
                       &      &       &     & ($\textrm{\AA}$) &   &     type   &   \\  
\hline
4FGL~J0258.1+2030      & 21.0 & -     & 60  &   0.80           & $>$0.3              & h      &  0.80   \\ 
4FGL~J0339.2$-$1736    & -    & 14.9  & 70  &    -             & 0.0655~$\pm$~0.0005   & e,g    &  0.9    \\ 
4FGL~J0545.0+0613c     & 21.1 & -     & 100 &    -             & 1.645~$\pm$~0.001     & e      &  -    \\  
4FGL J0658.6+0636      & 19.9 & -     & 60  &   0.45           & $>$0.5              & h      &  1.30  \\ 
4FGL~J0946.2+0104      & -    & 19.2  & 80  &    -             & 0.5768~$\pm$~0.0005   & g      &  1.0     \\ 
4FGL~J1003.4+0205      & 19.1 & -     & 130 &    -             & >0.4695~$\pm$~0.0005 & i      &  1.4\\%
4FGL~J1055.7$-$1807    & -    & 20.0  & 80  &    -             & >0.8465~$\pm$~0.0005 & i      &  0.40    \\ 
4FGL~J1124.0+2045      & 19.4 & -     & 100 &   -              & >0.5805~$\pm$~0.0005 & i      &  1.40   \\ 
4FGL~J1124.9+2143      & 19.4 & -     & 120 &   0.20           & $>$0.60             & h      &  1.25  \\  
4FGL~J1314.7+2348      & 17.5 & -     & 300 &   0.10           & $>$0.50             & h      &  1.10  \\ 
4FGL~J1321.9+3219      & 18.8 & -     & 180 &    -             & 0.8126~$\pm$~0.0005   & e,g,i  &  1.15 \\ 
4FGL~J1439.5$-$2525    & -    & 17.2  & 40  &    -             & 0.161~$\pm$~0.005     & g      &  0.5 \\ 
4FGL~J1507.3$-$3710    & -    & 17.7  & 100 &    -             & 0.239~$\pm$~0.005     & e,g    &  0.8 \\ 
4FGL~J1528.4+2004      & -    & 19.7  & 70  &    -             & 0.64~$\pm$~0.01       & e,g    &  1.00 \\  
4FGL~J1702.2+2642      & -    & 18.1  & 190 &    -             & 0.3197$\pm$0.0005   & e,g    &  1.00 \\ 
4FGL~J1808.2+3500      & -    & 16.7  & 250 &    -             & 0.282~$\pm$~0.005     & e,g    &  0.80 \\  
4FGL~J2030.9+1935      & 18.0 & -     & 150 &    -             & 0.3662~$\pm$~0.0005   & e,g    &  1.40 \\ 
4FGL~J2030.5+2235      &  -   & 19.7  & 40  &    0.90          & $>$0.50             &  h     &  1.35  \\
4FGL~J2133.1+2529c     & 18.8 & -     & 55  &    -             & 0.295~$\pm$~0.005     & g      &  0.80 \\ 
4FGL~J2350.6$-$3005    & -    & 17.4  & 85  &    -             & 0.233~$\pm$~0.001     & g      &  0.70 \\ 
4FGL~J2358.1$-$2853    & -    & 19.4  & 110 &    -             & $>$1.5425~$\pm$~0.0005 & i      &  0.20 \\ 
\hline
\end{tabular}
\label{tab:results}
\end{center}
\raggedright
\footnotesize \textit{Notes}. Column~2-3: Observed magnitude (g,r) measured from the acquisition image; Column~4: Median S/N of the spectrum; Column~5: Minimum equivalent width (EW$_{\rm min}$) derived in the 5500 - 6500 $\textrm{\AA}$ range (provided only in case of featureless spectrum); Column~6: Redshift. The error is evaluated as a combination of the uncertainty of the centroid of the features with the overall accuracy of the wavelength calibration; Column~7: Type of detected line to estimate the redshift: \textit{e} = emission line, \textit{g} = galaxy absorption line, \textit{i}= intervening absorption assuming Mg~II 2796,2803~$\textrm{\AA}$ identification, \textit{h}= lower limit based on the lack of detection of host galaxy absorption lines assuming an elliptical host galaxy with $M(R) = -22.9$. Since the distribution of BLL host galaxies has a dispersion (1$\sigma$) of $\sim$ 0.5 mag, these limits may change by 0.05-0.1 depending on the redshift limit \citep[see details in][]{paiano2017tev}. Column-8: Spectral index $\alpha$ of the continuum described by a power law F$_\lambda \sim \lambda^{-\alpha}$: for the cases with evident signature of the host galaxy (see Fig.~\ref{fig:decomposition}) the spectral index of the non-thermal component is estimated by the decomposition of the observed spectrum. \\
\end{table*}

For the targets with host galaxy signatures, we performed a spectral decomposition considering two components, namely: 1. a power law; 2. a template for the host galaxy \citep{mannucci2001}.  
The best fit of the decomposition was obtained from the variation of the parameters of the two components (nucleus-to-host ratio [N/H],
evaluated at $\sim$ 6000~$\textrm{\AA}$, and spectral slope of the non-thermal component) and imposing the normalization of the fit at about the central wavelength of each spectrum.
We found that in all above cases the decomposition is a good representation of the observed spectrum (see Fig.~\ref{fig:decomposition}). 
In some cases (e.g. 4FGL~J0339.2$-$1736, 4FGL~J1439.5$-$2525, and 4FGLJ~2350.6$-$3005) the optical spectrum is significantly dominated by the host galaxy while in others the non-thermal component is dominant (e.g. 4FGL~J1808.2+3500 and 4FGL~J2030.9+1935). 
The accuracy of these decompositions clearly depends on the quality of the spectra and on N/H, the targets with higher N/H having more uncertain host galaxy estimates. 
We estimated the absolute magnitude of the host galaxy in the R filter, k-corrected using the above galaxy template spectrum\footnote{This measurement of the absolute magnitude is derived from the flux through the slit and needs to be corrected for the flux slit loss.}.

We used the host galaxy luminosity to derive $M_{\rm BH}$ 
from the $M_{\rm BH}$-M(R) relationship \citep{labita2007}, as
described in Paper II. 
Since the total magnitude enters in the relationship, we 
need to correct for the fact that aperture photometry, 
based on the calibration image, misses some flux. 
To evaluate this correction we assume that the host galaxies of these targets are similar to other host galaxies of BL Lac objects (giant elliptical with $\langle R_{\rm eff} \rangle =8$~kpc; \citealt{falomo2014} ). Thus from the assumed shape and apparent size of the galaxy and the adopted aperture we derive the magnitude correction. This is in the range of $0.3-0.8$ mag and depends very little on R$_{\rm eff}$.

In order to test the above correction we performed an image analysis of 6 targets with a substantial contribution from the host galaxies (see Table 6).
We used our calibration images (Section~\ref{sec:obsdata}) to decompose the source into the nucleus (described by the point spread function [PSF]) and a galaxy model convolved with the PSF \citep[see e.g. ][and references therein for example of the adopted method]{liuzzo2016}. 
We then derived the total magnitude of the host galaxy, which was found to be in agreement with those obtained by spectroscopy described above.

Finally we also set the host galaxy luminosity to the present epoch assuming a passive stellar evolution for massive ellipticals \citep{bressan1994}. 
The corrected host galaxy absolute magnitudes, with a typical uncertainty of the order of 0.7, are reported in Table 6 and used to derive $M_{\rm BH}$.

\subsection{Host galaxies, emission lines and BH masses}\label{sect:host_mass}

A peculiar characteristic of the BLL class is the weakness (or absence) of spectral lines in the optical spectra. The absorption lines, detectable if the N/H is advantageous and if the spectrum is of good quality in terms of S/N and resolution, are typical of the old stellar population of the host elliptical galaxies.
The possible emission lines can arise from region of recent star formation or from the nuclear activity and their properties are crucial to determine their origin and for the source characterization. 
For 7 objects, of which 6 also with host galaxy signatures in the spectrum, weak (EW~$<$~3~$\textrm{\AA}$) and narrow emission lines of [O~II]~3727 and/or [O~III]~5007 and/or [N~II]~6583 are found. 
The average [O~II] and [O~III] line luminosity are $\sim$1.8$\times$10$^{41}$ erg s$^{-1}$ and $\sim$3.5$\times$10$^{40}$ erg s$^{-1}$, respectively.
See Table~\ref{tab:emission} for details on the properties of the emission lines.

From the spectral decomposition, for 10 objects with stellar spectral signature we are able to estimate the properties of the host galaxy component over the continuum.  
On average the host galaxy luminosity of the targets is somewhat fainter ($\langle M(R) \rangle = -22.5$) than the average value of BLLs as derived from HST images ($\langle M(R) \rangle = -22.9$). 
In particular, in two cases (4FGL~J1528.4+2004 and 4FGL~J1702.2+2642) the absolute magnitude is about 1 mag fainter than the average for the BLL class. 

The central black hole masses are in the interval $2 \times 10^8 - 8 \times10^{8}$ $M_{\odot}$ (median $M_{BH}$ $\sim$ $5 \times10^8 $ $M_{\odot}$) (see Table~\ref{tab:abs_mag}).
The uncertainty of an individual $M_{BH}$ depends on both the dispersion of the $M_{BH}$-M(R) relationship, $\sim 0.45$ dex, and the uncertainty (of about a factor of 2) on the estimate of the host galaxy luminosity. 

\setcounter{table}{3}
\begin{table*}
\begin{center}
\caption{Properties of the emission lines of BLL objects. } 
\begin{tabular}{llllll}
\hline
Object Name             &   z      & ~$\lambda$         &   EW             &  Line ID &  \textit{L} (line)   \\ 
                        &          &  ($\textrm{\AA}$) & ($\textrm{\AA}$) &          &   ($\times $10$^{40}$ erg s$^{-1}$)     \\ 
\hline
4FGL~J0339.2$-$1736     &  0.0655  &  7015             & 1.10~$\pm$~0.05             & [NII]~6583   &  4.2  \\ 
                        &          &  5335             & $<$0.8           & [OIII]~5007  &  $\lesssim$2.7\\
4FGL~J0946.2+0104       &  0.5768  & 5876              & $<$0.4           & [OII]~3727   &  $\lesssim$4.0 \\
4FGL~J1321.9+3219       &  0.8126  &  6755             & 1.5~$\pm$~0.1             & [OII]~3727   &  56  \\ 
4FGL~J1439.5$-$2525     &  0.161   &  5813             & $<$1.3           & [OIII]~5007  &  $\lesssim$3.5 \\
4FGL~J1507.3$-$3710     &  0.239   &  6203             & 3.2*~$\pm$~0.2            & [OII]~3727   &  14  \\%
                        &          &  6203             & 1.05~$\pm$~0.15            & [OIII]~5007  &  4.7 \\ %
                        &          &  8340             & 1.25~$\pm$~0.25            & [SII]~6731   &  5.0 \\ 
4FGL~J1528.4+2004       &  0.64    &  6110             & 2.80~$\pm$~0.1             & [OII]~3727   &  23   \\ 
4FGL~J1702.2+2642       &  0.3197  &  4918             & 0.55~$\pm$~0.05             & [OII]~3727   &  4.1  \\ 
                        &          &  6608             & 0.4~$\pm$~0.1             & [OIII]~5007  &  2.4 \\ %
4FGL~J1808.2+3500       &  0.282   &  4778             & 0.30~$\pm$~0.05             & [OII]~3727   &  5.6  \\ 
                        &          &  6419             & 0.30~$\pm$~0.05             & [OIII]~5007  &  4.8 \\ 
                        &          &  8440             & 0.15~$\pm$~0.05             & [NII]~6583   &  2.0  \\ 
4FGL~J2030.9+1935       &  0.3662  &  5092             & 0.50~$\pm$~0.05             & [OII]~3727   &  8.7 \\ 
                        &          &  6840             & 0.20~$\pm$~0.05             & [OIII]~5007  &  2.6 \\ %
4FGL~J2133.1+2529c      & 0.295    & 4826              & $<$3             & [OII]~3727   &  $\lesssim$15 \\
                        &          & 6484              & $<$1.5           & [OIII]~5007  &  $\lesssim$9.0 \\
4FGL~J2350.6$-$3005     & 0.233    & 6173              & $<$0.3           & [OIII]~5007  & $\lesssim$1.4 \\
\hline
\end{tabular}
\label{tab:emission}
\end{center}
\raggedright
\footnotesize \textit{Notes}. Column~2: Redshift; Column~3: Barycenter of the detected line; Column~4: Measured EW of the line; Column~5: Line identification; Column~6: Line luminosity.\\
(*) The [O~II] emission line is detected at the very blue edge of the spectral range and a fraction of the signal is lost.\\
\end{table*}

\setcounter{table}{4}
\begin{table*}
\begin{center}
\caption{Equivalent width of the main host galaxy absorption lines.}\label{tab:table_abs}
\begin{tabular}{llllll}
\hline
Object Name  &   Redshift  & Ca II (H,K)  &   G-band &   Mg I 5175 & Na I 5892  \\ 
             &  &  ($\textrm{\AA}$) &  ($\textrm{\AA}$) &  ($\textrm{\AA}$)  & ($\textrm{\AA}$)  \\ 
\hline 
 4FGL~J0339.2$-$1736 & 0.0655 & -            &  -          &  2.6$\pm$0.2  & -          \\
 4FGL~J0946.2+0104   & 0.5768 & 7.5$\pm$0.3  & 3.8$\pm$0.3 &  -            & -          \\ 
 4FGL~J1439.5$-$2525 & 0.161  &  -           & 4.9$\pm$0.5 & 4.0$\pm$0.3   & 2.8$\pm$0.2\\
 4FGL~J1507.3$-$3710 & 0.239  & 4.7$\pm$0.2  & 2.4$\pm$0.2 & 2.0$\pm$0.2   & -          \\ 
 4FGL~J1528.4+2004   & 0.64   & 2.0$\pm$0.4  & -           & -             &            \\ 
 4FGL~J1702.2+2642   & 0.3197 &0.30$\pm$0.05 & -           & -             &            \\ 
 4FGL~J1808.2+3500   & 0.282  & 0.34$\pm$0.1 & -           & -             &            \\
 4FGL~J2030.9+1935   & 0.3662 & 0.90$\pm$0.05& -           & -             &            \\
 4FGL~J2133.1+2529c  & 0.295  & 6.5$\pm$0.4  & 1.3$\pm$0.4 &  2.5$\pm$0.4  & -          \\
 4FGL~J2350.6$-$3005 & 0.233  & 6.2$\pm$0.4  & 2.2$\pm$0.4 & 2.8$\pm$0.2  &  3.6$\pm$0.2 \\
\hline
\end{tabular}
\end{center}
\raggedright
\footnotesize Column~1: Name of the target; Column~2: Redshift; Column~3: EW of the doublet CaII 3934, 3968 \AA; Column~4: EW of G-band 4304 \AA; Column~5: EW of Mg I 5175 \AA~blend; Column~6: EW of Na I 5892 \AA. 
EW measurements and errors follow the procedure outlined in Section 4 for the emission lines. \\ 
\end{table*}

\begin{table*}
 \begin{center}
 \caption{Absolute magnitudes of the host galaxies and black hole mass estimates.} \label{tab:abs_mag}
\begin{tabular}{llllc}
 \hline
 Object Name         & $z$ & N/H & $M(R)$  &  log$(M_{\rm BH}/M_{\odot}$) \\
\hline 
4FGL~J0339.2$-$1736$^*$  & 0.0655  & 0.5   &  $-$22.6  &    8.7  \\
4FGL~J0946.2+0104$^*$    & 0.5768  & 11    &  $-$23.0  &    8.9  \\
4FGL~J1439.5$-$2525$^*$  & 0.161   & 0.25  &  $-$22.8  &    8.8  \\
4FGL~J1507.3$-$3710$^*$  & 0.239   & 3     &  $-$22.3  &    8.6  \\
4FGL~J1528.4+2004$^*$    & 0.64    & 11    &  $-$22.1  &    8.4  \\
4FGL~J1702.2+2642        & 0.3197  & 11    &  $-$21.9  &    8.3  \\
4FGL~J1808.2+3500        & 0.282   & 8     &  $-$22.6  &    8.7  \\
4FGL~J2030.9+1935        & 0.3662  & 14    &  $-$22.6  &    8.7  \\
4FGL~J2133.1+2529c       & 0.295   & 2.2   &  $-$22.8 &     8.8  \\ 
4FGL~J2350.6$-$3005$^*$  & 0.233   & 1.0   &  $-$22.7  &    8.7  \\
  \hline
   \end{tabular}
  \end{center}
  \raggedright
\footnotesize \textit{Notes}. Column~2: Redshift; Column~3: N/H estimated through the slit, i.e. the ratio of the fluxes of the nucleus 
to the host galaxy at 6000~\AA; Column~3: Absolute magnitude of the host galaxy in the R filter; Column~5: Logarithm of $M_{\rm BH}$ in 
solar units derived through the $M_{\rm BH}$-M(R) relationship. The absolute magnitude of the host galaxy is estimated by the decomposition of the
optical spectra as the sum of a power law and an elliptical galaxy
template. The associated error is dominated by the dispersion ($\sim$0.45 dex) of the $M_{\rm BH}$-M(R) relation \citep{labita2007}.\\
For the sources marked with the asterisk (*), we performed the decomposition of the acquisition image in order to compare the host galaxy magnitude obtained with the imaging and with the spectral decomposition plus the slit loss correction (see text in Sec 4.)
\end{table*}


\section{Notes on individual sources} 
\label{sec:notes}

\begin{itemize}
\item[] \textbf{4FGL~J0258.1+2030}: 
This source was proposed as a BLL in the BZCAT catalogue \citep{BZCAT} and is quoted as a plausible association with two neutrino events: IceCube-191231A \citep{GCN26620} and IceCube-211125A \citep{GCN31126}.
The spectrum is heavily reddened (E(B-V)=1.1) and appears featureless (see Fig.~\ref{fig:spectra}). 
The continuum is described by a power-law shape ($\alpha\sim$0.8), which confirms the BLL classification of the target. 
No emission lines with EW$>$0.8~$\textrm{\AA}$ are found and from the absence of the absorption lines of the host galaxy, we can set a redshift lower limit z~$>$~0.3, consistent with the previous result of \citet{shaw2013}. \\

\item[] \textbf{4FGL~J0339.2$-$1736}: 
In our spectrum we clearly detect absorption lines (H$_{\beta}$, Mg~I, Ca+Fe, and H$_{\alpha}$) from the host galaxy at z~=~0.0656 confirming the redshift proposed by \citet{jones2009} from the 6dF Survey spectrum based on Ca~II absorption. 
In addition we also detect a faint emission line at 7015~$\textrm{\AA}$ (EW$\sim$1.15~$\textrm{\AA}$) attributed to [N~II]~6584 at the same redshift. 
The decomposition of the spectrum into an elliptical galaxy and nucleus described by a power-law ($\alpha$=0.9) yields an N/H~=~0.5 (see Fig.~\ref{fig:decomposition}). \\

\item[] \textbf{4FGL~J0545.0+0613c}: 
The optical spectrum exhibits two prominent and broad emission lines at $\sim$5030~$\textrm{\AA}$ (EW$\sim$70~$\textrm{\AA}$) and $\sim$7400~$\textrm{\AA}$ (EW$\sim$17.5~$\textrm{\AA}$) and a weaker emission line at $\sim$5485~$\textrm{\AA}$.  
We identify these lines as C~III] 1909 (FWHM~=~125~$\textrm{\AA}$) and Mg~II 2796,2803 (FWHM~=~60~$\textrm{\AA}$) at z~=~1.645. 
In addition, a broad absorption feature (EW$\sim$20~$\textrm{\AA}$) is present on the blue side of C~III] emission line.
The spectral characteristics are typical of low ionization broad absorption lines (LoBAL) quasars \citep{chen2022}.
\\

\item[] \textbf{4FGL~J0658.6+0636}: 
This source is classified as blazar candidate of uncertain type (BCU) in the 4FGL-DR3 catalogue and with a possible detection up to very high energy (E$>$100 GeV) \citep[Atel\#14200, ][]{ATel14200}.
It was proposed as a plausible association to the neutrino event IceCube 201114A \citep{GCN28887}. 
This object is significatly reddened (E(B-V)~=~0.22). 
The optical spectrum is featureless and well described by a power law ($\alpha\sim$1.3) typical of a BLL. 
We set an EW limit $<$0.45~$\textrm{\AA}$ for any emission/absorption lines. 
From the absorption line limit we can set a redshift lower limit z~$>$~0.5 assuming a standard elliptical host galaxy. 
No previous optical spectra are found in literature. \\

\item[] \textbf{4FGL~J0946.2+0104}:
A noisy spectrum of the source is obtained by the SDSS survey where the Ca~II doublet is detected at z~=~0.577. 
In our better quality spectrum (S/N$\sim$80), in addition to Ca~II, we clearly detect the G-band absorption line yielding a redshift of z~=~0.5768 (see Fig.~\ref{fig:spectra}).
The spectrum is dominated by a non-thermal component (N/H = 11: see Fig.~\ref{fig:decomposition}).
\\

\item[] \textbf{4FGL~J1003.4+0205}: 
The optical spectrum  secured by SDSS is featureless, although the automatic procedure proposed z~=~1.4.
Our high quality (S/N$\sim$190) spectrum obtained at the VLT appears featureless and well described by a power law with $\alpha\sim$1.4. No emission/absorption
features are apparent at level of EW~=~0.20~$\textrm{\AA}$ and we can set a redshift lower limit $>$0.75 from the non-detection of the absorption lines of the host galaxy. 
We also obtained another spectrum at GTC that covers better the blue part of the spectral range down to 3900$\textrm{\AA}$. At $\sim$4115$\textrm{\AA}$ we can detect an intervening absorption system due to Mg~II~2796,2803 (see Fig.~\ref{fig:closeup}) allowing us to set a spectroscopic redshift lower limit $\geq 0.469$. 
In the following we conservatively adopt this value, as
it requires no assumption on the presence of emission/absorption features. \\

\item[] \textbf{4FGL~J1055.7$-$1807}: 
In the optical spectrum (S/N$\sim$80) there are two absorption doublets at 4774, 4801~$\textrm{\AA}$ and 5163, 5176~$\textrm{\AA}$, which are attributed to intervening Fe~II~2586,2600 and Mg~II~2796,2803 systems, respectively, yielding a robust spectroscopic lower limit $\geq$~0.8465.
The spectral shape is well described by a power law ($\alpha$~=~0.4) confirming the BLL classification.
No other spectra are found in the literature. 
\\

\item[] \textbf{4FGL~J1124.0+2045}: 
In our spectrum (S/N$\sim$100), well described by a power law with $\alpha\sim$1.40, we detect a faint absorption doublet system (EW$\sim$0.6~$\textrm{\AA}$) at 4419,4430~$\textrm{\AA}$ that is attributed to intervening Mg~II~2796,2803 cold gas, yielding a spectroscopic redshift lower limit $\geq$0.5805. 
The same line system appears in the SDSS spectrum, but it was not previously identified.
\\

\item[] \textbf{4FGL~J1124.9+2143}: 
Neither absorption nor emission features are present in our
good spectrum (S/N$\sim$120) that is well described by a power law ($\alpha\sim$1.25) typical of BLL.
From the minimum EW~=~0.20~$\textrm{\AA}$ of absorption lines of the host galaxy, we set a redshift lower limit $>$~0.6.
There is an SDSS spectrum of the source that appears featureless. 
\\

\item[] \textbf{4FGL~J1314.7+2348}: 
Optical spectra reported in \citet{massaro2014,paggi2014, shaw2013} and in the SDSS database are featureless, although \cite{massaro2014} quoted an SDSS uncertain redshift of 0.15, which we used in Paper II. 
Also our very high quality (S/N$\sim$300) spectrum does not exhibit emission/absorption lines and shows the typical BLL power law shape ($\alpha\sim$1.1)
From the estimated minimum EW~=~0.10~$\textrm{\AA}$, we set a redshift lower limit $>$0.50. 
\\

\item[] \textbf{4FGL~J1321.9+3219}: 
We obtained a high quality S/N$\sim$180 spectrum from which 
we clearly detect an intervening absorption system at 4243,4254 \AA~that is identified with Mg~II~2796,2803 at z~=~0.5175.
In addition an emission line at 6755~$\textrm{\AA}$ identified as [O~II]~3727 yields  z~=~0.8126. 
Note that based on a lower S/N featureless SDSS spectrum and the overall spectral energy distribution (SED) we proposed z$\sim$0.4 in Paper II. \\

\item[] \textbf{4FGL~J1439.5$-$2525}: 
No emission lines are found in our spectrum. We detect absorption lines due to the host galaxy, confirming the redshift of 0.161 reported in \citet{desai2019}. 
From the spectral decomposition in Fig.~\ref{fig:decomposition}, the contribution of the nucleus is very marginal with N/H$\sim$0.25.
\\

\item[] \textbf{4FGL~J1507.3$-$3710}: 
We obtain a spectrum with S/N$\sim$100 which exhibits clear absorption lines attributed to the host galaxy (Ca~II~3934,3968, G-band~4305, Mg~I~5175, and Na I 5892) and emission lines at 6144,6203~$\textrm{\AA}$ attributed to [O~III]~4959,5007, yielding z~=~0.239. In addition we detect very faint emission lines due to [N~II] and [S~II]. 
The [O~II]~ emission line is detected at the very blue edge of the spectral range and a fraction of the line is lost.
No previous spectra are available in literature.
\\

\item[] \textbf{4FGL~J1528.4+2004}: 
This $\gamma$-ray source is unassociated with a lower energy counterpart and unclassified in the 4FGL-DR3 catalogue. 
However, the source 3HSP~J152835.8+200420 \citep{3HSP}, located well within the 0.005 square degrees error ellipse of the \textit{Fermi}-LAT source, is the likely counterpart since its $\gamma$-ray flux and spectral shape are perfectly consistent with the SED of an HSP blazar \citep[see also][]{Paiano2017sedtemp}.
Our LBT and GTC spectra are the first ones obtained for the optical counterpart.
Our LBT spectrum is characterized by a power-law continuum with  $\alpha$=1  superimposed on a significant contribution of the host galaxy in the redder part of the spectrum (see Fig.~\ref{fig:decomposition}), yielding N/H$\sim$11. 
This confirms the BLL classification for the source.
We clearly detect an emission line (EW$\sim$2.80~$\textrm{\AA}$) at 6110~$\textrm{\AA}$ attributed to [O~II]~3727.
The same line is found in a second spectrum taken at the GTC (see the closeup around the line in Fig.~\ref{fig:closeup}).
The redshift of the source is z~=~0.64.
Other three weak absorption features are detected due to the Ca~II doublet and the G-band from the stellar component of the host galaxy at the same redshift. 
\\

\item[] \textbf{4FGL~J1702.2+2642}: 
This source was proposed as a plausible association to the neutrino event IceCube 200530A (\citet{GCN27865} and \citet{GCN27879}).
Our high quality (S/N$\sim$190) optical spectrum exhibits faint emission lines attributed to [O~II]~3727 (at 4918~$\textrm{\AA}$, EW~=~0.55~$\textrm{\AA}$) and [OIII]~5007 (at 6608~$\textrm{\AA}$, EW~=~0.40~$\textrm{\AA}$), together with stellar absorption lines of its host galaxy (CaII~(3934,3968) at 5191,5237~$\textrm{\AA}$ and marginally the G-band~4304 at 5681~$\textrm{\AA}$), 
yielding z~=~0.3197. 
From the spectral decomposition (see Fig.~\ref{fig:decomposition}) of the two components (nucleus and host galaxy), we get N/H = 11.
No previous spectra are available in literature.
\\

\item[] \textbf{4FGL~J1808.2+3500}: 
No spectra are present in the literature for the source.
We obtained a S/N$\sim$250 optical spectrum at the LBT that exhibits emission lines at 4778, 6419, and 8440 ~$\textrm{\AA}$ attributed to [O~II]~3727, [O~III]~5007, and [N~II]~6583 respectively (see Fig.~\ref{fig:spectra} and Table~\ref{tab:emission}). 
In addition we found weak absorption lines of the Ca~II doublet at 5043,5088~$\textrm{\AA}$.
The redshift of the source is z=0.282.
\\

\item[] \textbf{4FGL~J2030.9+1935}: 
In our good (S/N$\sim$150) optical spectrum, showing the typical BLL non-thermal continuum, we found one emission line at 5092~$\textrm{\AA}$ (EW$\sim$0.50~$\textrm{\AA}$) attributed to [O~II] and another very weak (EW$\sim$0.2~$\textrm{\AA}$) emission feature at 6840~$\textrm{\AA}$ consistent with [O~III], yielding z~=~0.3662.
The G-band is overlapped to the Galactic Na~I absorption line.
From the spectral decomposition analysis (see Fig.~\ref{fig:decomposition}) we get N/H~=~14. In addition, at the same redshift we detect the absorption Ca~II doublet due to stellar population of its host galaxy. 
However, from the analysis of the GTC acquisition image, we are not able to detect the host galaxy ($M(R) > -22.6$).
A spectrum of the source was published by \citet{kasai2023} after we performed our observation and our redshift is compatible with theirs. \\

\item[] \textbf{4FGL~J2030.5+2235}:
This $\gamma$-ray source is unassociated in the 4FGL-DR3 catalogue. The likely association with the $\gamma$-ray detection is the source 3HSP~J203031.6+223439, well located inside the \textit{Fermi}-LAT error ellipse and with a broad-band SED typical of an HSP blazar.
The optical spectrum (S/N$\sim$40), obtained at LBT, is featureless and well described by a power law continuum ($\alpha$ $\sim$ 1.35). Based on the absence of host galaxy lines we can set a redshift lower limit $>$0.5.\\

\item[] \textbf{4FGL~J2133.1+2529c}: 
Our optical spectrum clearly shows absorption lines due to the elliptical galaxy stellar population, in particular the Ca~II~3934,3968 doublet, G-band~4305, and Mg~I~5157 at z~=~0.295.
This value is in agreement with that reported by \citet{massaro2015} and adopted by \cite{3HSP}.
From our decomposition analysis (see Fig.~\ref{fig:decomposition}), we get N/H$\sim$2 and $\alpha$~=~0.8 for the nuclear power-law component. \\

\item[] \textbf{4FGL~J2350.6$-$3005}: 
The optical spectrum displays the typical absorption lines (Ca~II, G-band, 
H$_{\beta}$ and Mg~I, Ca+Fe, Na~I) due to the old stellar population of the
host galaxy at z = 0.233. The decomposition analysis (Fig.~\ref{fig:decomposition}) 
shows the presence of the host galaxy component with a moderate non-thermal nucleus emission (N/H$\sim$1) described by a power law with index $\alpha$~=~0.7.
\\

\item[] \textbf{4FGL~J2358.1$-$2853}: 
The spectrum exhibits a non-thermal continuum described by a power law ($\alpha\sim$0.20).
We detect a strong absorption doublet (EW$\sim$10~$\textrm{\AA}$) at 7106, 7124~$\textrm{\AA}$ that corresponds to a Mg~II~2796,2803 intervening absorption systems at z~=~1.5415 (see Fig.~\ref{fig:spectra}), with other two absorption features found at 6054 and 6608~$\textrm{\AA}$ due to Fe~II at the same redshift.
Moreover, we find three additional absorption systems at 5270~$\textrm{\AA}$, $\sim$5730~$\textrm{\AA}$, and $\sim$6200~$\textrm{\AA}$ corresponding to Fe~II and Mg~II intervening at z~=~1.2123. 

\end{itemize}

\section{Source characterization}\label{sec:characterisation}

We now characterize  the new sources belonging
to the G20 sample using the data discussed above
and in Paper II, Section 3 (where details on the radio and $\gamma$-ray data and \nup\ derivation can be found). The extra sources are discussed in Section \ref{sec:extra}.
One of the issues we want to address is that of the so-called ``masquerading'' BLLs. \cite{Padovani_2019} showed that TXS\,0506+056, the first plausible non-stellar neutrino source is, despite appearances, {\it not} a blazar of the BL Lac type but instead a masquerading BLL.
This class was introduced by \cite{giommibsv1,giommibsv2} (see also
\citealt{Ghisellini_2011}) and includes sources which are in reality Flat Spectrum Radio Quasars 
(FSRQs) whose emission lines are washed out by a very bright, Doppler-boosted jet
continuum, unlike ``real'' BLLs, which are {\it intrinsically}
weak-lined objects. This is extremely important for two reasons: (1) ``real'' 
BLLs and FSRQs turn out to belong to two very different physical classes, 
i.e., objects without and with high-excitation emission lines in their optical 
spectra, referred to as low-excitation (LEGs) and high-excitation galaxies (HEGs),
respectively \cite[e.g.][and references therein]{Padovani_2017}; (2) 
masquerading BLLs, being HEGs, benefit from several radiation fields 
external to the jet (i.e., the accretion disc, photons reprocessed in the 
broad-line region (BLR) or from the dusty torus), which, by providing more 
targets for the protons, might enhance neutrino production as compared to LEGs. 

\cite{Padovani_2019} and Paper II, to which we refer the reader for more 
details, used the following four parameters for this classification, in decreasing order of relevance: (1) location on the radio power 
-- [\ion{O}{II}] emission line power, $P_{\rm 1.4GHz}$ -- $L_{\rm [\ion{O}{II}]}$, diagram, which defines the locus of jetted (radio-loud) quasars; (2) a radio power $P_{\rm 1.4GHz} > 10^{26}$ W Hz$^{-1}$, since HEGs become the dominant population in the radio sky above this value; (3) an Eddington ratio, i.e, the ratio between the (accretion-related) observed luminosity, $L_{\rm acc}$, and the Eddington luminosity\footnote{The Eddington luminosity is 
$L_{\rm Edd} = 1.26 \times 10^{46}~(M/10^8 \rm M_{\odot})$ erg s$^{-1}$, where 
$\rm M_{\odot}$ is one solar mass.}, $L_{\rm acc}/L_{\rm Edd} \gtrsim 0.01$, which is typical of HEGs \citep{Padovani_2017}; (4) a $\gamma$-ray Eddington ratio $L_{\gamma}/L_{\rm Edd} \gtrsim 0.1$, where $L_{\gamma}$ is the rest-frame, k-corrected, $\gamma$-ray power between 0.1 and 100 GeV, derived by integrating the {\it Fermi} best-fit spectra, together with  $1\,\sigma$ statistical uncertainties, as described in Sect. 3.3 of Paper II. 
The derivation of the thermal, accretion-related bolometric luminosity is not easy for BLLs and it involves $L_{\rm [\ion{O}{II}]}$ and $L_{\rm [\ion{O}{III}]}$ (see \citealt{Punsly_2011}, \citealt{Padovani_2019} and Paper II). 
For two sources (down from five in Paper II)  
for which we have no line information we got an upper limit on $L_{\rm disc}$ from \cite{Paliya_2021}, which
implies $L_{\rm acc} < 2 \times L_{\rm disc}$. 

We anticipate that no object is classified as masquerading based only on $L_{\rm acc}/L_{\rm Edd}$ and/or $L_{\gamma}/L_{\rm Edd}$, which are the least certain parameters given their dependence on the accretion power, 
$L_{\rm acc}$, and $M_{\rm BH}$ (see Section \ref{sect:host_mass}). Moreover, the LEGs -- HEGs dividing lines for these two parameters are somewhat blurry and not clear-cut. In practice, then, the Eddington ratios are used mostly as a consistency check.  

\begin{table*}
\caption{New and updated source properties.}
 \begin{center}
 \begin{tabular}{lllrrrrrrrr}
   \hline
    Name & 4FGL name & \magg$z$~~& \nup & $P_{\rm 1.4GHz}$ & $L_{\rm [\ion{O}{II}]}$ & $L_{\rm [\ion{O}{III}]}$ & $M_{\rm BH}$ & $L_{\rm acc}/L_{\rm Edd}$ & $L_{\gamma}$~~ & $L_{\gamma}/L_{\rm Edd}$\\
         &  &  & [Hz] & [W Hz$^{-1}$] & [erg s$^{-1}$] & [erg s$^{-1}$] & [$M_{\odot}$] &  & ~~~~[erg s$^{-1}$] & \\
      \hline
 3HSP J033913.7$-$173600$^*$ & 4FGL~J0339.2$-$1736 &  \magg0.0655 &  15.6 &  24.22 &   ...~~~& <40.4 &  8.7 & <-2.2 &  43.9 &  -2.9 \\
 3HSP J094620.2$+$010451$^*$ & 4FGL~J0946.2+0104 & \magg0.5768 & >18.2 &  25.11 & <40.6 & <40.5 &  8.9 & <-2.0 & 45.7 &  -1.3 \\
 3HSP~J100326.6+020455 & 4FGL~J1003.4+0205 & >0.469 &  16.0 & >24.53 &   ...~~~  &   ...~~~  &  ...~~~&   ...~~~  & 
 >45.1 & >-1.8 \\
 VOU J105603$-$180929   & 4FGL~J1055.7$-$1807 & >0.8465 &  14.4 & >25.36 &   ...~~~  &   ...~~~  &  ...~~~&   ...~~~  & >45.3 & >-1.6 \\
 3HSP~J112405.3+204553 & 4FGL~J1124.0+2045 & >0.5805 &  15.5 & >24.89 &   ...~~~  &   ...~~~  &  ...~~~&   ...~~~  & >45.4 & >-1.5 \\
 3HSP~J112503.6+214300 & 4FGL~J1124.9+2143 & >0.6 &  16.0 & >24.88 &   ...~~~  &   ...~~~  &  ...~~~&   ...~~~ & >45.2 & >-1.7 \\ 
 5BZB J1314+2348$^*$ & 4FGL~J1314.7+2348 & >0.5 & >14.2 & >26.07 &   ...~~~  &   ...~~~  &  ...~~~&   ...~~~  & >46.0 & >-0.9 \\
 5BZB J1322+3216$^*$ & 4FGL~J1321.9+3219 &  \magg0.8126 &  14.8 &  26.83 &  41.8 &   ...~~~&  ...~~~&  -1.3 &  45.7 &  -1.2 \\
 VOU J143934$-$252458$^*$ & 4FGL~J1439.5$-$2525 &  \magg0.161 &  14.1 &  24.33 &   ...~~~& <40.5 & 8.8 & <-2.2 &  44.1 &  -2.8 \\
 VOU J150720$-$370902   & 4FGL~J1507.3$-$3710 &  \magg0.239 &  14.6 &  25.04 &  41.2 &  40.7 &  
 8.6 &  -1.6 & 44.7 &  -2.0 \\ 
 3HSP J152835.7+20042$^*$ & 4FGL~J1528.4+2004 &  \magg0.64 &  16.4 &  24.57 &  41.4 &   ...~~~&  8.4 &  -1.0 &  45.3 &  -1.2 \\
 CRATES~J180812+350104 & 4FGL~J1808.2+3500 &  \magg0.282 &  14.6 &  25.26 &  40.8 &  40.7 &  8.7 &  -1.8 &  45.1 &  -1.7 \\  
 3HSP J203031.6+223439 & 4FGL~J2030.5+2235 & >0.5 &  16.4 & >24.53 &   ...~~~  &   ...~~~  &  ...~~~&   ...~~~&  >45.1 & >-1.8 \\
 3HSP J203057.1+193612 & 4FGL~J2030.9+1935 &  \magg0.3662 &  15.9 &  25.28 &  40.9 &  40.4 &   8.7 &  -1.8 & 45.4 &  -1.4 \\ 
 3HSP J213314.3+252859$^*$ & 4FGL~J2133.1+2529 &  \magg0.295 &  15.3 &  24.93 &  <41.3 &  <40.9 &  8.8 &  <-1.8 & 44.9 &  -2.0 \\
 3HSP J235034.3$-$300604$^*$ & 4FGL~J2350.6$-$3005 &  \magg0.233 &  15.8 &  24.70 &   ...~~~&  <40.4 &  8.7 &  <-2.2 &  44.8 &  -2.0 \\
 CRATES~J235815$-$285341 & 4FGL~J2358.1$-$2853 & >1.5425 &  14.4 & >27.01 &   ...~~~&   ...~~~  &  ...~~~&   ...~~~& >46.7 & >-0.2 \\
\hline
   \end{tabular}
  \end{center}
\footnotesize \textit{Notes.} Column 1: name; column 2 {\it Fermi-}4FGL name;
column 3: redshift; column 4: rest-frame \nup; column 5: $P_{\rm 1.4GHz}$;
column 6: $L_{\rm [\ion{O}{II}]}$; column 7: $L_{\rm [\ion{O}{III}]}$; 
column 8: $M_{\rm BH}$; column 9: $L_{\rm acc}/L_{\rm Edd}$; column 10: $L_{\gamma}$; column 11: $L_{\gamma}/L_{\rm Edd}$.
All values, apart from redshift, are in logarithmic scale. An asterisk denotes sources, which have updated/improved data (redshifts, line powers, masses) as
compared to Tab. 1 of Paper II. 
\label{tab:sample}
\end{table*}

Table \ref{tab:sample} gives the main properties of the new objects and
of the sources for which we have better or new estimates for some parameters (the
latter marked by an asterisk). Namely, name
(column 1), {\it Fermi-}4FGL  name (column 2), redshift (column 3), 
rest-frame \nup~(column 4), $P_{\rm 1.4GHz}$ (column
5), $L_{\rm [\ion{O}{II}]}$ and $L_{\rm [\ion{O}{III}]}$ (columns 6 and 7),
$M_{\rm BH}$ (column 8), $L_{\rm acc}/L_{\rm Edd}$,
$L_{\gamma}$, and $L_{\gamma}/L_{\rm Edd}$
(columns 9, 10, and 11). When $M_{\rm BH}$ is missing we assumed a value of 
$6.3 \times 10^8
M_{\odot}$ (e.g., Paper II and references therein). Note that, as mentioned in Paper II, 5BZB J1322+3216 
(a.k.a. 4C+32.43) has a very steep ($\nu^{-0.8}$ between 130
MHz and 4.8 GHz) radio spectrum. Its large radio power is therefore also 
due to extended emission and not only to a strong core, unlike the other sources, 
which are characterized by a flat radio spectrum, and we made a correction for this\footnote{This was done by using the peak flux at 1.4 GHz from the FIRST survey \citep{White_1997}, which is $\sim 40$ per cent smaller than the NVSS value \citep{Condon_1998}. We remark that this
source appears to be anomalous as a blazar based on its steep radio spectrum and 
compact radio image \citep{Gordon2020}. Still, it is strongly variable, 
namely by a factor of 2 in the 3.4 $\mu$m band in about six months, 
based on NEOWISE data \citep{Mainzer2014}.}.

Table \ref{tab:masq} lists all masquerading BLLs in the G20's sample, including 
for completeness also those discussed in Paper II, with the conditions they fulfil. 
All sources satisfy at least two criteria. 

\begin{table*}
\caption{Whole sample masquerading BLL properties.}
 \begin{center}
 \begin{tabular}{lclcccc}
   \hline
    Name & IceCube Name & ~~~~~$z$ & $P_{\rm 1.4GHz} - L_{\rm [\ion{O}{II}]}$ & $P_{\rm 1.4GHz}$ & 
    $L_{\rm acc}/L_{\rm Edd}$ & $L_{\gamma}/L_{\rm Edd}$ \\
      \hline
 3HSP J010326.0+15262 & IC160331A &  \magg0.2461 & \checkmark & I & \checkmark & \xmark \\
 5BZU J0158+0101      & IC090813A &  \magg0.4537 & \checkmark & I & \checkmark & \checkmark \\
 TXS 0506+056         & IC170922A &  \magg0.3365 & \checkmark & \checkmark & \checkmark & \checkmark \\
 CRATES~J052526$-$201054 & IC150428A &  \magg0.0913 & \checkmark &  I & \checkmark & \xmark \\
 GB6 J1040+0617       & IC141209A &  \magg0.74   & \checkmark & I & \checkmark & \checkmark \\
 3HSP J111706.2+20140 & IC130408A &  \magg0.138  & \checkmark & I  & \checkmark & \xmark \\
 5BZB J1314+2348      & IC151017A & >0.5    &  --  & \checkmark &  -- & \checkmark \\
 5BZB J1322+3216      & IC120515A &  \magg0.8126 & \checkmark & \checkmark & \checkmark & \xmark \\
 3HSP J143959.4$-$23414 & IC170506A &  \magg0.309  & \checkmark &  I & \checkmark & \xmark \\
 VOU J150720$-$370902   & IC181014A &  \magg0.239  & \checkmark & I  & \checkmark & \xmark \\
 3HSP J152835.7+20042 & IC110521A &  \magg0.64   & \checkmark & I & \checkmark & \xmark \\
 CRATES~J180812+350104 & IC110610A &  \magg0.282  & \checkmark & I  & \checkmark & \xmark \\
 3HSP J203057.1+193612 & IC100710A &  \magg0.3662 & \checkmark &  I & \checkmark & \xmark \\
 5BZB J2227+0037      & IC140114A & >1.0935 &  --  & \checkmark &  -- & \checkmark \\
 CRATES~J232625+011147 & IC160510A &  \magg1.595 & -- & \checkmark & \xmark & \checkmark \\
 CRATES~J235815$-$285341 & IC190104A & >1.5425 & --  & \checkmark & -- & \checkmark \\
  \hline
  \end{tabular}
  \end{center}
\footnotesize {\textit{Notes.} `\checkmark' implies that the condition is met, `I' that the condition is not met but this does not mean this is not a masquerading BLL,  `\xmark' that the condition is not met}, and `--' that no information is available.
 \label{tab:masq}
\end{table*}

\subsection{Statistical Analysis}\label{sec:stats}

We now present updated versions of Figs. 1 -- 3 of Paper II, which 
were used to discuss the multi-wavelength properties of the G20's sample. 
We anticipate that the figures confirm with larger statistics the 
trends highlighted in Paper II. We used the following software: 
{\tt fit} \citep{Press_1992} to fit data to a straight line (linear 
regression), with the correlation significance given by the Spearman 
test ({\tt spear}: \citealt{Press_1992}; note that exactly the same 
results were obtained using the Pearson test [{\tt pears}]). The issue
of a possible dependence of the correlations between powers on the 
common redshift dependence was dealt with by using a partial correlation
analysis \citep[e.g.][]{Padovani_1992}; {\tt ASURV} 
\citep{la92}, the Survival Analysis package, which employs the routines 
described in \cite{fei85} and \cite{iso86} and, amongst
other things, 
evaluates mean values by 
trying to deal properly with limits using the Kaplan-Meier estimator. The 
probability that two samples are drawn from the same parent population was
estimated using a Kolmogorov-Smirnov (KS) test ({\tt kstwo}: \citealt{Press_1992}) 
and the Peto-Prentice test \citep{la92} in {\tt ASURV}, which 
is the least affected by differences in the censoring patterns, which might be 
present in the two samples. We did not use {\tt ASURV} for linear regression
fits in the presence of lower limits because all its routines require binning 
and the resulting best fit turned out to depend on the chosen bin size. We have instead 
tested how the limits affected our results by increasing artificially the lower
limits on the powers (see Section \ref{sec:P14_Lgamma}).

\subsection{The radio power -- emission line diagram ($P_{\rm 1.4GHz}$ -- 
$L_{\rm [\ion{O}{II}]}$)}\label{sec:P14_LOII}

\begin{figure}
\vspace{-2.2cm}
\hspace{-0.6cm}
\includegraphics[width=0.55\textwidth]{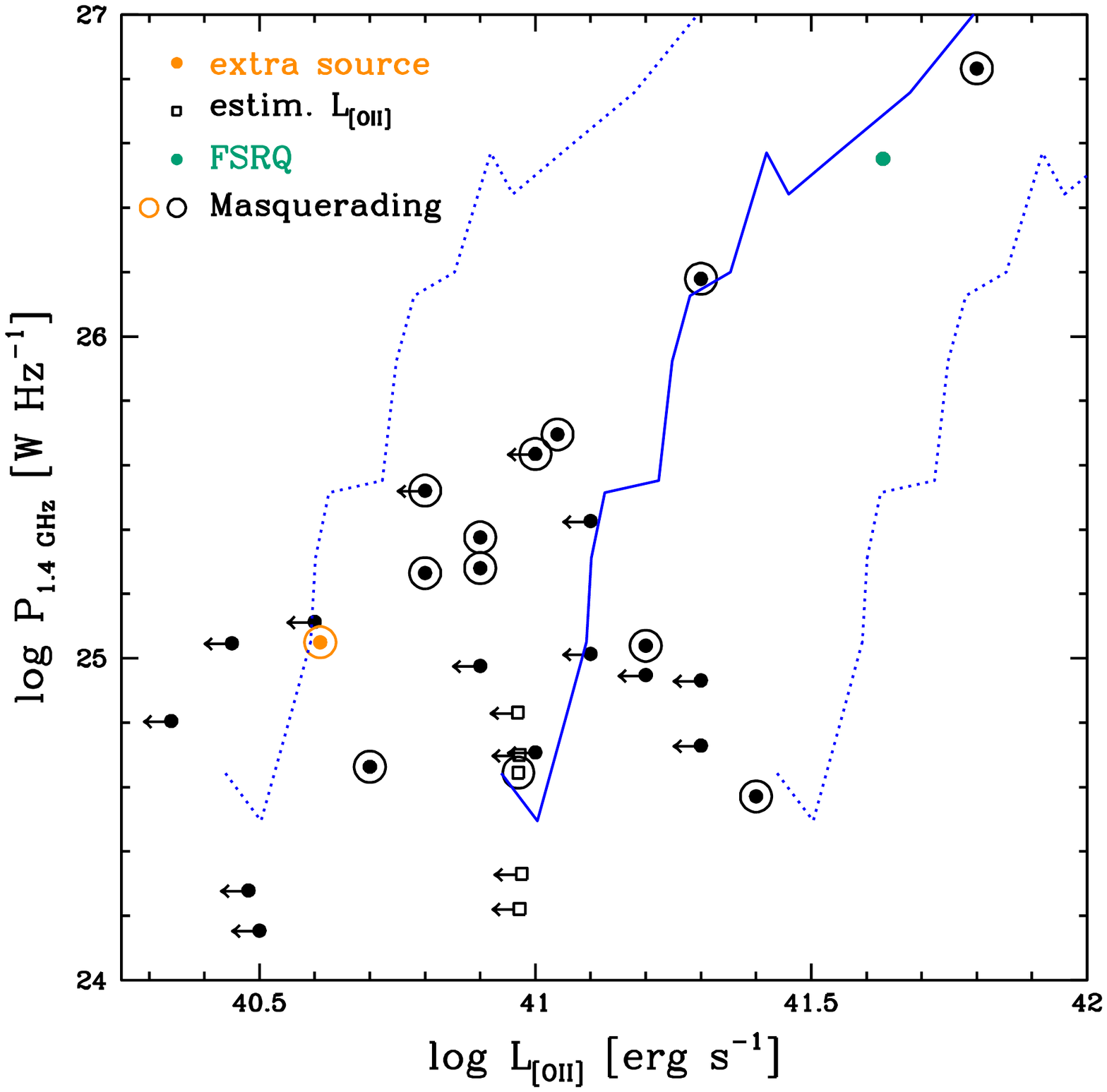}
\vspace{-3.1cm}
\caption{$P_{\rm 1.4GHz}$ vs. $L_{\rm [\ion{O}{II}]}$ for 
 the objects in our sample 
 with [O~II] (or [O~III]) information (black
  filled circles), with masquerading sources highlighted (larger empty
  circles). Sources for which $L_{\rm [\ion{O}{II}]}$ has been estimated
  from $L_{\rm [\ion{O}{III}]}$ are denoted by black empty squares. The
  green filled circle indicates the FSRQ in our sample, while the orange point is 
  one of the extra sources (Section \ref{sec:extra}). 
  The solid blue line is the locus of jetted (radio-loud) quasars, with
  the two dotted lines indicating a spread of 0.5 dex, which includes most
  of the points in Fig. 4 of \protect\cite{Kalfountzou_2012} (converted from 
  radio powers in W Hz$^{-1}$ sr$^{-1}$ and line powers in W). 
  Arrows denote upper limits on $L_{\rm [\ion{O}{II}]}$.}
\label{fig:Lr_LOII}
\end{figure}

Fig. \ref{fig:Lr_LOII} shows the location of the sources with [\ion{O}{II}]
information on the $P_{\rm 1.4GHz}$ -- $L_{\rm [\ion{O}{II}]}$ diagram. As 
done in Paper II we have also added five objects for which only the
[\ion{O}{III}] flux was available, converting $L_{\rm [\ion{O}{III}]}$ to
$L_{\rm [\ion{O}{II}]}$ using Fig. 7 of \cite{Kalfountzou_2012}, marking them differently [open squares] in the figure. 
Twelve objects\footnote{Two of these objects 
have upper limits on their $L_{\rm [\ion{O}{II}]}$, although
quite close to the locus, but were still classified as masquerading in Paper II
because they also have an [\ion{O}{III}] detection and $L_{\rm acc}/L_{\rm Edd} > 0.01$, 
with one of them also having $L_{\gamma}/L_{\rm Edd} \sim 0.8$. Note that the 
$L_{\rm [\ion{O}{II}]}$ value for VOU J150720$-$370902 
(4FGL~J1507.3$-$3710) has to be considered a lower 
limit (Section \ref{sec:notes}) but even so the source is very close to the quasar locus (see Tab. \ref{tab:sample}).} are close to the locus of jetted quasars and are therefore
``bona fide'' masquerading BLLs. All have
also $L_{\rm acc}/L_{\rm Edd} \ge 0.01$. 
CRATES J024445+132002, the FSRQ in
our sample, is also (by definition) very close to the locus. Five more
sources have quite stringent $L_{\rm [\ion{O}{II}]}$ upper limits, two more
have $L_{\rm [\ion{O}{II}]}$ upper limits plus low $P_{\rm 1.4GHz}$, while
eight more have an upper limit not too far from (or to the right of) the 
locus. However, given
that these latter sources have no other HEG-like property and no
[\ion{O}{III}] detection, and that we want to keep the selection conservative,
we are not including them with the masquerading sources (although some
of them are borderline). 

\subsection{The $\gamma$-ray power -- radio power diagram 
($L_{\gamma} - P_{\rm 1.4GHz}$)}\label{sec:P14_Lgamma}

\begin{figure}
\vspace{-1.65cm}
\hspace{-0.6cm}
\includegraphics[width=0.55\textwidth]{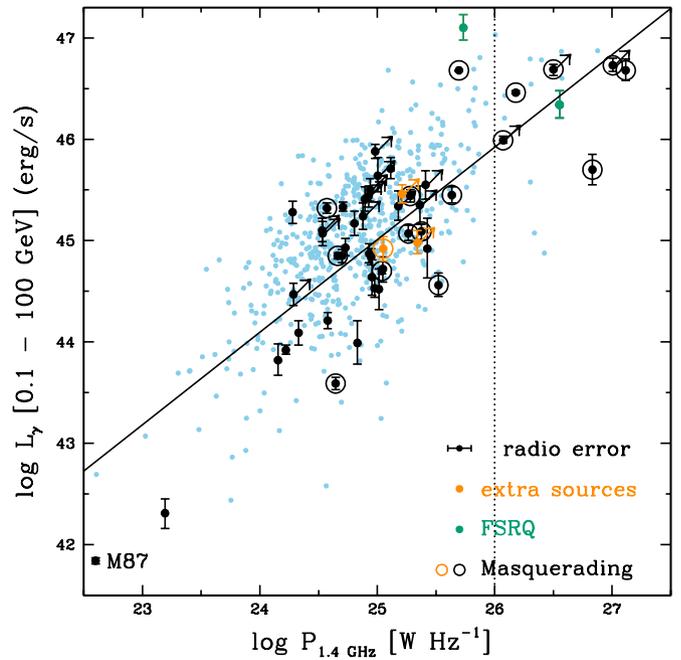}
\vspace{-3.1cm}
\caption{$L_{\gamma}$ vs. $P_{\rm 1.4GHz}$ for our sample (black filled
  circles), with masquerading sources highlighted (larger empty
  circles). The green filled circle indicates the two FSRQs (the top-left one being also an extra source), while the orange points are the extra sources (Section \ref{sec:extra}). 
  The radio power for M87, which is labelled, is derived from a time-averaged 
  VLBA core flux at 15 GHz \protect\citep{Kim_2018}. Vertical error 
  bars denote the $1\sigma$ uncertainties on the individual $\gamma$-ray powers, while the horizontal error bar represents the typical uncertainty on the radio power due to variability, based on
  \protect\cite{Richards_2011} (see text for details). The solid line is the linear best fit $L_{\gamma} 
  \propto P_{\rm 1.4GHz}^{0.90}$, while the vertical dotted line marks 
  the $10^{26}$ W Hz$^{-1}$ power above which a source is classified as 
  masquerading (see text for details). Arrows denote 
  lower limits on redshift and therefore powers. The
  small light blue points are the control sample of IBLs and HBLs (Section \ref{sec:diff_blazar}).}
\label{fig:Lr_Lgamma}
\end{figure}

Fig. \ref{fig:Lr_Lgamma} shows $L_{\gamma}$ vs. $P_{\rm 1.4GHz}$ for our
sources (black circles). 
Vertical error bars denote the $1\sigma$ uncertainties on the individual $\gamma$-ray powers, while the horizontal error bar represents the typical $1\sigma$ error for the radio band due to variability, derived using the results of 
\cite{Richards_2011} by converting the mean value of their 
intrinsic modulation index for $\gamma$-ray loud blazars to 
an uncertainty on radio power.
We also include the time-averaged very long baseline 
array (VLBA) core power at 15 GHz for M87, which, given its
substantially flat radio spectrum between $15 - 129$ GHz, should be
representative of its 1.4 GHz core power as  well 
(this source is also part of the G20 sample, as discussed in Paper II). 
Four more sources get
classified as masquerading BLLs thanks to their $P_{\rm 1.4GHz} >
10^{26}$ W Hz$^{-1}$, all of them with $L_{\gamma}/L_{\rm Edd} >
0.1$. For these objects we have no [\ion{O}{II}] or [\ion{O}{III}] information.  

Fig. \ref{fig:Lr_Lgamma} shows a very strong linear correlation between the
two powers, significant at the $> 99.99$ per cent level\footnote{We exclude
from the fit the FSRQ and M87; this has no influence on the significance or 
the slope.}, 
with $L_{\gamma} \propto
P_{\rm 1.4GHz}^{0.90\pm0.11}$. Removal of the commmon 
redshift dependence using a partial correlation 
analysis (Section \ref{sec:stats}) still gives a 99.2 per 
cent level significance. The sample includes also fifteen 
sources with lower limits on their redshifts and therefore on their
powers. Since the {\tt ASURV} routines require binning, with 
a resulting best fit, which depends on the chosen bin size (Section 
\ref{sec:stats}), we have 
tested how this affects our results by increasing artificially the lower
limits on the powers by 0.75 dex, which corresponds, for example, to a
(large) increase from $z=0.7$ to $z=1.4$. The new best fit is $L_{\gamma}
\propto P_{\rm 1.4GHz}^{1.03\pm0.09}$ with the same significance, fully
consistent with the previous slope. These slopes are slightly flatter but
still consistent with those of Paper II ($L_{\gamma} \propto
P_{\rm 1.4GHz}^{1.04\pm0.13}$ and $L_{\gamma} \propto
P_{\rm 1.4GHz}^{1.14\pm0.11}$ respectively).

The issue of how the properties
of masquerading sources might differ from those of the other sources is 
discussed in Section \ref{sec:masq_diff}. 

\subsection{The synchrotron peak -- $\gamma$-ray power diagram 
(\nup~-- $L_{\gamma}$)}\label{sec:sequence}

\begin{figure}
\vspace{-2.0cm}
\hspace{-0.6cm}
\includegraphics[width=0.55\textwidth]{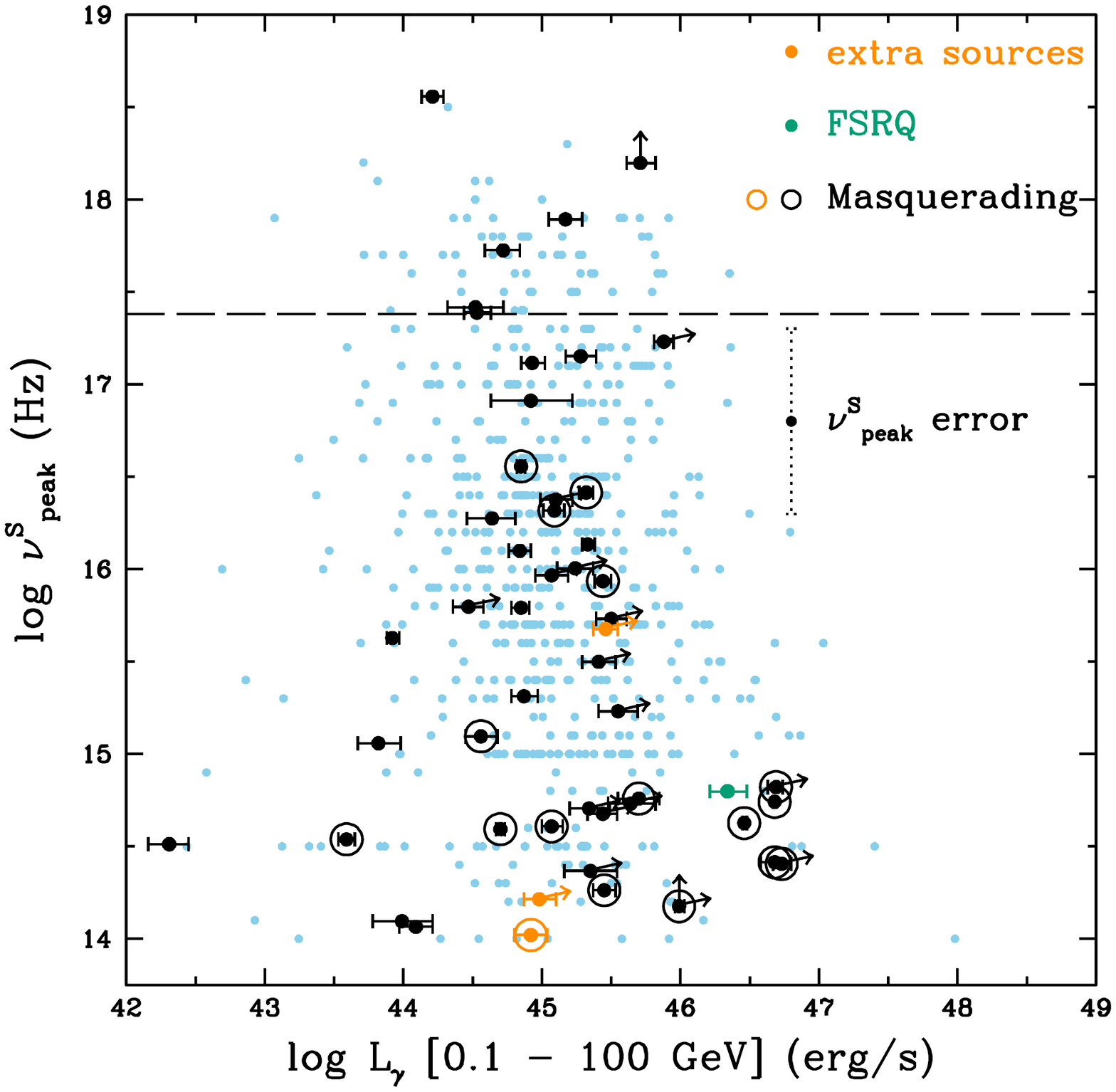}
\vspace{-3.1cm}
\caption{\nup~vs. $L_{\gamma}$ for our sample (black filled circles),
  with masquerading sources highlighted (larger empty circles). The 
  green filled circle indicates one FSRQ (as the other one has no \nup~estimate), 
  while orange points are the extra sources (Section \ref{sec:extra}). Error bars 
  denote the uncertainties. The typical \nup~uncertainty is also
  shown (vertical dotted line). Arrows denote lower limits on \nup~(vertical) and
  redshifts (diagonal), which also affect the rest-frame \nup~values.
  Sources above the dashed line are extreme blazars (defined as having \nup~$> 2 
  \times 10^{17}$ Hz). The small light blue points
  are the control sample of IBLs and HBLs (see
  Section \ref{sec:diff_blazar} for details).}
\label{fig:nu_peak_L_Gamma}
\end{figure}

Fig. \ref{fig:nu_peak_L_Gamma} plots the location of our sources (black circles) 
on the \nup~-- $L_{\gamma}$ diagram (note that by definition our sample 
includes only sources with \nup~$> 10^{14}$ Hz). IBLs cover a relatively 
broad power range, i.e. $L_{\gamma} \sim 10^{42} - 10^{47}$ erg s$^{-1}$, 
while HBLs occupy the narrower $\sim 10^{44} - 10^{46}$ 
erg s$^{-1}$ region all the way up to \nup $\sim 3 \times 10^{18}$ Hz, although 
there are also some lower limits on redshift, which means higher $L_{\gamma}$ are
very likely. Six sources have \nup~$> 1$ keV ($2.4 \times 10^{17}$ Hz) and are therefore 
classified as extreme, with one very close at \nup~$> 1.7 \times 10^{17}$ Hz, given
also its redshift lower limit. Note that masquerading BLLs appear
to have somewhat lower \nup~values; this issue is discussed in 
Section \ref{sec:masq_diff}. 

\section{The extra sources}\label{sec:extra}

Some targets that were not in the G20 list were also observed. 
These are still blazars without a redshift determination, which 
were associated to IceCube tracks after the G20 work 
was completed and which satisfy in most cases the G20 criteria,
as explained in Section \ref{sec:sample}. 

We discuss here these four sources, which are not included in the 
statistical analysis of Section \ref{sec:characterisation} nor in that of
Section \ref{sec:discussion}, with the aim of 
characterizing them and looking for other masquerading BLLs. Table \ref{tab:extra_sample} gives the main properties of the extra sources, 
where the columns are the same as in Table \ref{tab:sample}. 
In the case of NVSS J054341+062553, the FSRQ at $z = 1.6458$, we used its \ion{Mg}{II} and \ion{C}{III]} luminosities to estimate the BLR
      luminosity, $L_{\rm BLR}$, from the composite spectrum of
      \cite{van01} (we then derived $L_{\rm disc} \sim 10 \times L_{\rm
        BLR}$ assuming a standard covering factor of $\sim 10$ per cent and
      $L \sim 20 \times L_{\rm BLR}$: see \citealt{Padovani_2019}). Its $M_{\rm BH}$
was obtained by applying the virial technique to the 
\ion{Mg}{II} line using the calibration given by \cite{Paliya_2021}. 

One extra source fulfils two criteria for being a masquerading BLL, as shown in Table \ref{tab:masq_extra}.

\begin{table*}
 \caption{Extra sample properties.}
 \begin{center}
 \begin{tabular}{lllrrrrrrrr}
   \hline
    Name & 4FGL name & \magg$z$~~& \nup & $P_{\rm 1.4GHz}$ & $L_{\rm [\ion{O}{II}]}$ & $L_{\rm [\ion{O}{III}]}$ & $M_{\rm BH}$ & $L_{\rm acc}/L_{\rm Edd}$ & $L_{\gamma}$~~ & $L_{\gamma}/L_{\rm Edd}$\\
         &  &  & [Hz] & [W Hz$^{-1}$] & [erg s$^{-1}$] & [erg s$^{-1}$] & [$M_{\odot}$] &  & ~~~~[erg s$^{-1}$] & \\
    \hline
 5BZB J0258+2030        & 4FGL J0258.1+2030 & >0.3 &  14.2 & >25.34 &   ...~~~  &   ...~~~  &  ...~~~&   ...~~~& >45.0  & >-1.9 \\
 NVSS J054341+062553  & 4FGL J0545.0+0613 &  \magg1.645 &   ...~~~ &  25.73 &   ...~~~ &   ...~~~ &  9.0 &   0.0 &  47.1 & 0.0 \\
 3HSP J065845.0+063711 & 4FGL J0658.6+0636 & >0.5 &  15.7 & >25.21 &   ...~~~  &   ...~~~  &  ...~~~&   ...~~~& >45.5 & >-1.4 \\
 5BZB J1702+2643      & 4FGL J1702.2+2642 &  \magg0.3197 &  14.0 &  25.05 &  40.6 &  40.4 &  8.3 &  -1.5 & 44.9 & -1.5 \\
 \hline
\multicolumn{7}{l}\footnotesize{\textit{Notes.} All values, apart from redshift, are in logarithmic scale.}\\
  \end{tabular}
  \end{center}
 \label{tab:extra_sample}
\end{table*}

\begin{table*}
 \caption{Masquerading BLL properties: extra sample.}
 \begin{center}
 \begin{tabular}{lclcccc}
   \hline
    Name & IceCube Name & ~~~~~$z$ & $P_{\rm 1.4GHz} - L_{\rm [\ion{O}{II}]}$ & $P_{\rm 1.4GHz}$ & 
    $L_{\rm acc}/L_{\rm Edd}$ & $L_{\gamma}/L_{\rm Edd}$ \\
      \hline
 5BZB J1702+2643      & IC200530A &  0.3197 & \checkmark  & I & \checkmark  & \xmark  \\      
  \hline
  \end{tabular}
  \end{center}
\footnotesize {\textit{Notes.}  `\checkmark' implies that the condition is met, `I' that the condition is not met but this does not mean this is not a masquerading BLL, and `\xmark' that the condition is not met.}
 \label{tab:masq_extra}
\end{table*}

\section{Discussion}\label{sec:discussion}

\subsection{Are our sources different from the rest of the blazar population?}\label{sec:diff_blazar}

To check if our sources are any different from the rest of the blazar
population we use a control sample of 783 IBLs and HBLs with $|b_{\rm II}| > 
30^{\circ}$, 630 of which have redshift, detected at the 100 and 97 per 
cent level in the $\gamma$-ray and radio bands, respectively, based on the 
3HSP sample \citep{3HSP} and an IBL sample put together for this purpose 
(see Paper II for details). 

The location of these objects on the $L_{\gamma} - P_{\rm 1.4GHz}$ plane is
shown in Fig. \ref{fig:Lr_Lgamma} (small light blue points), where they appear 
to populate roughly the same region as our sample. The density of high-power
BLLs appears in fact to be larger for our sources, which is likely due to
their larger mean redshift (see below). 
Moreover, the control sample displays a very
strong linear correlation between the two powers, significant at the $>
99.99$ per cent level with $L_{\gamma} \propto P_{\rm 1.4GHz}^{0.83\pm0.04}$, 
consistent with the correlation found for our sample ($L_{\gamma} \propto P_{\rm
  1.4GHz}^{0.90\pm0.11}$: Section \ref{sec:P14_Lgamma}). 
As for the \nup~-- $L_{\gamma}$ plane, Fig. \ref{fig:nu_peak_L_Gamma} shows that the 
IBL plus HBL control sources, as is the case for Fig. \ref{fig:Lr_Lgamma}, 
appear to populate the same region as our sample. 

\begin{figure}
\vspace{-2.2cm}
\hspace{-0.6cm}
\includegraphics[width=0.55\textwidth]{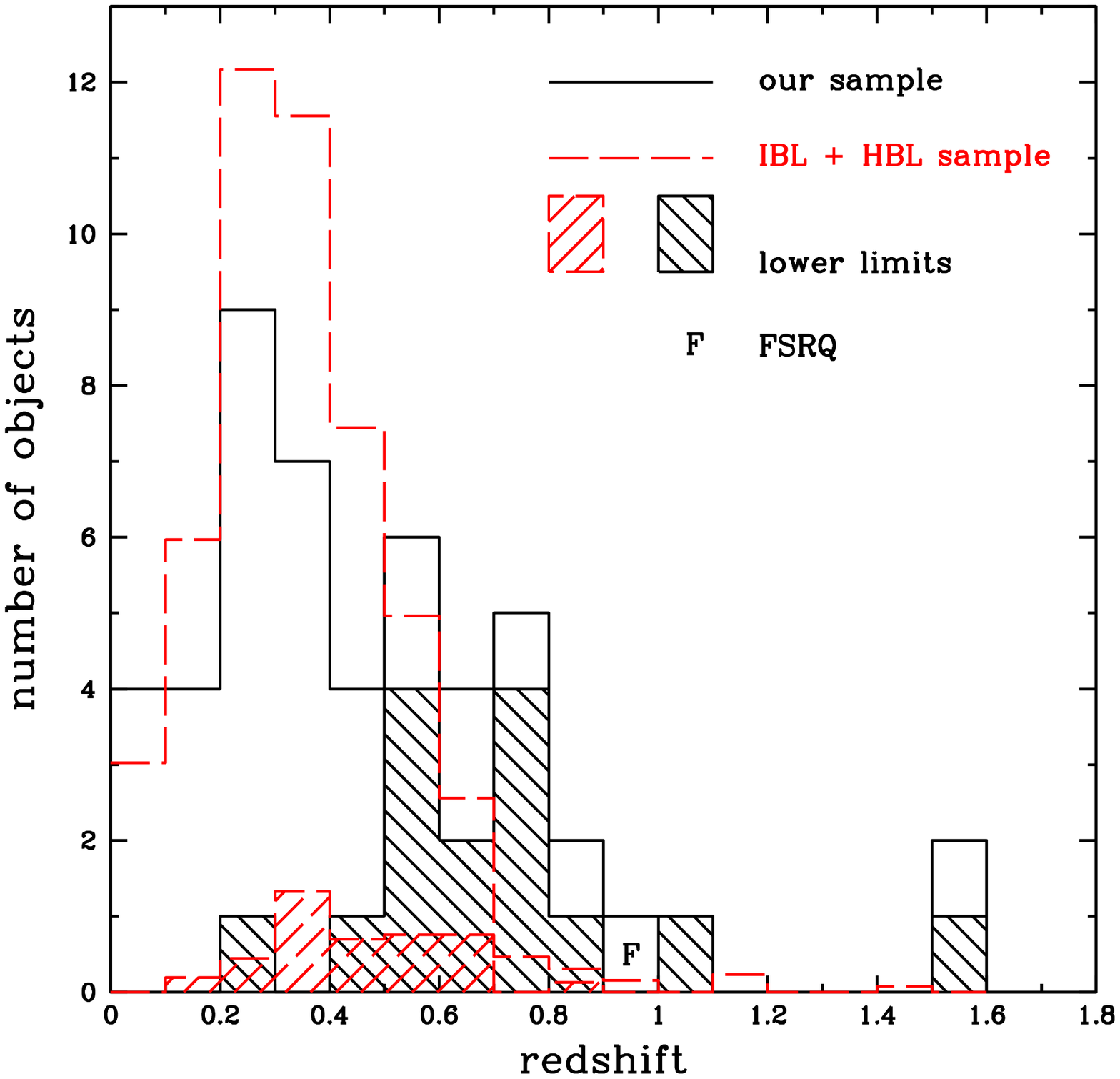}
\vspace{-3.1cm}
\caption{The redshift distribution for our sample (black solid line), with
  lower limits (denoted by the dashed areas). The FSRQ is
  denoted by an ``F''. The red long-dashed line indicates the control 
  sample of IBLs and HBLs scaled to the size of our sample.}
\label{fig:Nz}
\end{figure}

We also compared the redshift distribution, $N(z)$, of our sources
  (excluding M87 and the FSRQ), to that of the control sample, as shown
  in Fig. \ref{fig:Nz} (black solid and red long-dashed line respectively), taking into account the redshift
  lower limits using the Kaplan-Meier estimator in {\tt ASURV} (Section \ref{sec:stats}). 
  The mean redshift is $\langle z \rangle =
  0.66\pm0.09$ and $\langle z \rangle = 0.37\pm0.01$ for our sample and 
  the control sample respectively, different at the $\sim 99.8$ per cent level. The control sample is missing $\sim 20$
  per cent of the redshifts, while we have looked very carefully for even
  weak features and obtained a redshift or a lower limit for all the
  sources we took a spectrum of. Therefore, the fact that the control sample 
  mean redshift is lower than ours is to be expected. This is indeed confirmed
  by the comparison done in Paper II between our newly determined redshifts and 
  the corresponding values in the \cite{3HSP}'s sample for sources with a previous 
  redshift estimate, which showed that our redshifts were typically higher by 0.1 
  and up to $\gtrsim 0.2$. 

\subsection{What is the fraction of masquerading BLLs?}\label{sec:fract_masq}

We find sixteen masquerading BLLs (Table \ref{tab:masq}) and eleven 
sources which fulfil none of our criteria and therefore we consider ``bona fide'' 
non-masquerading objects. For twenty sources, however, we do not have the relevant
information to make a decision. Therefore, the fraction of masquerading BLLs is in the
range $34 - 77$ per cent but should be well above the lower bound given 
our conservative selection (Section \ref{sec:P14_LOII}). The corresponding range in 
Paper II was $24 - 79$ per cent, so we have increased by about 40 per cent the
value of our lower limit on this fraction.

\subsection{Do masquerading BLLs have different properties from 
non-masquerading ones?}\label{sec:masq_diff}

All sources in our sample with $L_{\gamma} \gtrsim 10^{46}$ erg s$^{-1}$ 
in Fig. \ref{fig:Lr_Lgamma} are masquerading BLLs, as are all objects 
with $P_{\rm 1.4GHz} \gtrsim 10^{25.5}$ W Hz$^{-1}$. In general, 
it appears that masquerading BLLs tend to be more powerful than 
non-masquerading ones (where here we include all 31 sources which are
not classified as masquerading). This is confirmed for the radio band, where radio 
powers are on average $\sim 9$ times larger and for which a KS test 
shows that the two samples are significantly different ($P \sim 99.98$ per 
cent). A similar result ($P \sim 98.9$ per cent) is obtained using
the Peto-Prentice test (Section \ref{sec:stats}). 
It could be argued that a high $P_{\rm 1.4GHz}$ was one of the criteria 
for selecting masquerading BLLs but not a single source has been classified 
as such only on the basis of its radio power. $L_{\gamma}$ is also on average $\sim 5$ 
times larger for masquerading BLLs but a KS and a Peto-Prentice test show 
that the two luminosity distributions are not significantly different.
The significant $P_{\rm 1.4GHz}$ difference is not surprising, as discussed by
\cite{Padovani_2019} and in Paper II, as masquerading BLLs need to have relatively 
high powers, besides having \nup $\gtrsim 10^{14}$ Hz, to be able to dilute 
the quasar-like emission lines, and the radio band is closer to the optical/UV 
one than the $\gamma$-ray band. 
Masquerading BLLs are also at higher redshift than the rest of the sample  
but not significantly so ($P \sim 18$ per cent): $\langle z \rangle = 
0.66\pm0.14$ vs. $\langle z \rangle = 0.54\pm0.06$, where again we have used 
{\tt ASURV} (Peto-Prentice and Kaplan-Meier tests). This is likely to be related to their higher powers. 

Fig. \ref{fig:nu_peak_L_Gamma} suggests that \nup~might be smaller for 
masquerading BLLs. Indeed, using 
the Kaplan-Meier and Peto-Prentice tests, given the two lower limits on
this parameter plus the redshift lower limits, which affects also
the rest-frame \nup, we derive $\langle \log$(\nup)$\rangle = 
15.21\pm0.22$ vs. $\langle \log$(\nup)$\rangle = 16.73\pm0.26$, e.g. a 1.5
dex difference, with the two distributions being different at the 
$\sim 99.8$ per cent level. As discussed in Paper II, this difference, again, 
can be explained by a selection effect.

\section{Conclusions}\label{sec:conclusions}

We have presented optical spectroscopy of 21 extragalactic sources that 
are candidates for being the astronomical counterparts of IceCube neutrino events. 
Apart from one target, which has a quasar-like spectrum, all of the 
remaining  objects have a spectral shape that is fully consistent with a 
BLL classification. We have provided a firm redshift for twelve sources 
($0.07 < z <1.6$), while for the 
others we set a lower limit on it. Ten objects have their continuum dominated 
by non-thermal emission, while in the other ten the spectrum shows the 
contribution of the elliptical host galaxy.  

We have then carefully characterized these sources to determine their 
real nature, extending to the whole G20 sample the work done in Paper II,  
to quantify the presence of masquerading BL Lacs, i.e., FSRQs in disguise 
whose emission lines are swamped by a very strong jet, and to check if these 
sources were any different from the rest of the blazar population. 
This was done by assembling a set of ancillary data and by measuring and 
estimating, in many cases for the first time, $L_{\rm [\ion{O}{II}]}$ and
$L_{\rm [\ion{O}{III}]}$,~and $M_{\rm BH}$, respectively. 

Our main conclusions in this respect are as follows:

\begin{enumerate}
    \item we do not find significant systematic differences between 
    the sources studied in this paper and other blazars of the same type, i.e. 
    IBLs and HBLs, in terms of their radio and $\gamma$-ray powers, and \nup, 
    while we do for redshift. The lack of such differences might be due to 
    our relatively small sample, given also that less than half of our sources 
    is expected to be associated with IceCube tracks. The redshift difference
    is instead due to the fact that our control sample is missing $\sim$ 20 
    per cent of the redshifts, while we have looked very carefully for even 
    weak features and obtained a redshift or a lower limit for all the sources 
    we took a spectrum of;
    \item the fraction of masquerading BL Lacs in our sample is $>$ 34 per cent and 
    possibly as high as 77 per cent; 
    \item masquerading BL Lacs turn out to be exactly as expected: more 
    powerful than the rest in the radio and $\gamma$-ray band, with a 
    smaller \nup. These are the properties, which allow them to effectively
    dilute their strong, FSRQ-like emission lines. 
\end{enumerate}

In subsequent papers we will collect all available multi-wavelength data for the 
G20 sources and put together their SEDs, and then start modelling them 
using a lepto-hadronic code to subsequently estimate the expected neutrino 
emission from each blazar.

\section*{Acknowledgments}

We acknowledge Martina Karl and Theo Glauch for estimating the $\gamma$-ray powers for the {\it Fermi} sources. 
This work is based on observations collected at the European Southern Observatory under ESO programme 106.213R, at the Gran Telescopio Canarias under the programme GTC2720B, and at the Large Binocular Telescope during the semester 2021A and 2022A.
Funding for the Sloan Digital Sky Survey IV has been provided by the Alfred P. Sloan Foundation, the U.S. Department of Energy Office of Science, and the Participating 
Institutions. SDSS-IV acknowledges support and resources from the Center for High Performance Computing  at the University of Utah. The SDSS website is www.sdss.org.

\noindent We thank the referee for her/his useful comments and suggestions that allow to improve our manuscript.

\section*{Data Availability}
The flux-calibrated and de-reddened spectra are available in our online data base ZBLLAC: \texttt{https://web.oapd.inaf.it/zbllac/}.



\appendix
\section{}\label{sec:appendix}
We show here flux calibrated and de-reddened spectra for our sources (Fig. 
\ref{fig:spectra}), some examples of close-ups around the detected spectral lines
(Fig. \ref{fig:closeup}), and spectral decompositions of the observed optical spectra
into a power-law and an elliptical template (Fig. \ref{fig:decomposition}). 

\setcounter{figure}{0}
\begin{figure*}
\includegraphics[width=0.33\textwidth, angle=-90]{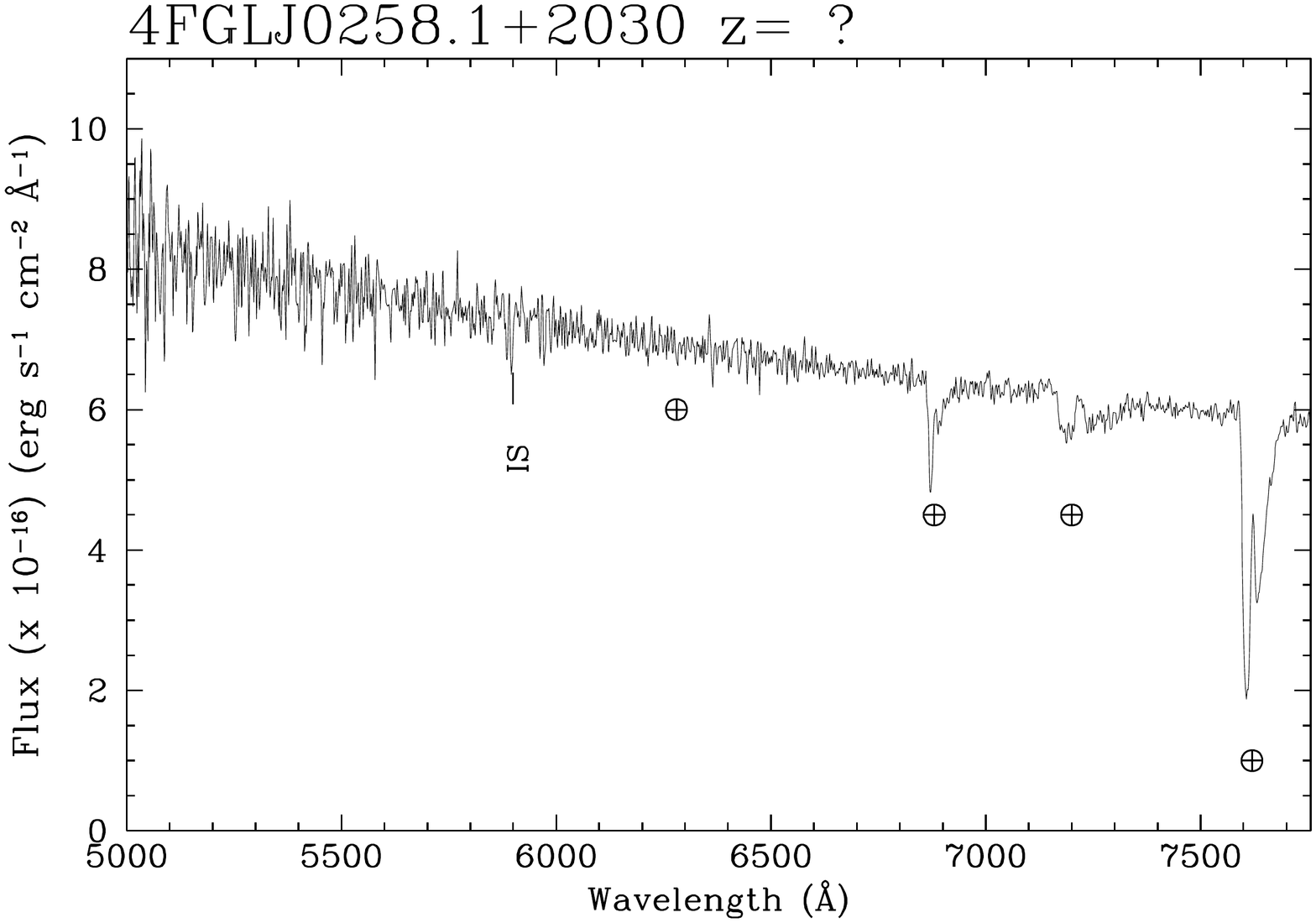}
\includegraphics[width=0.33\textwidth, angle=-90]{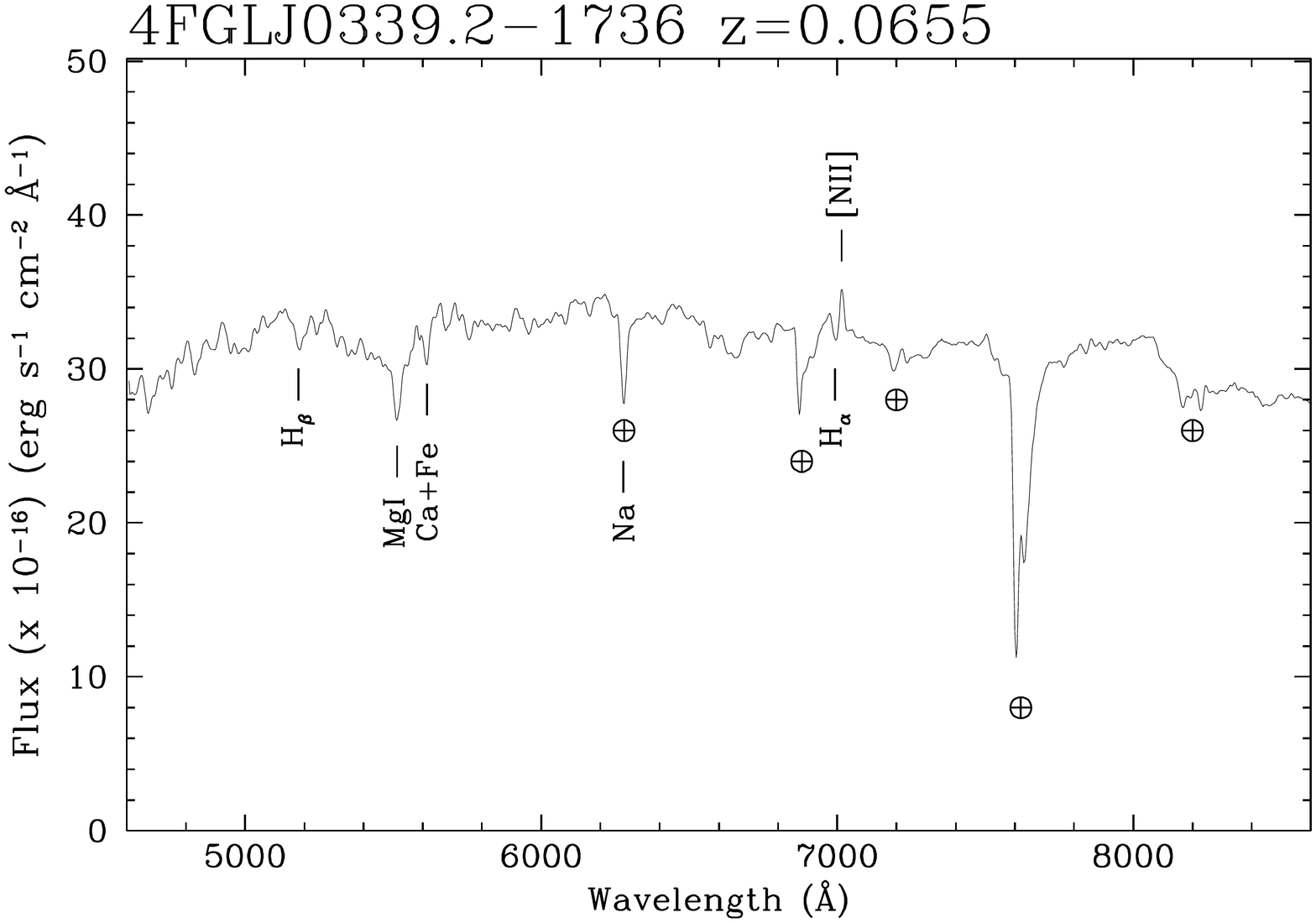}
\includegraphics[width=0.33\textwidth, angle=-90]{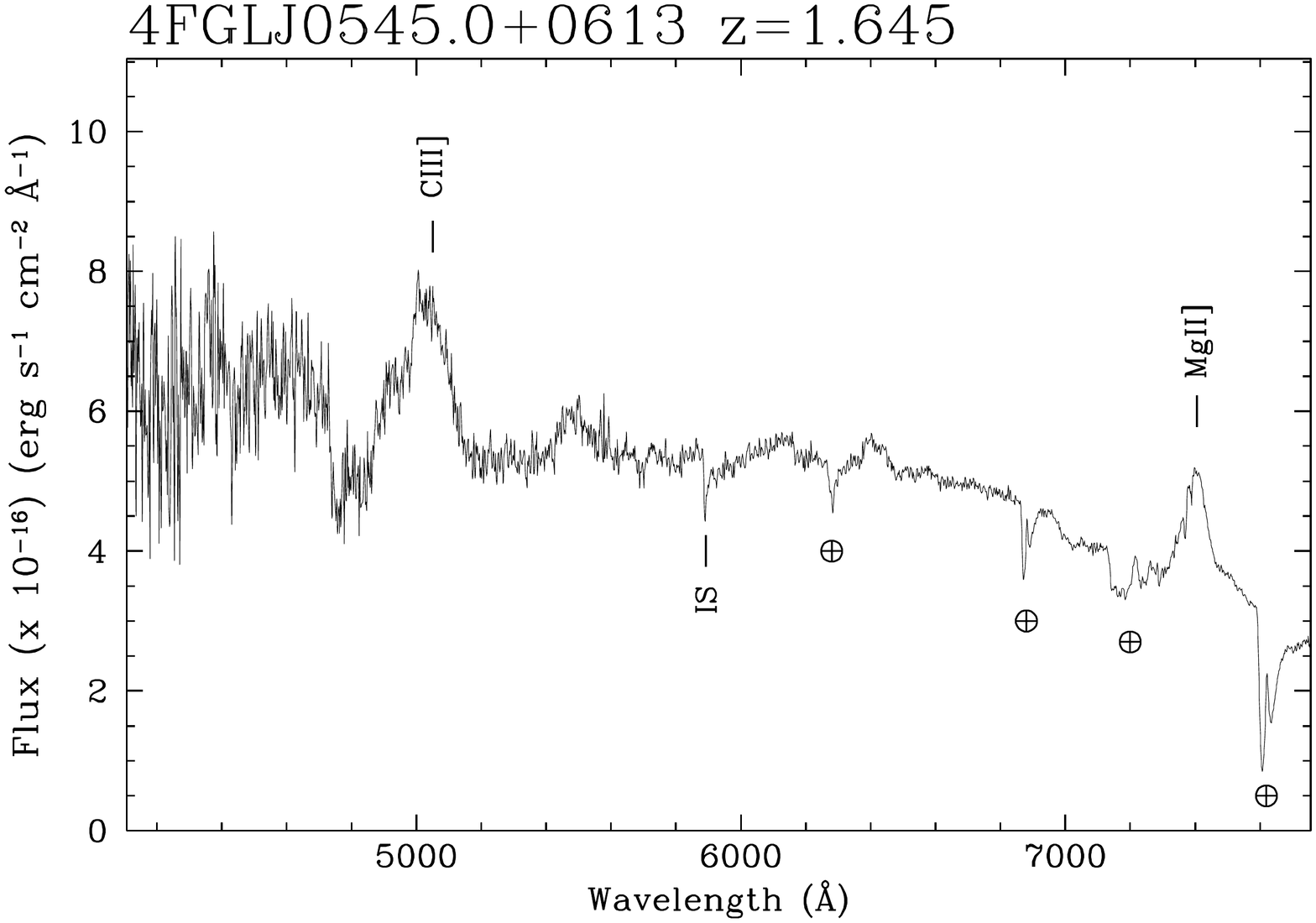} 
\includegraphics[width=0.33\textwidth, angle=-90]{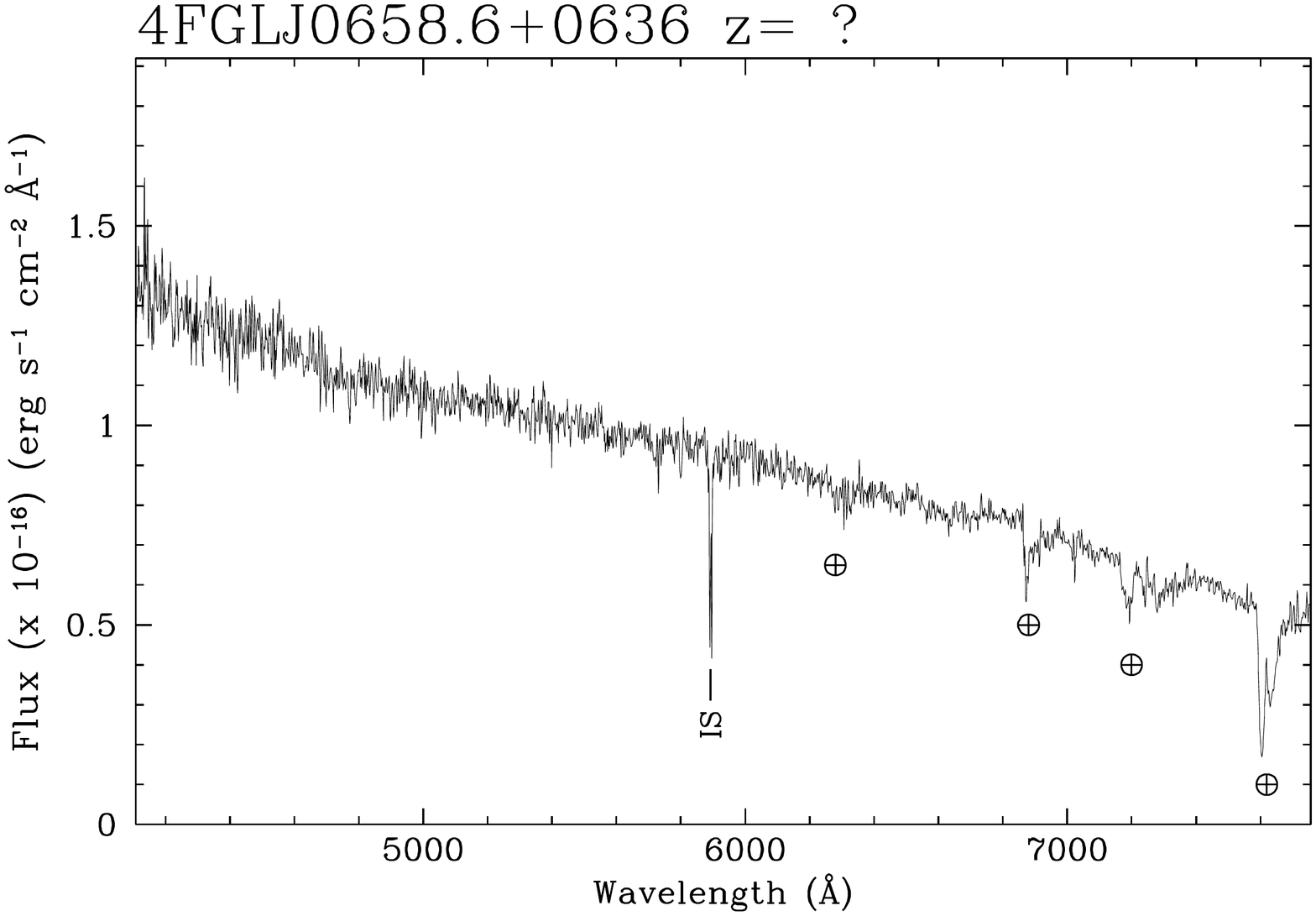}
\includegraphics[width=0.33\textwidth, angle=-90]{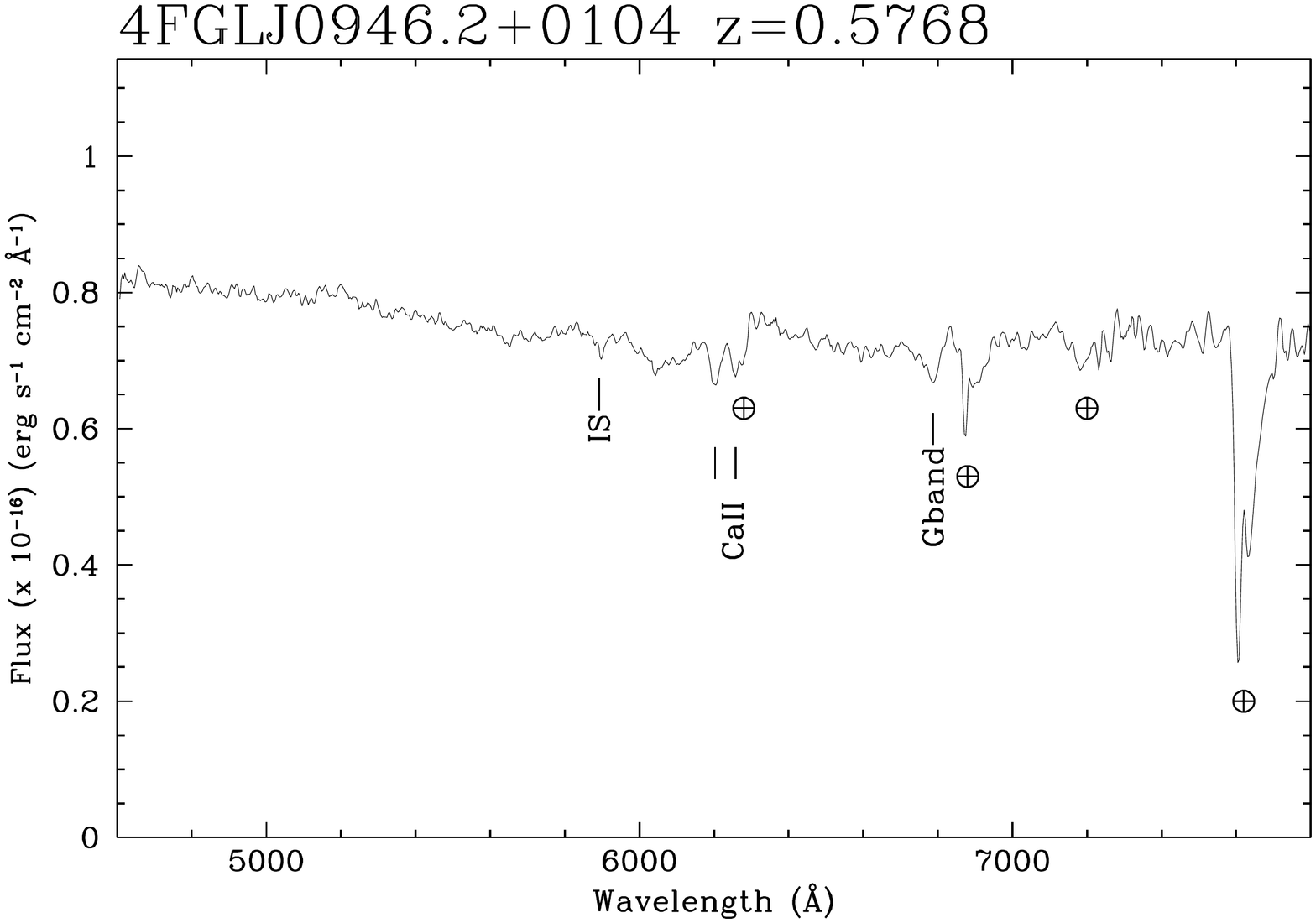}
\includegraphics[width=0.33\textwidth, angle=-90]{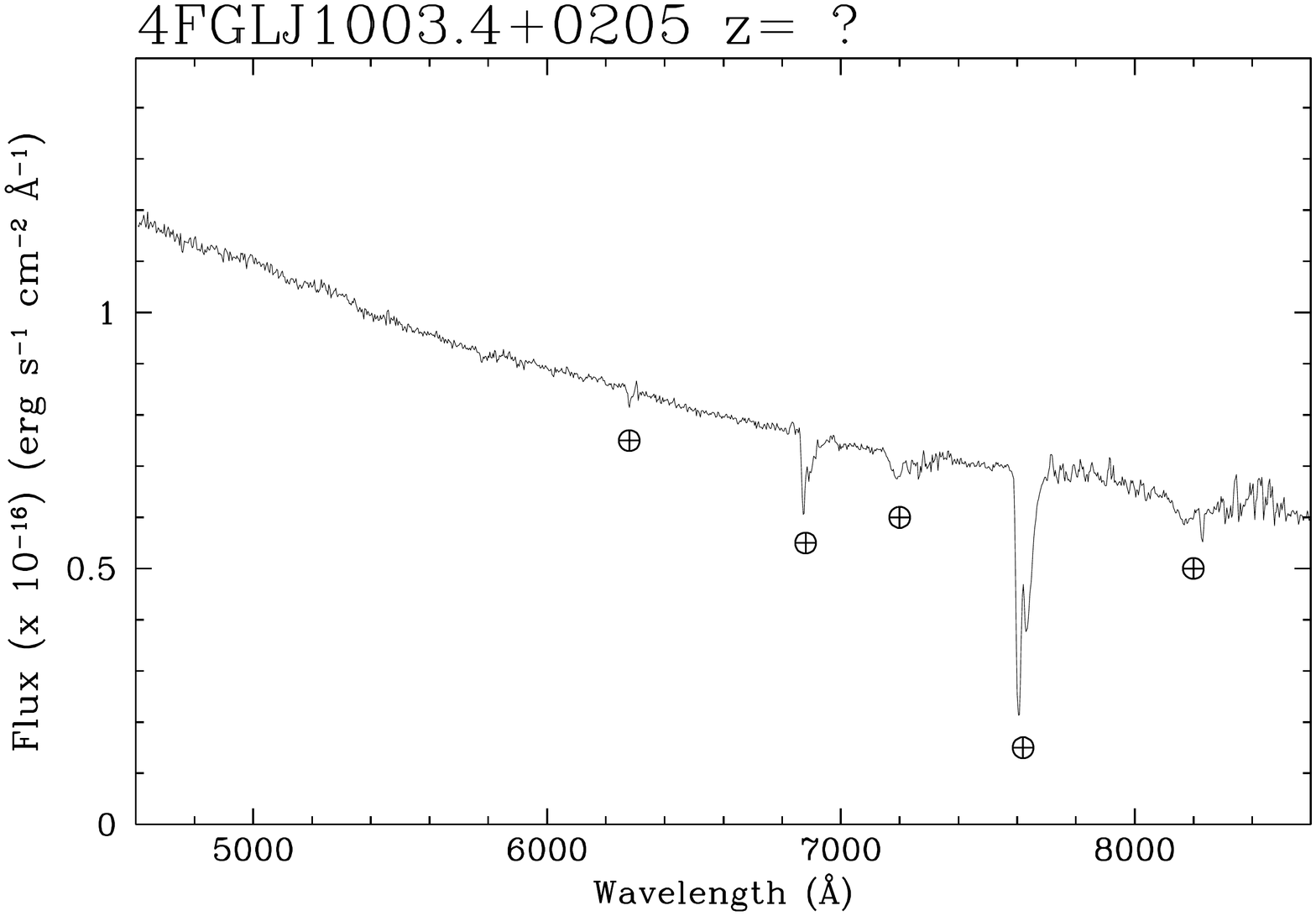} 
\includegraphics[width=0.33\textwidth, angle=-90]{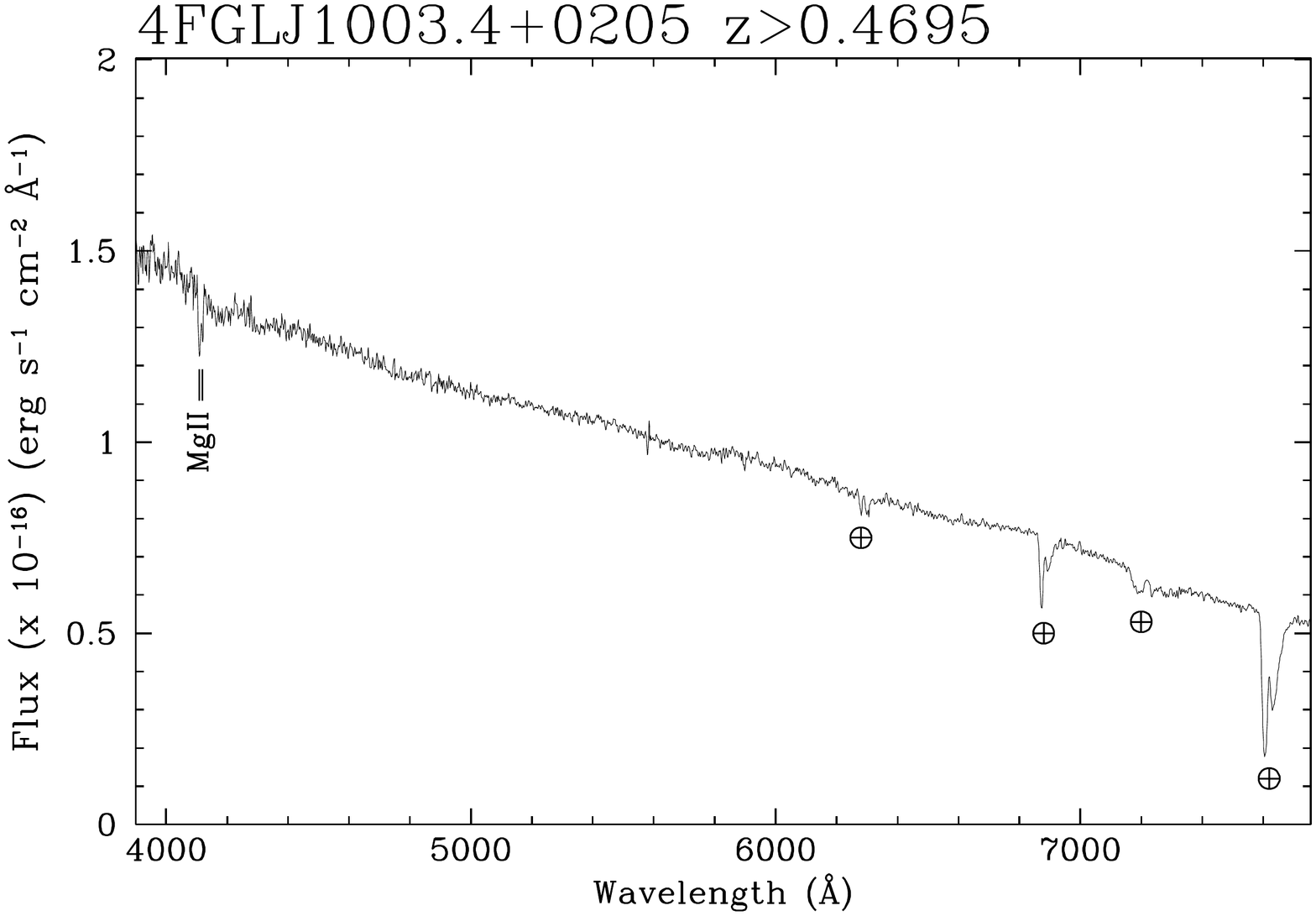} 
\includegraphics[width=0.33\textwidth, angle=-90]{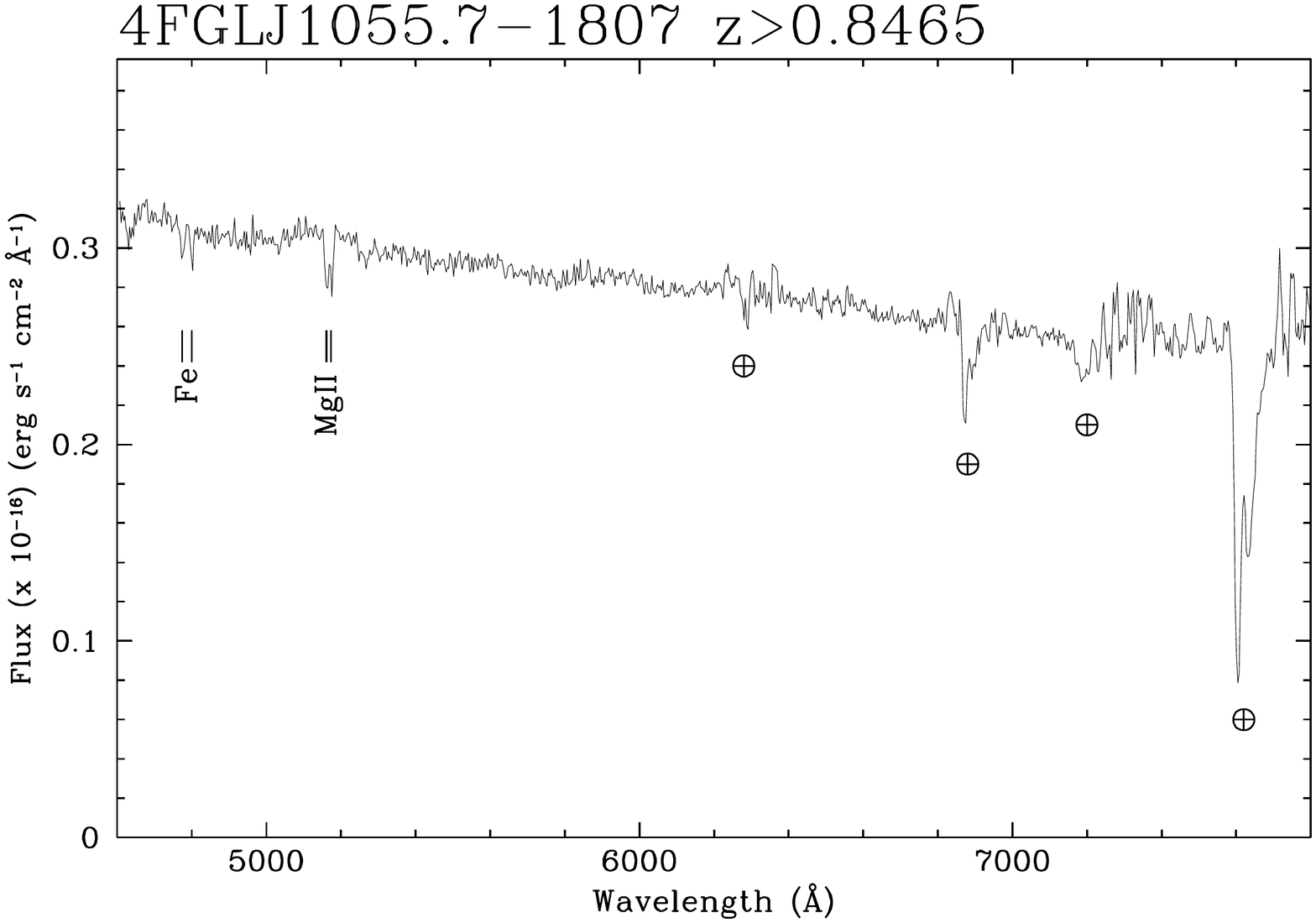}
\caption{Flux calibrated and de-reddened spectra of the neutrino candidate blazars obtained at GTC, VLT and LBT (see Table \ref{tab:observations} for details). The main telluric bands are indicated by $\oplus$, the absorption features from interstellar medium of our galaxies are labelled as IS (Inter-Stellar).} 
\label{fig:spectra}
\end{figure*}

\setcounter{figure}{0}
\begin{figure*}
\includegraphics[width=0.33\textwidth, angle=-90]{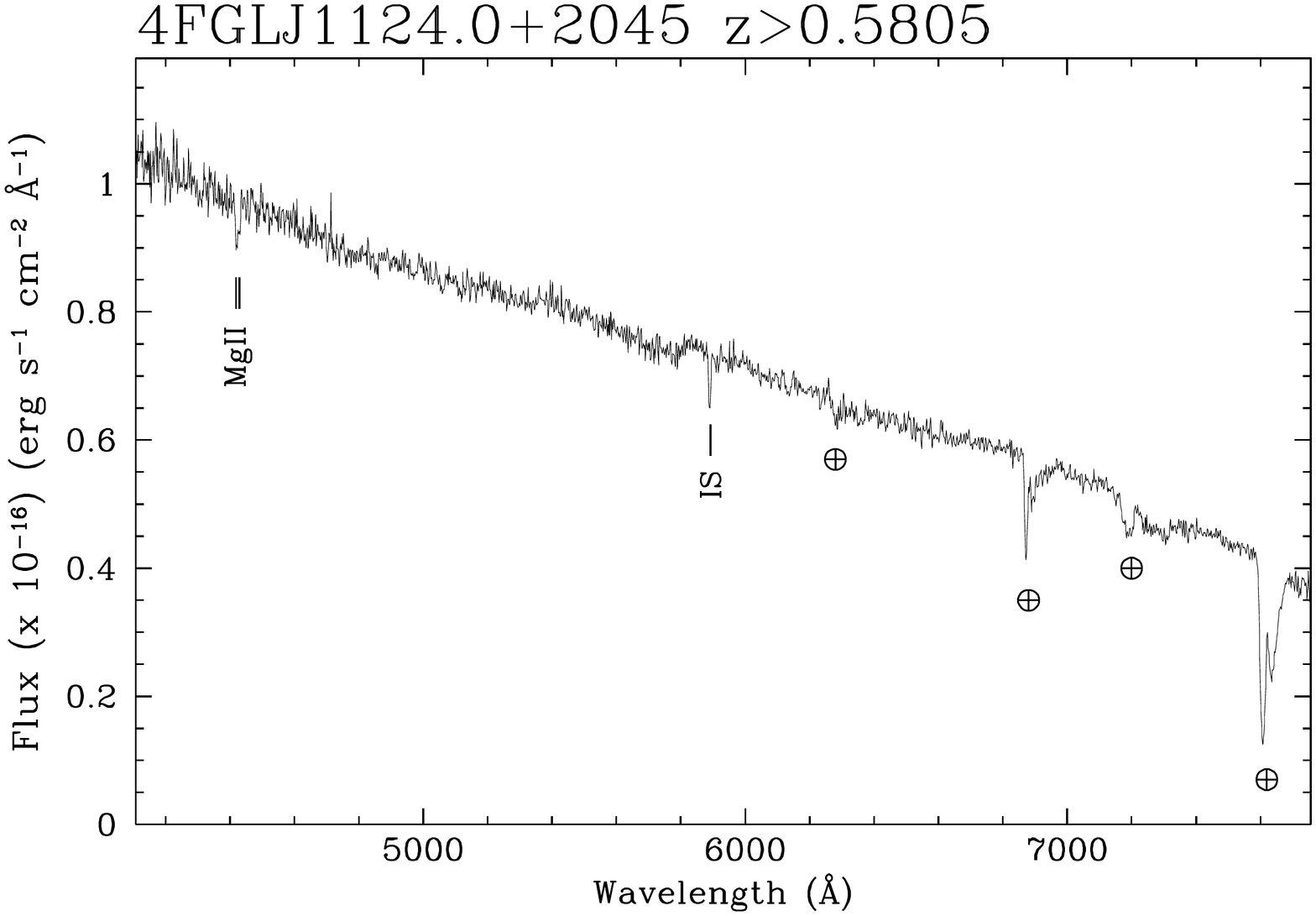}
\includegraphics[width=0.33\textwidth, angle=-90]{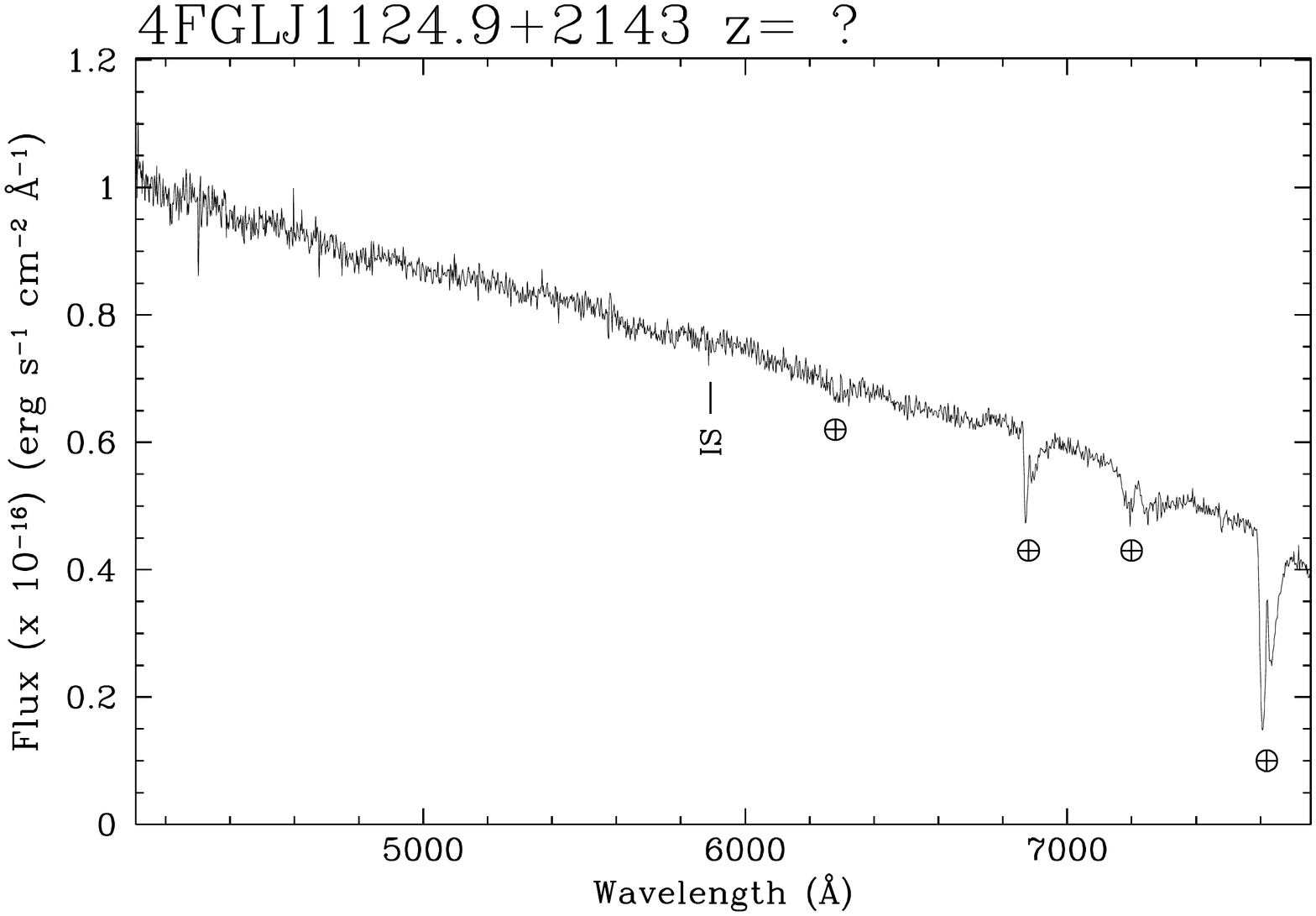}
\includegraphics[width=0.33\textwidth, angle=-90]{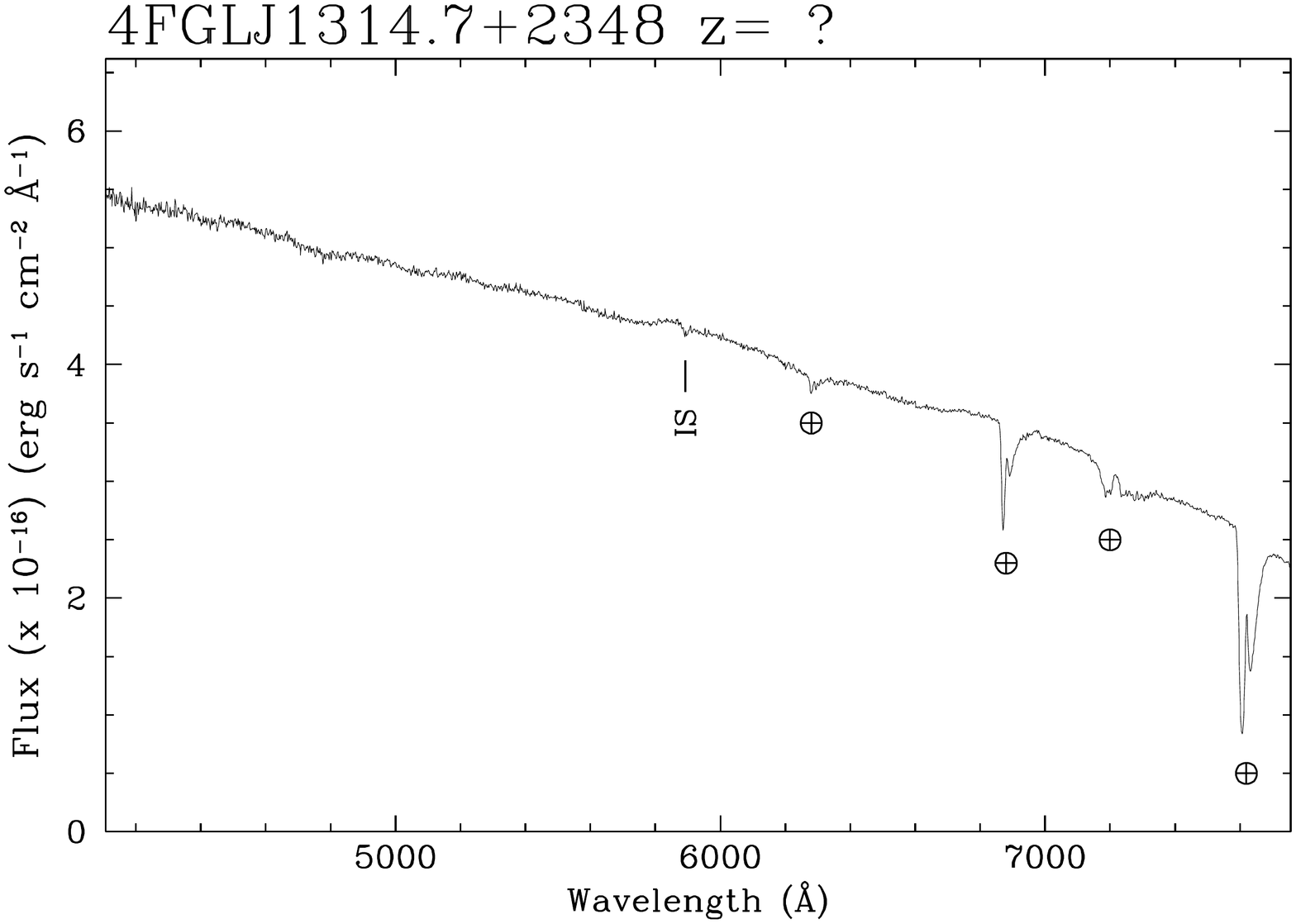}
\includegraphics[width=0.33\textwidth, angle=-90]{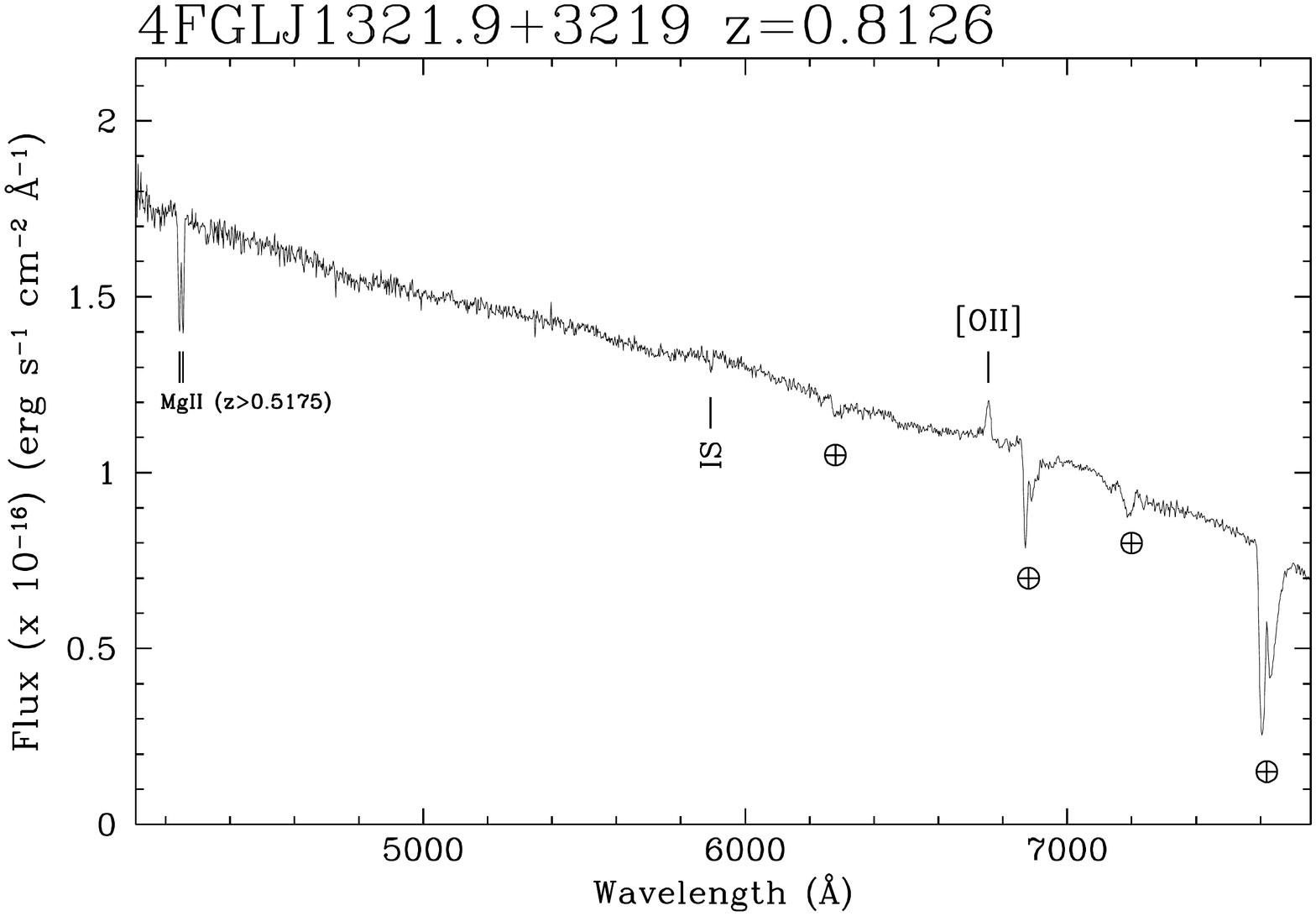}
\includegraphics[width=0.33\textwidth, angle=-90]{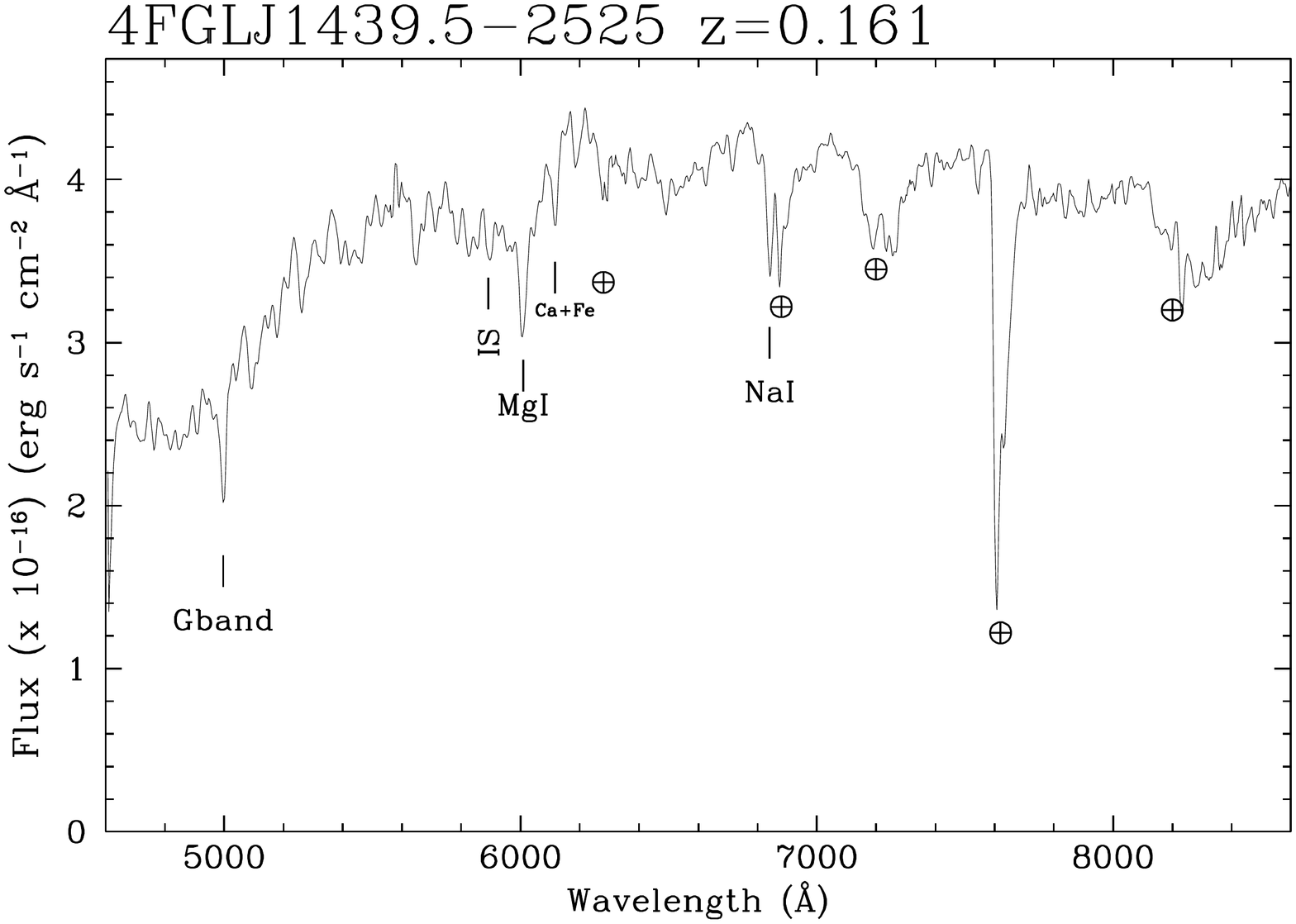}
\includegraphics[width=0.33\textwidth, angle=-90]{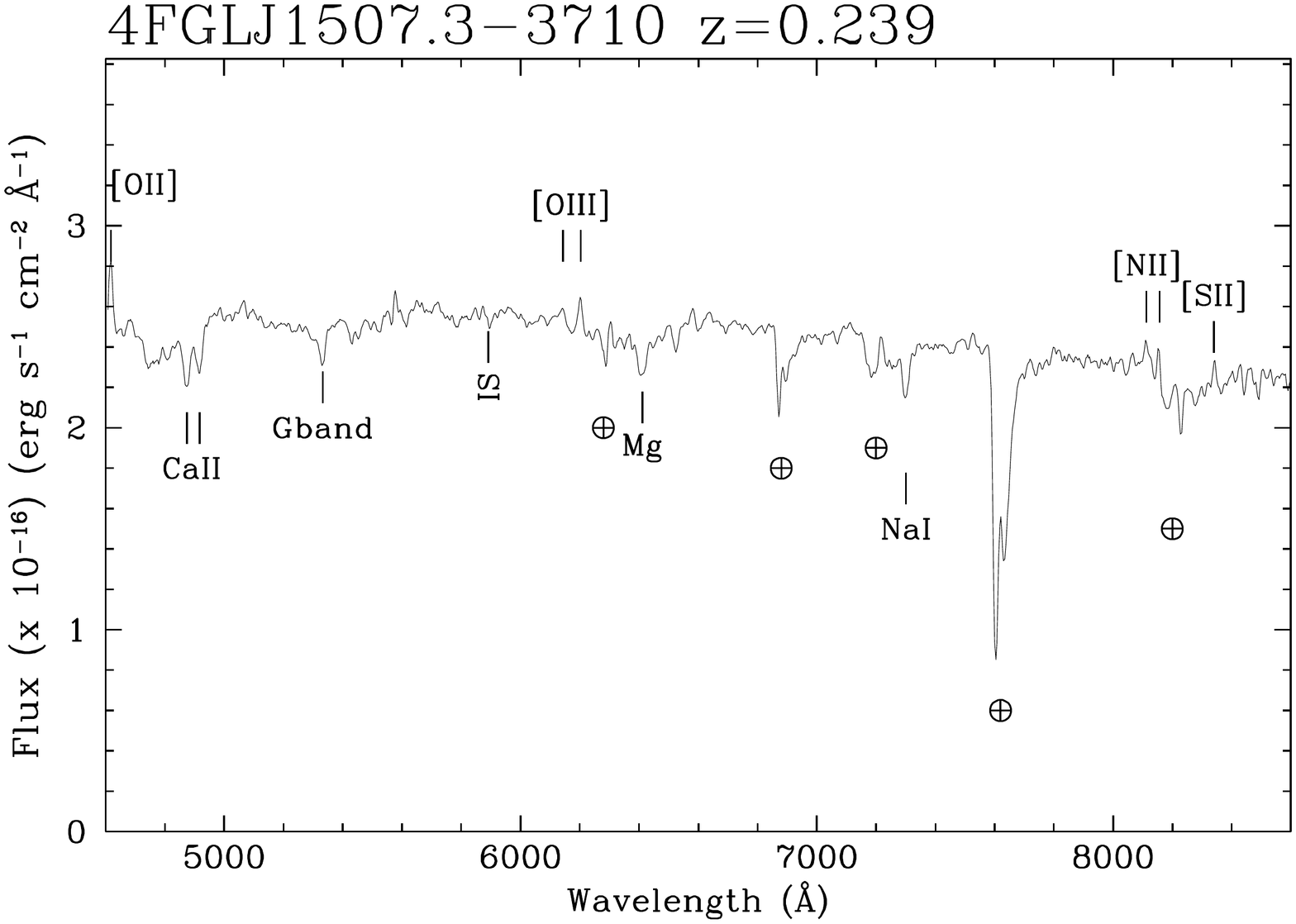}
\includegraphics[width=0.33\textwidth, angle=-90]{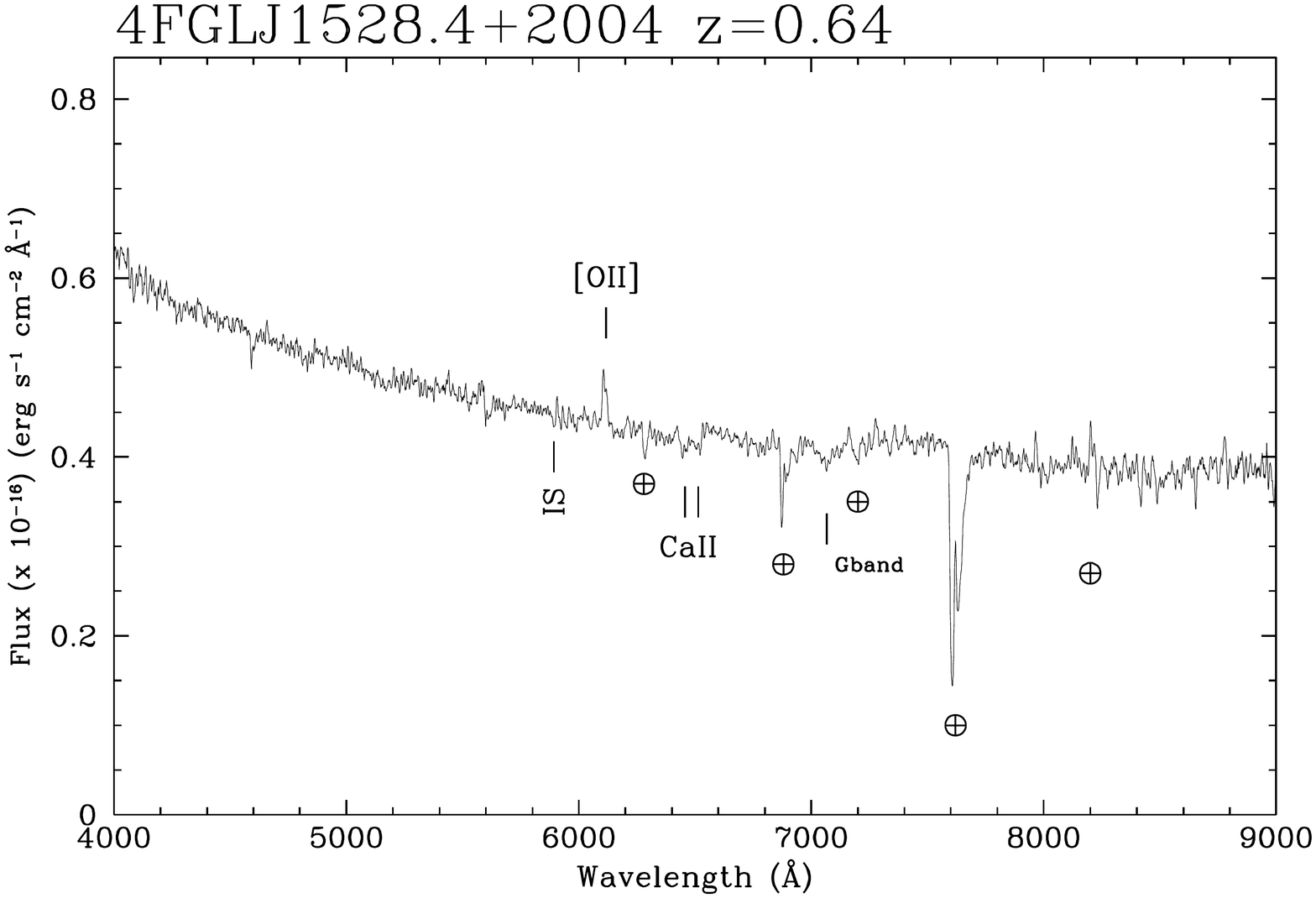}
\includegraphics[width=0.33\textwidth, angle=-90]{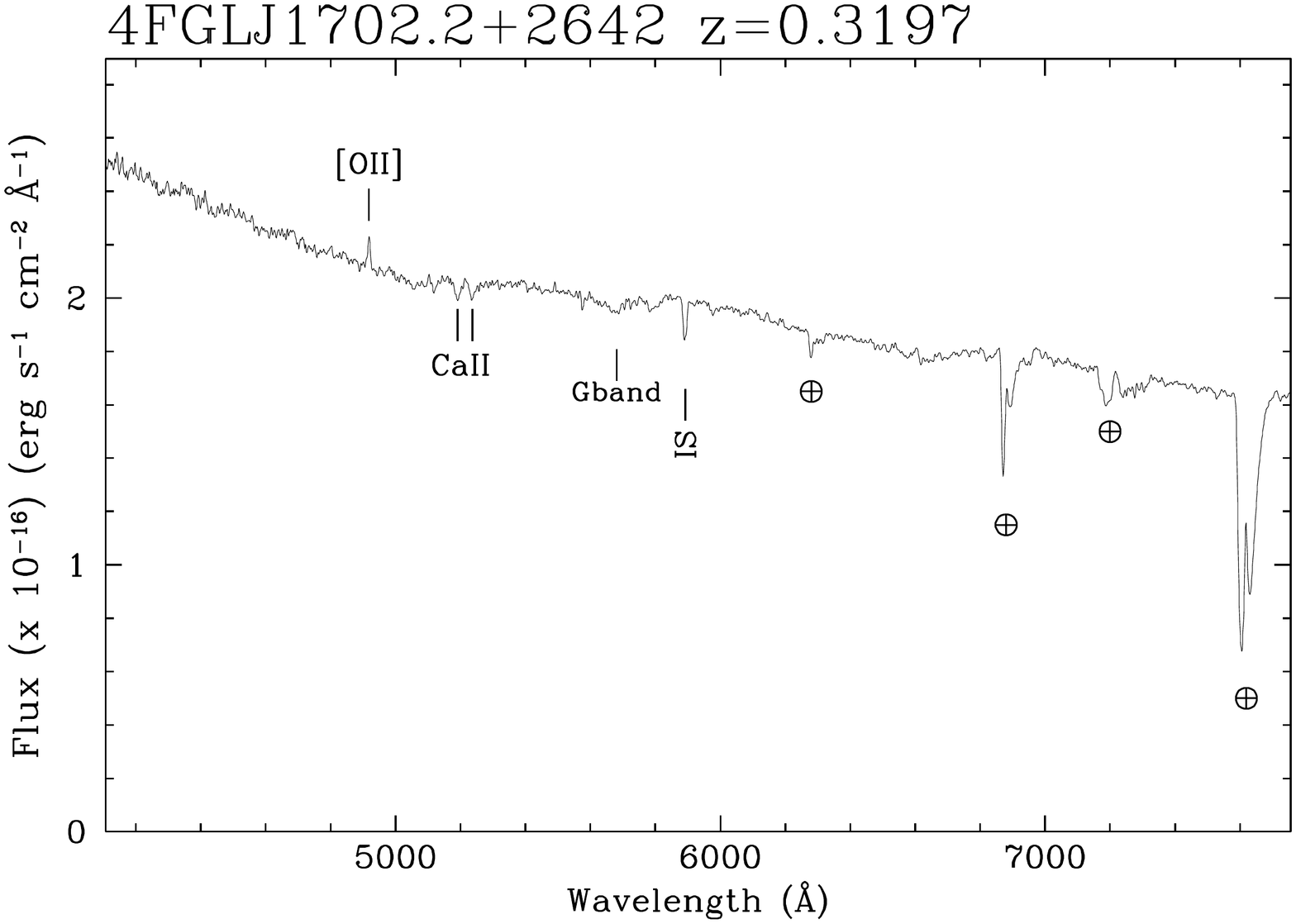}
\caption{- \textit{Continued}}
\end{figure*}

\setcounter{figure}{0}
\begin{figure*}
\includegraphics[width=0.33\textwidth, angle=-90]{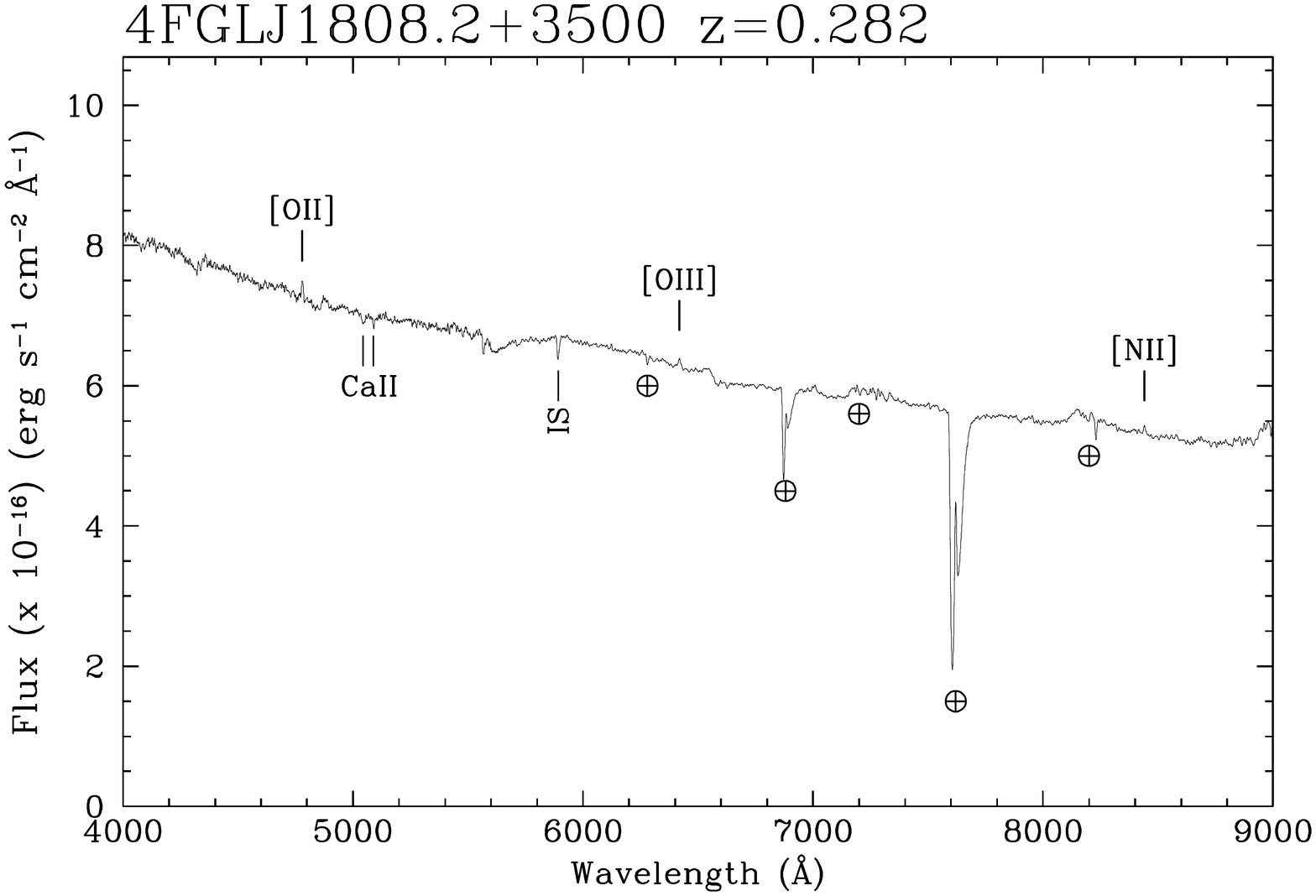}
\includegraphics[width=0.33\textwidth, angle=-90]{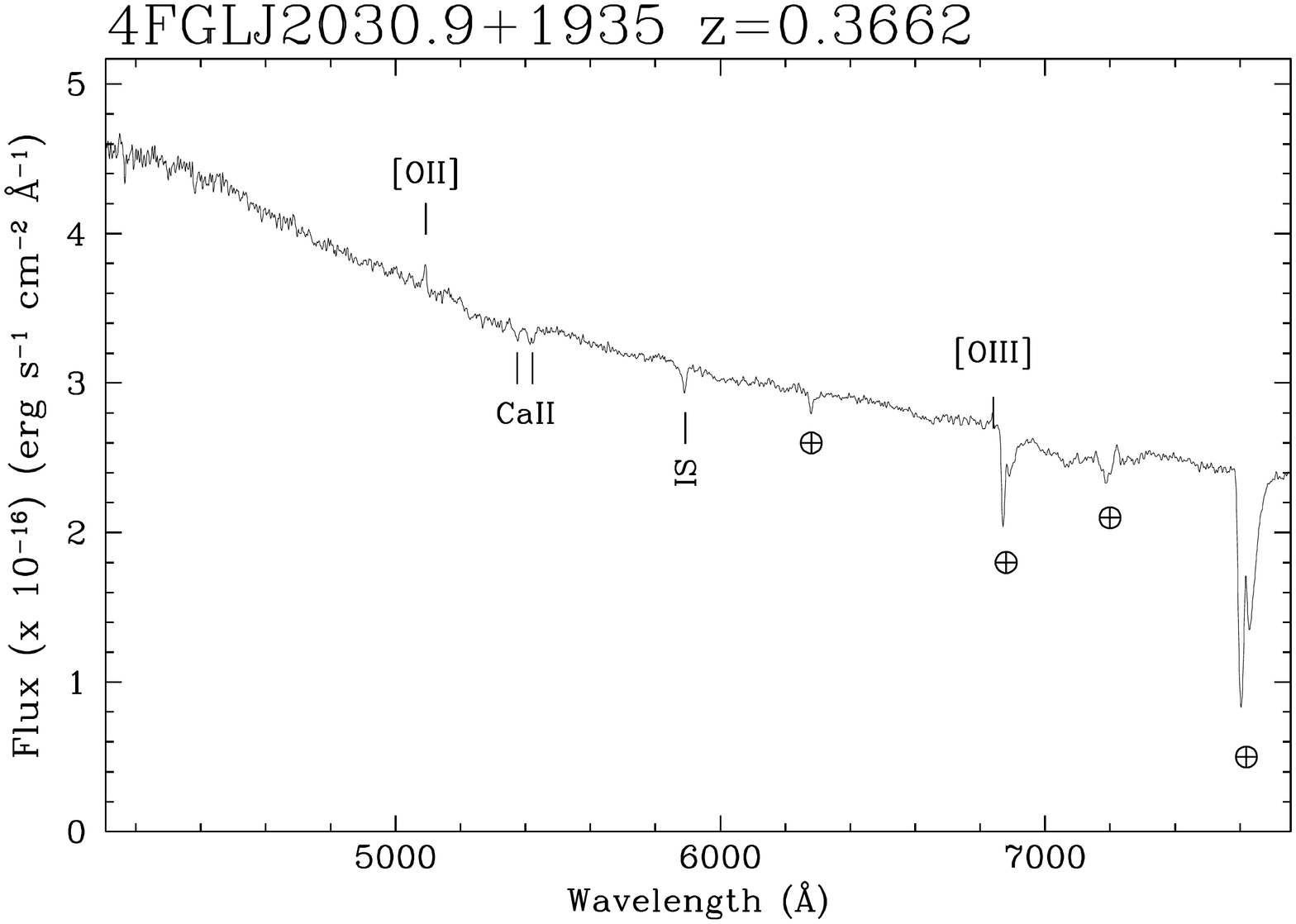}
\includegraphics[width=0.33\textwidth, angle=-90]{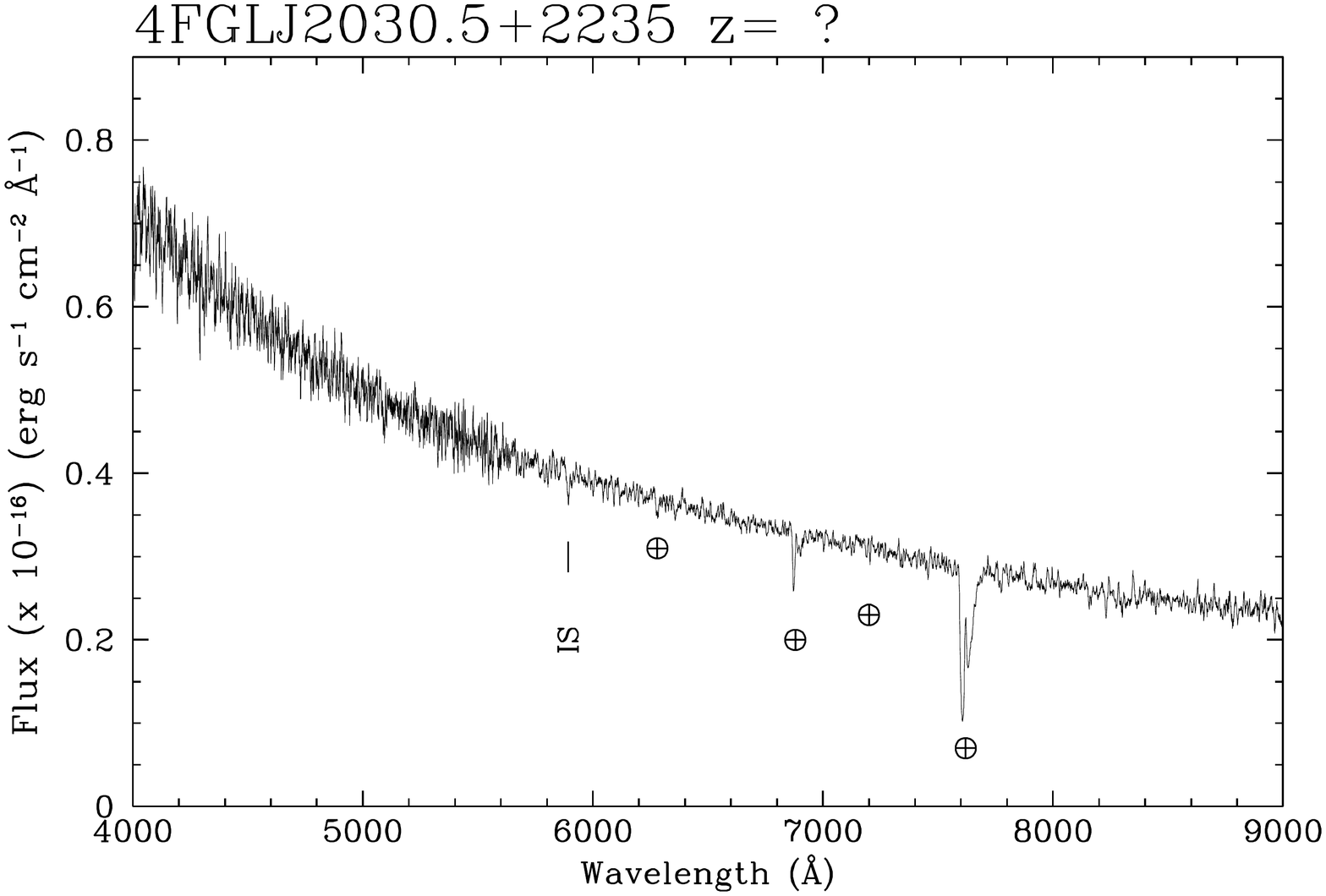}
\includegraphics[width=0.33\textwidth, angle=-90]{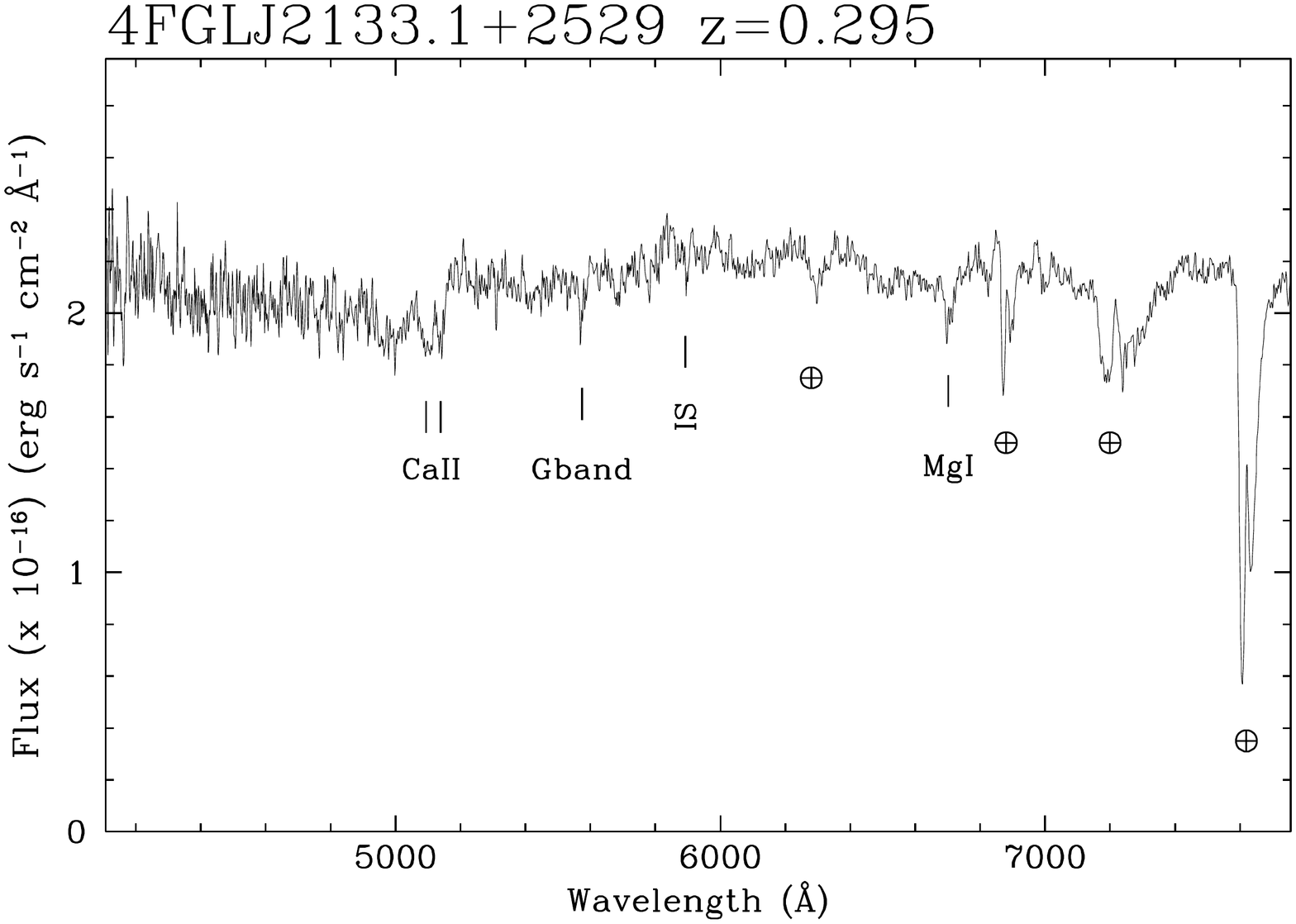}
\includegraphics[width=0.33\textwidth, angle=-90]{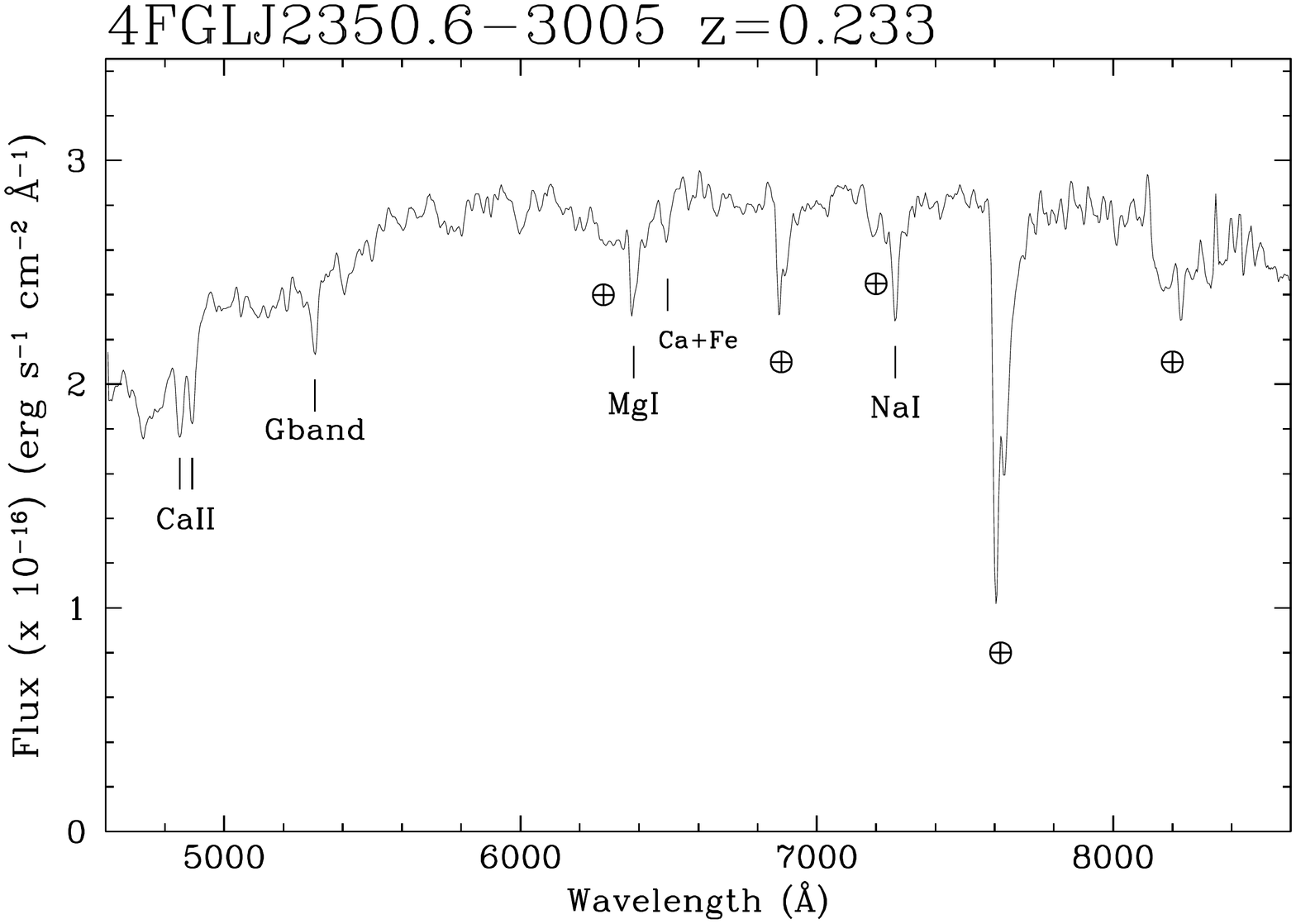}
\includegraphics[width=0.33\textwidth, angle=-90]{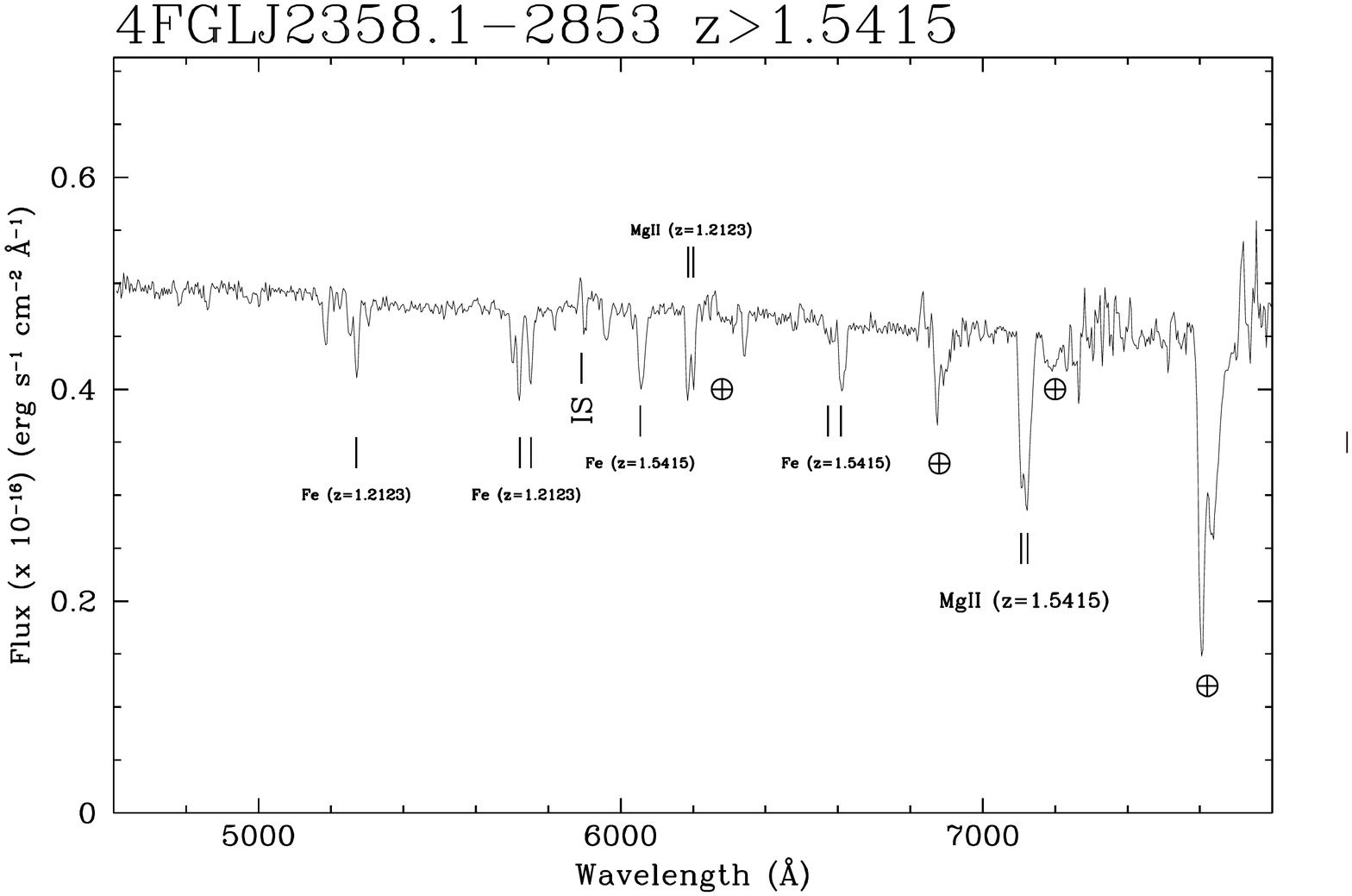}
\caption{- \textit{Continued}}
\end{figure*}

\setcounter{figure}{1}
\begin{figure*}
\includegraphics[width=0.38\textwidth, angle=-90]{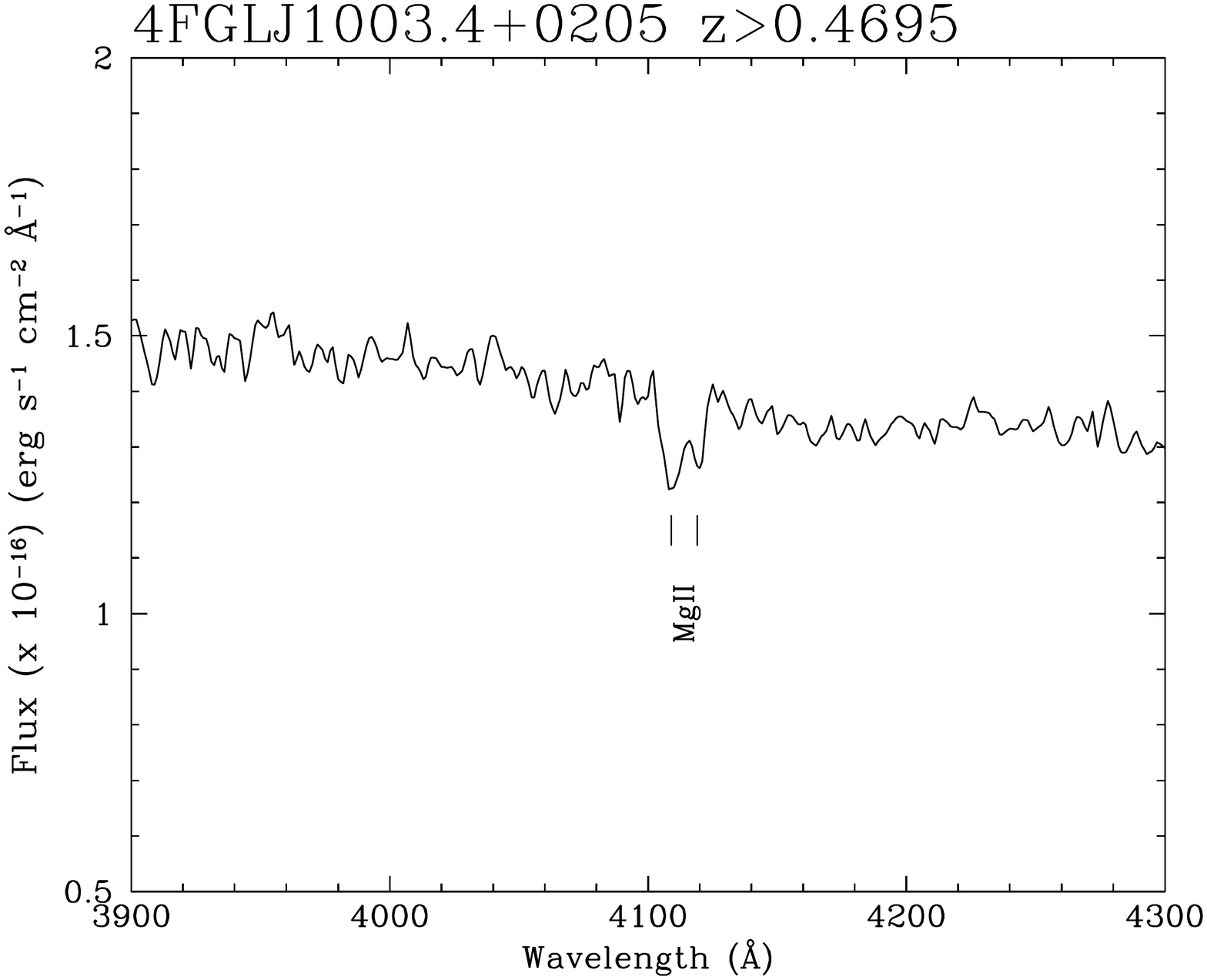}
\includegraphics[width=0.38\textwidth, angle=-90]{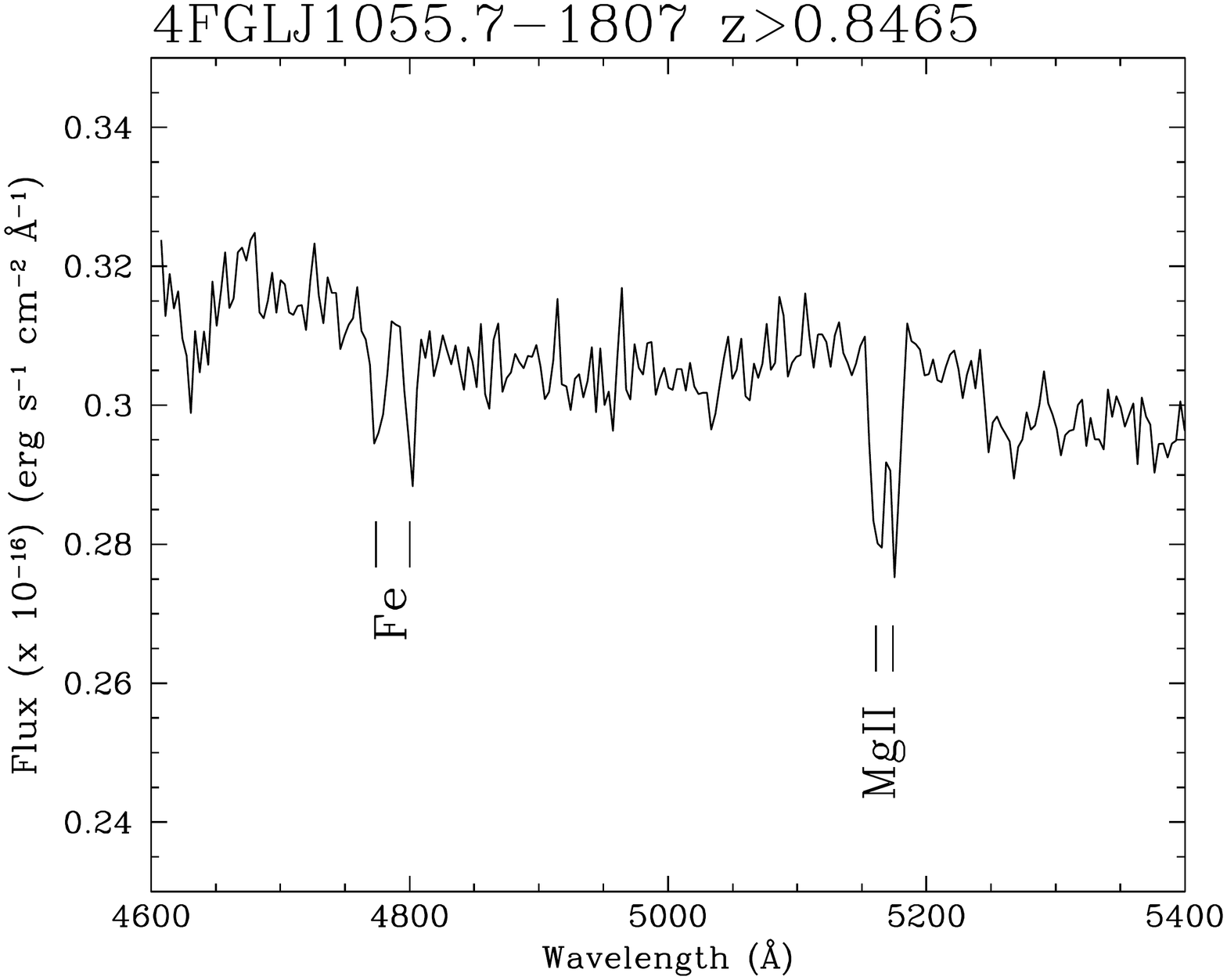}
\includegraphics[width=0.38\textwidth, angle=-90]{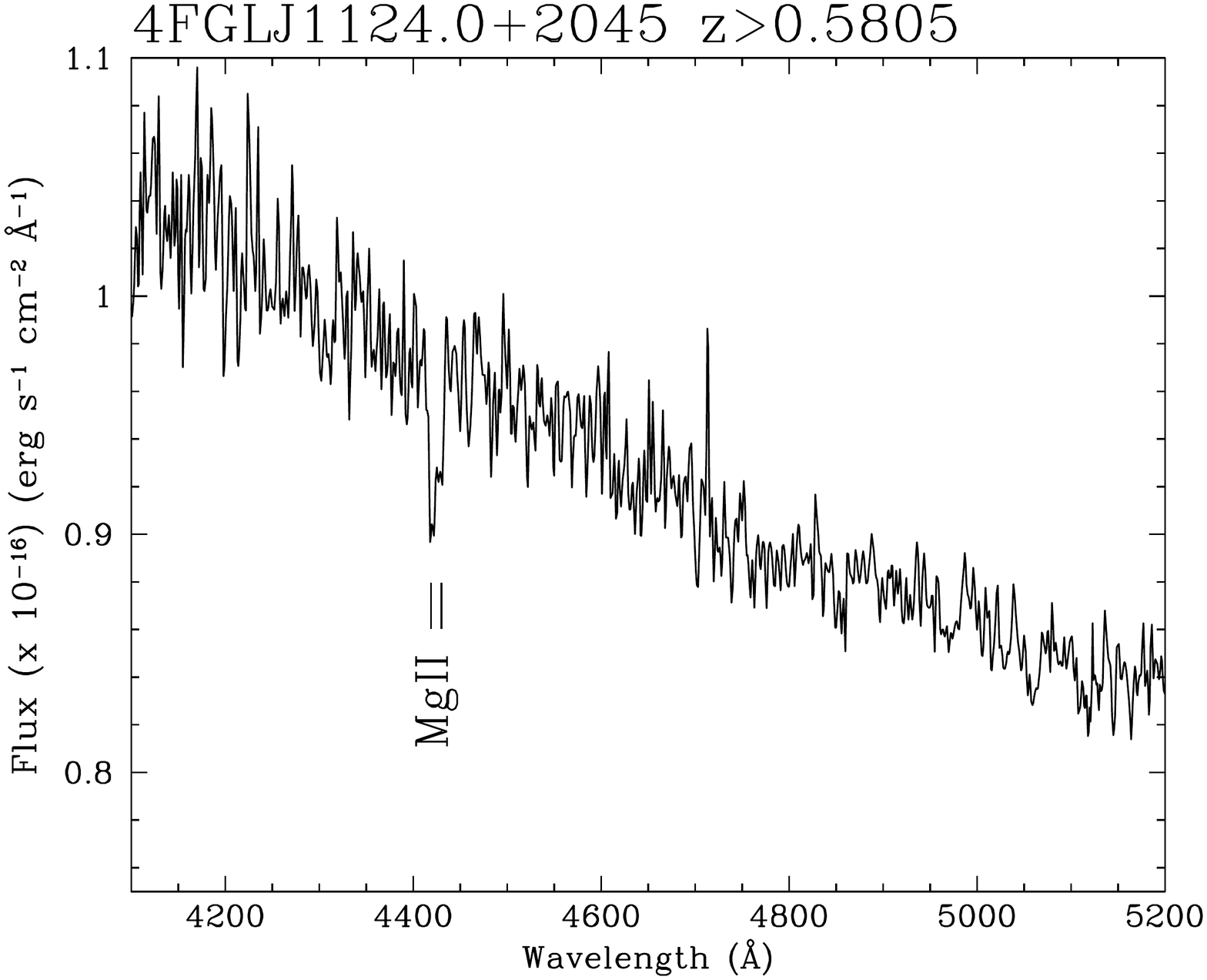}
\includegraphics[width=0.38\textwidth, angle=-90]{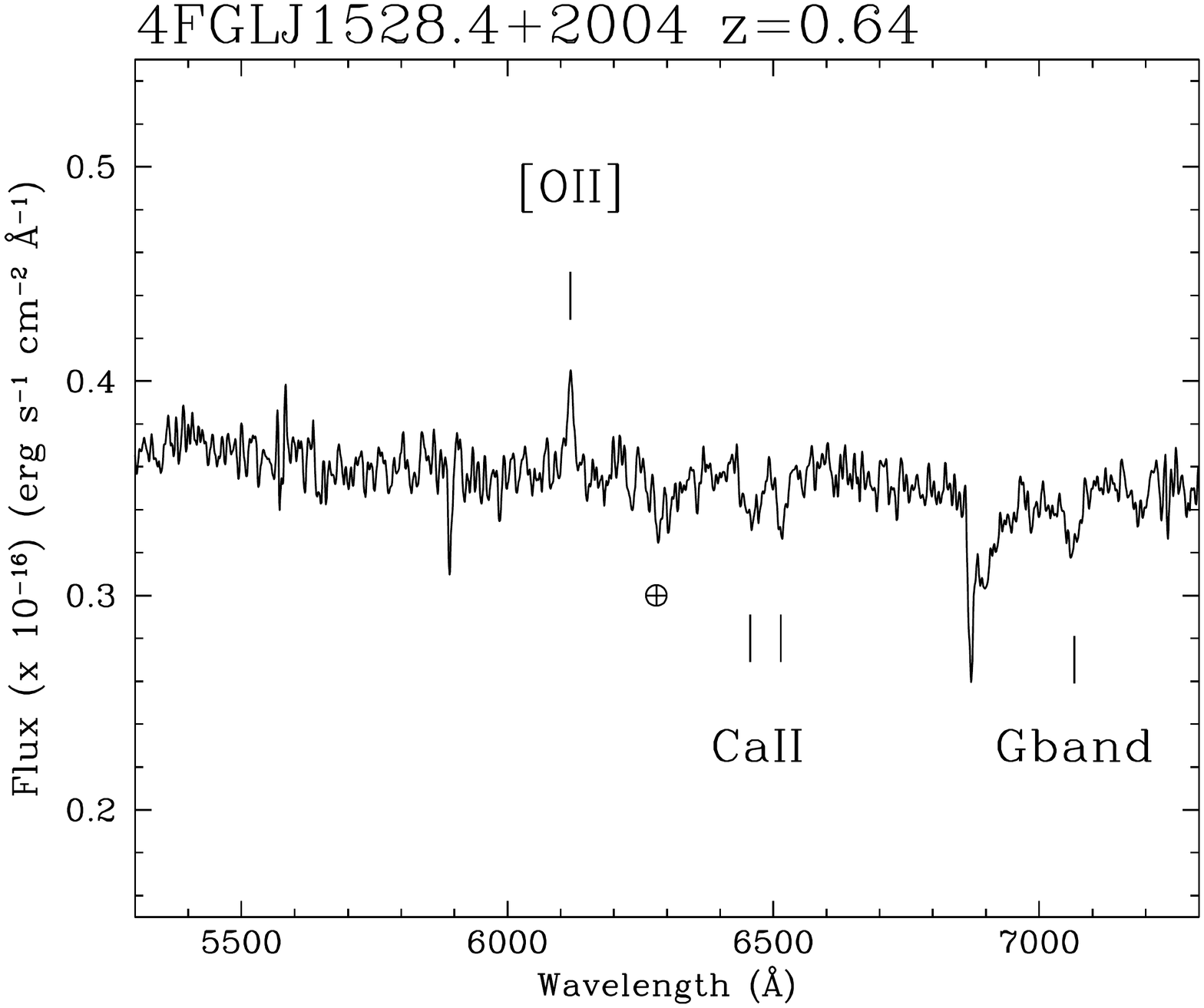}
\includegraphics[width=0.38\textwidth, angle=-90]{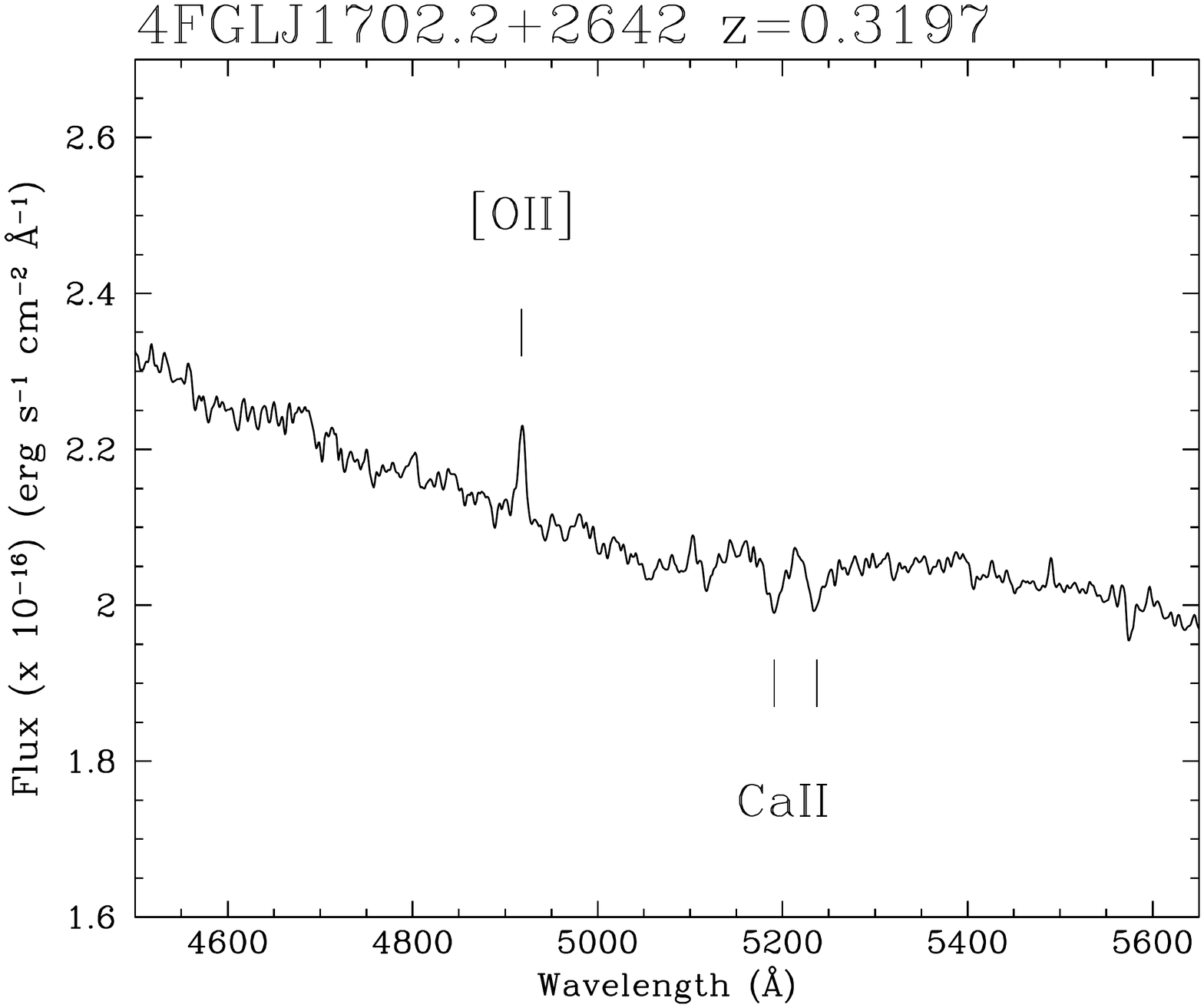}
\includegraphics[width=0.38\textwidth, angle=-90]{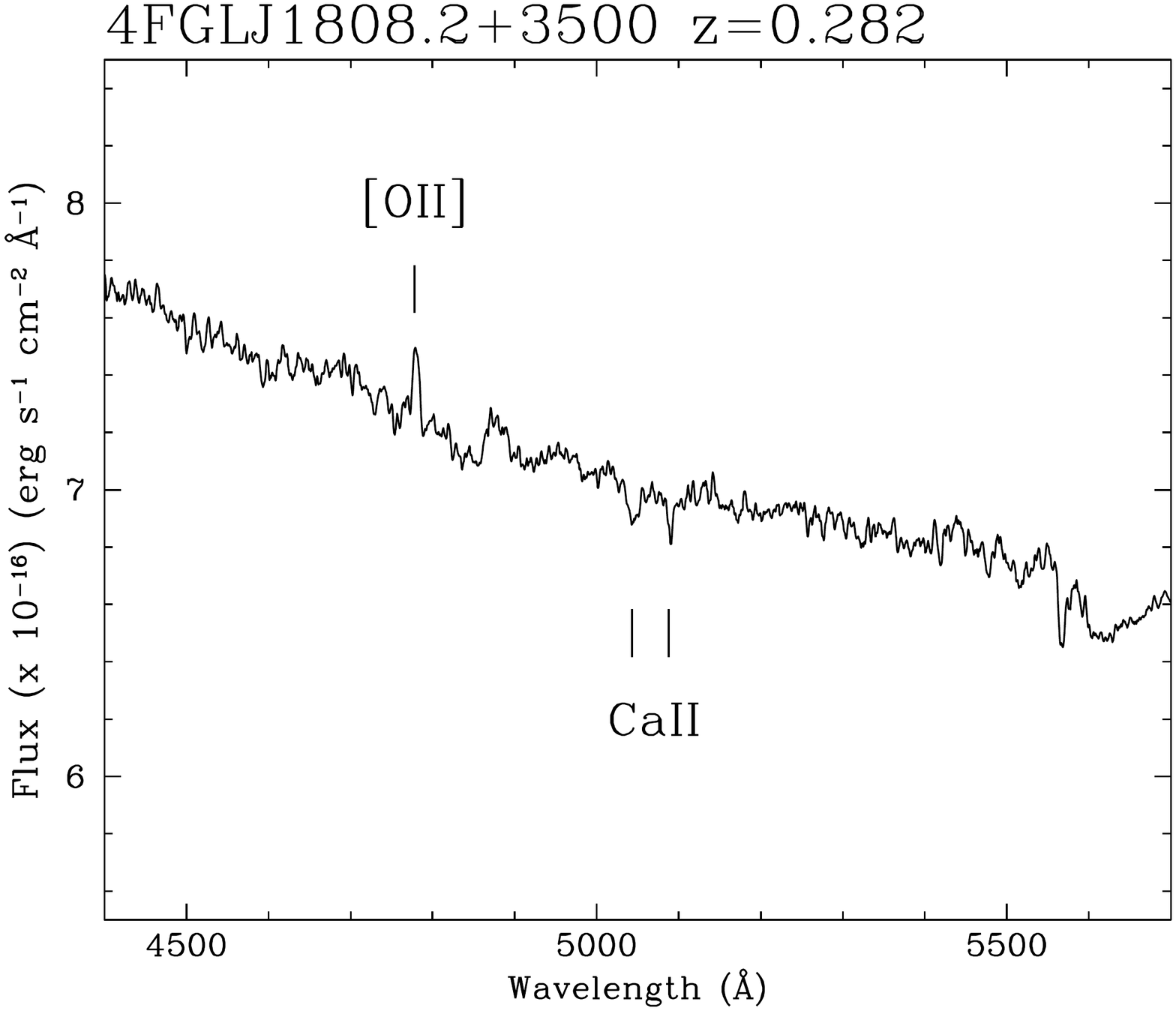}
\caption{Some examples of close-ups around the detected spectral lines of the spectra of the neutrino candidate blazars obtained at GTC, VLT and LBT. 
Main telluric bands are indicated by $\oplus$, spectral lines are marked by line identification.}
\label{fig:closeup}
\end{figure*}

\setcounter{figure}{1}
\begin{figure*}
\includegraphics[width=0.38\textwidth, angle=-90]{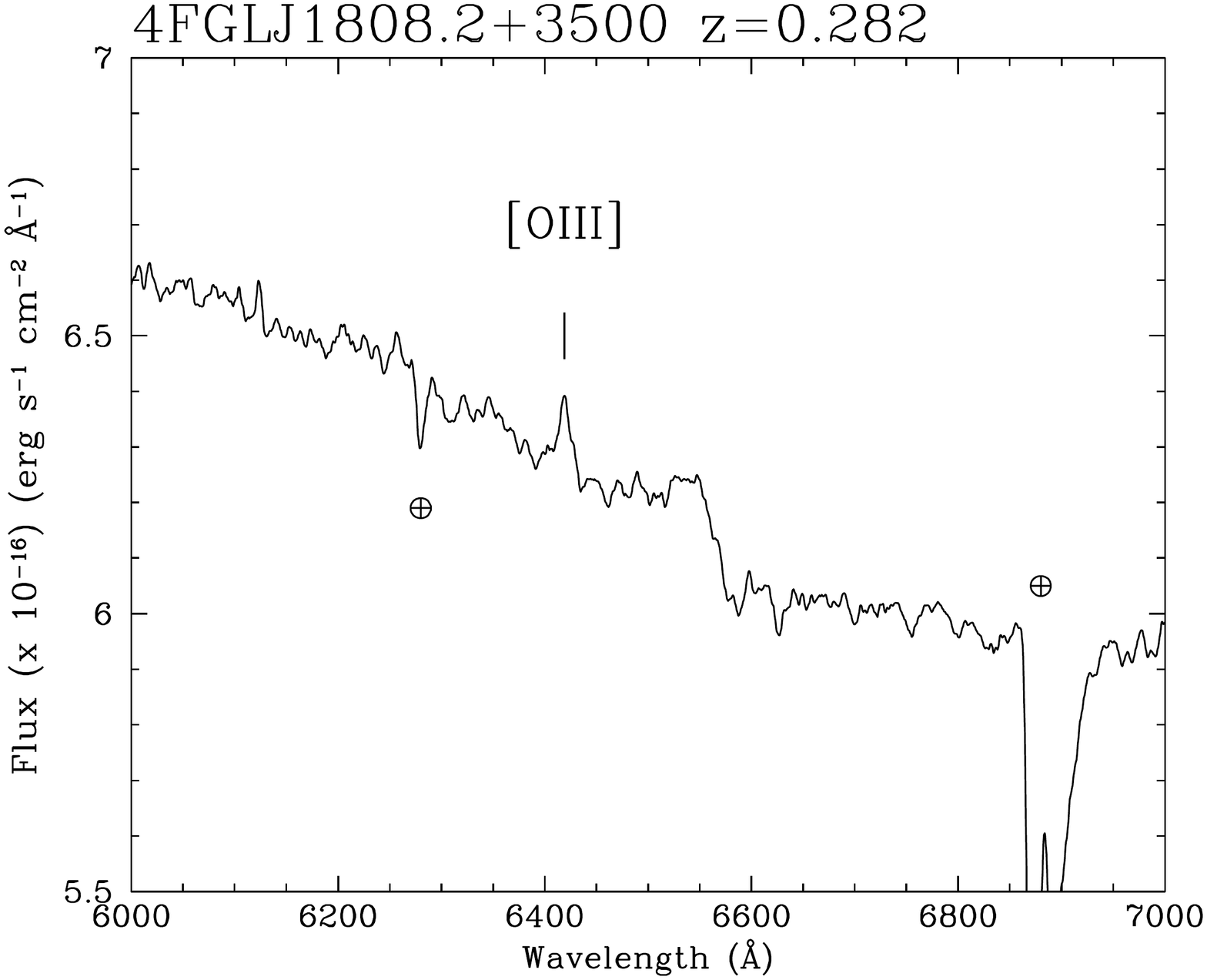}
\includegraphics[width=0.38\textwidth, angle=-90]{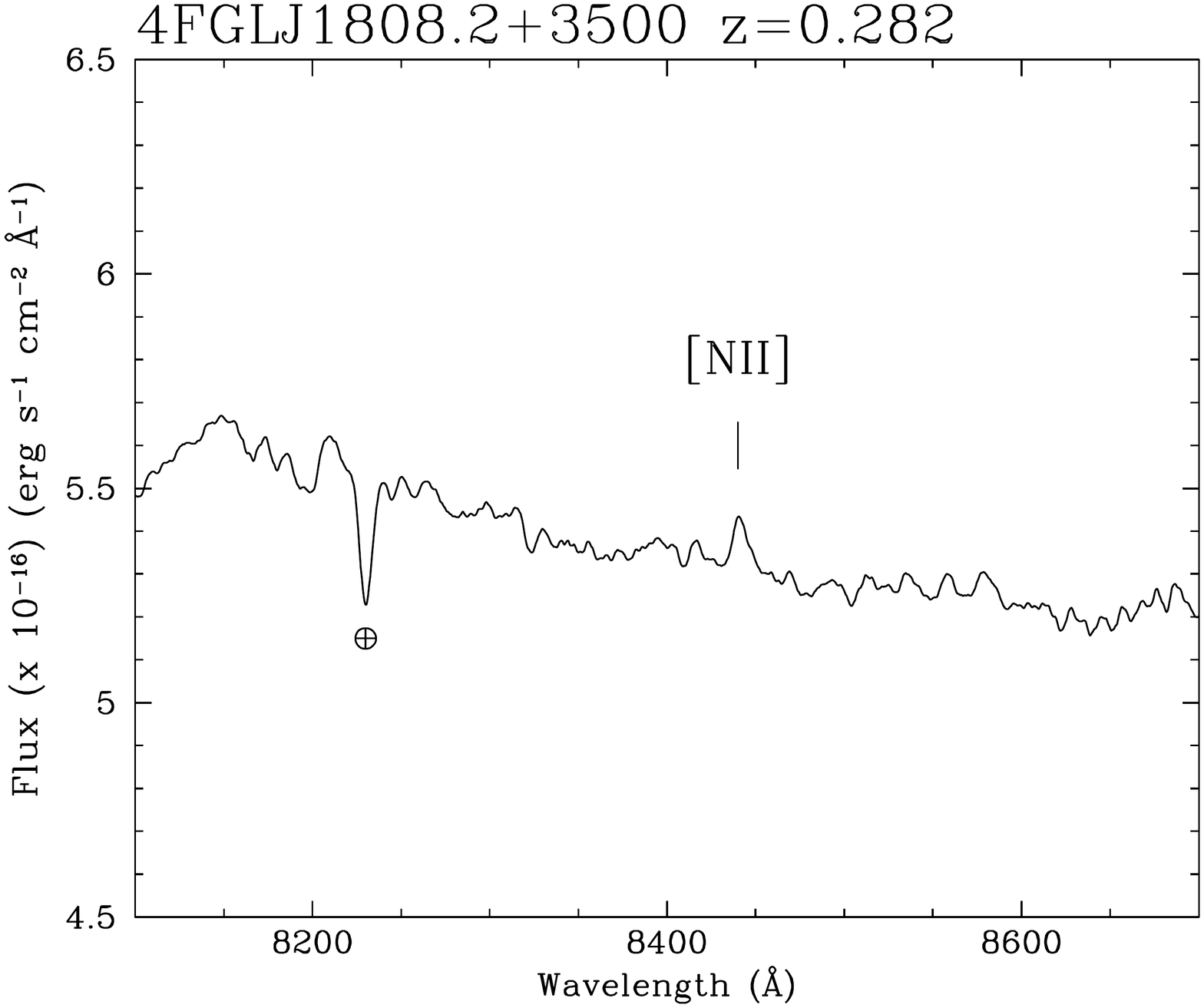}
\includegraphics[width=0.38\textwidth, angle=-90]{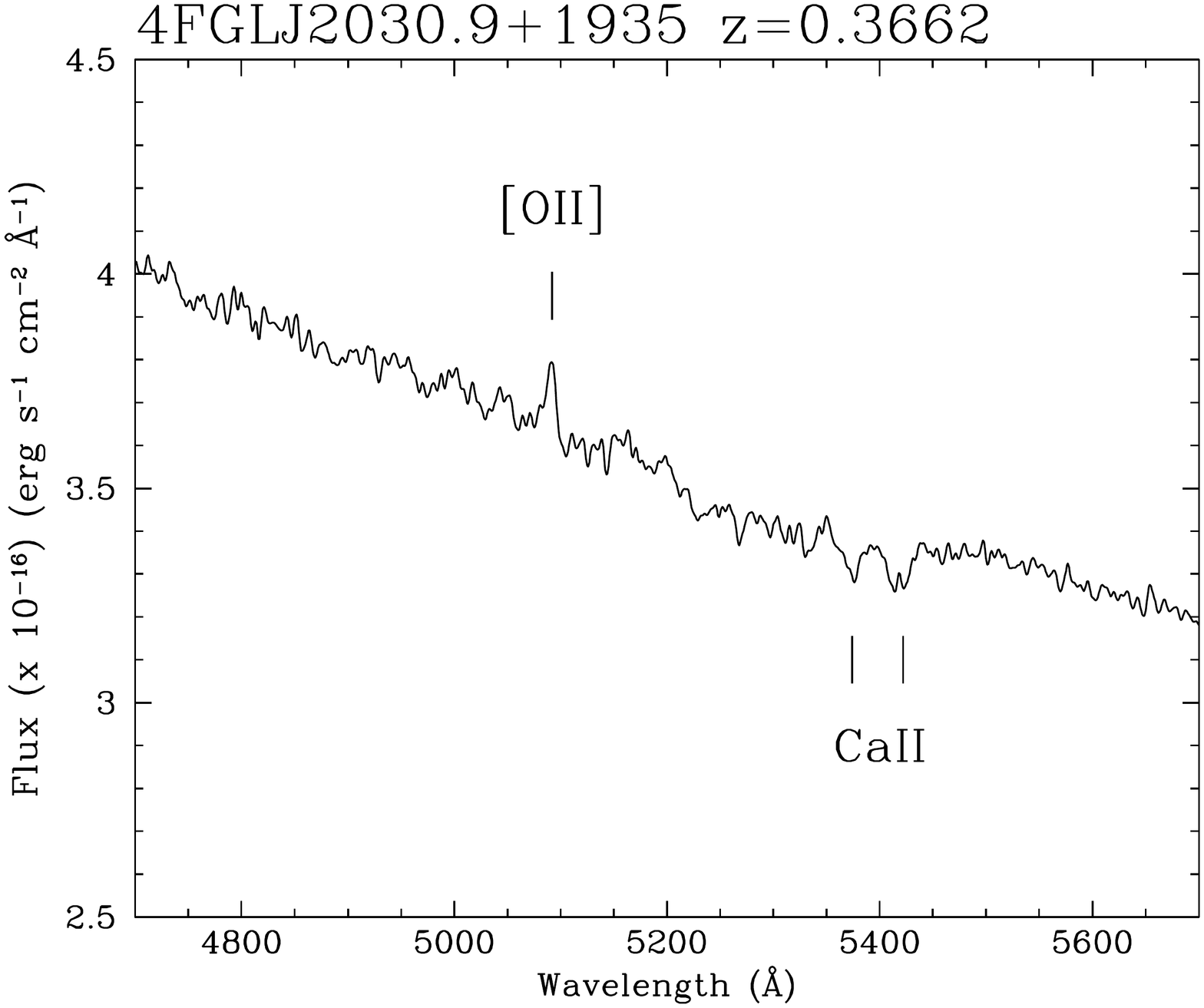}
\includegraphics[width=0.38\textwidth, angle=-90]{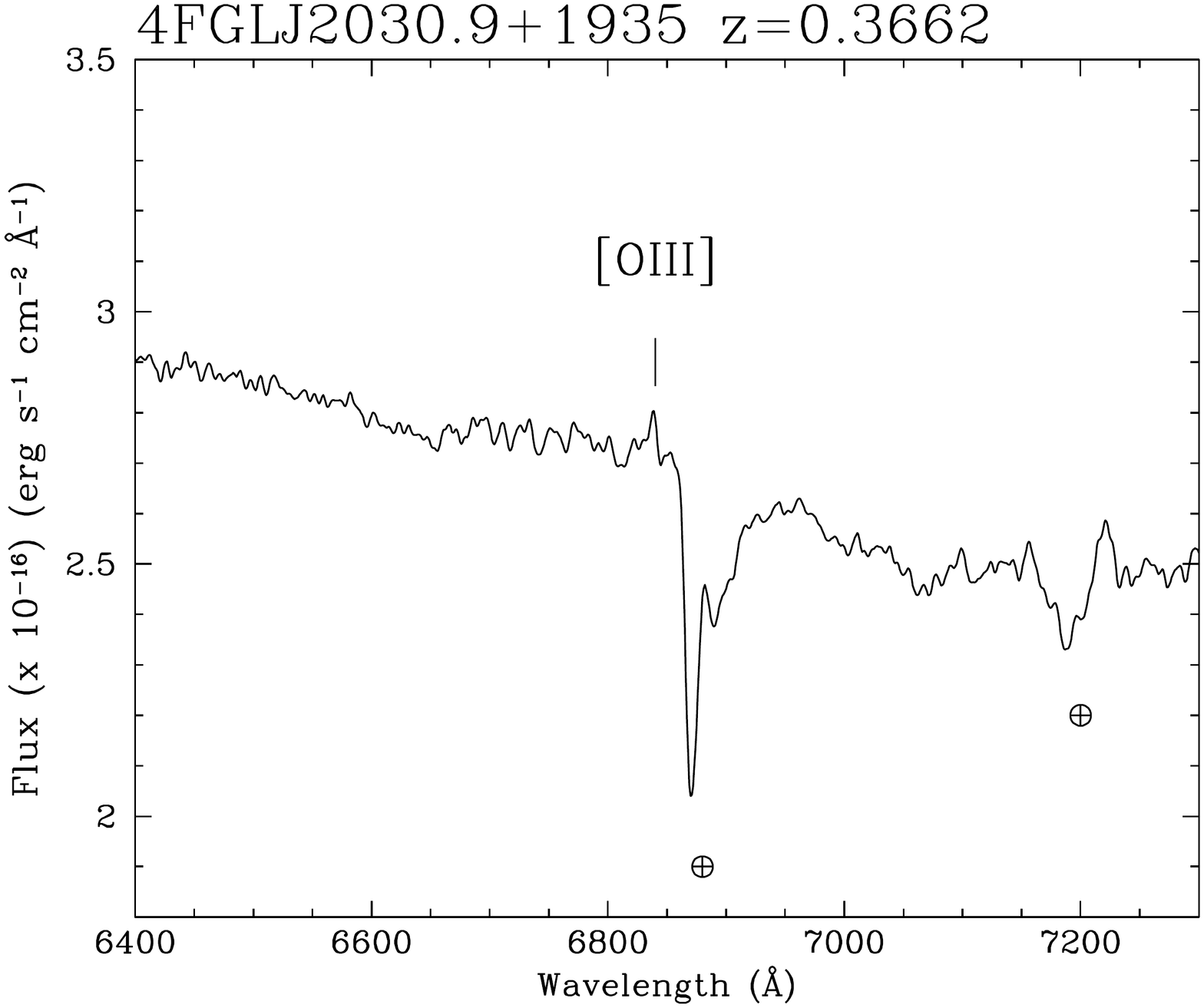}
\caption{- \textit{Continued}}
\end{figure*}

\setcounter{figure}{2}
\begin{figure*}
\includegraphics[width=1.1\textwidth, angle=0]{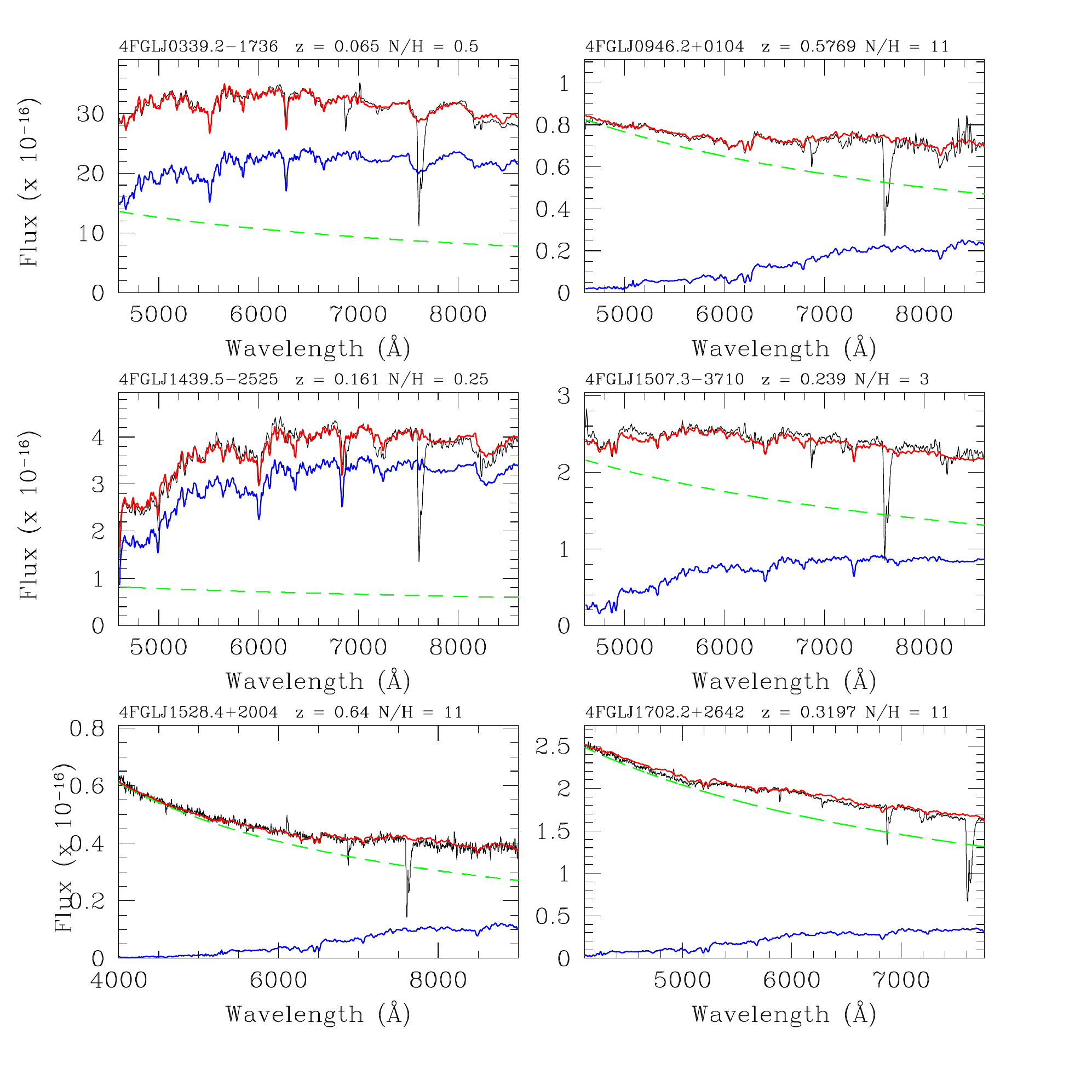}
\caption{Spectral decomposition of the observed optical spectrum (black line) of some targets of our sample into a power law (green dashed line) and an elliptical template  for the host galaxy (blue line). The fit is given by the red solid line (see Section~\ref{sec:results} for details). On each panel the nucleus to host ratio is given.}
\label{fig:decomposition}
\end{figure*}

\setcounter{figure}{2}
\begin{figure*}
\includegraphics[width=1.1\textwidth, angle=0]{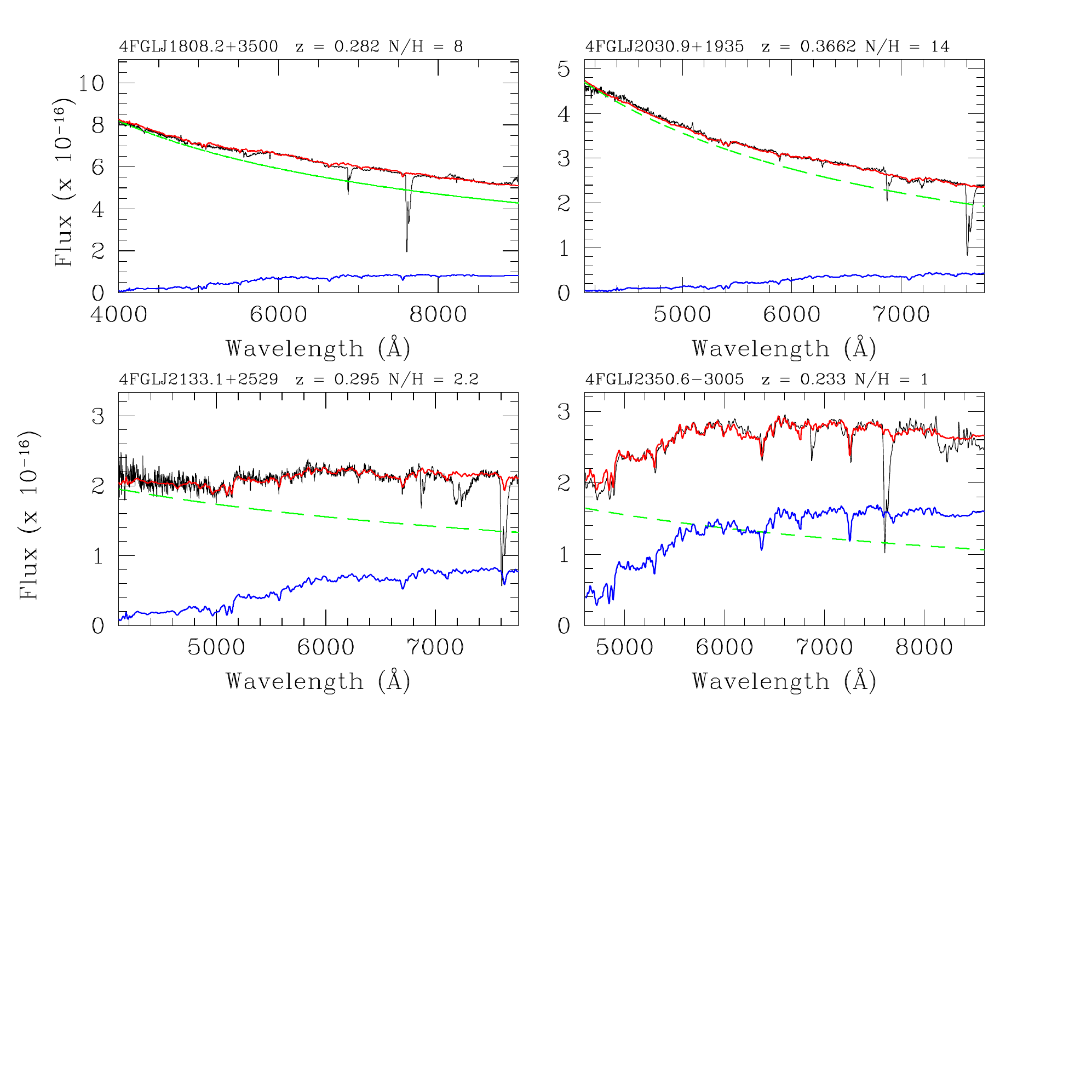}
\caption{- \textit{Continued}}
\end{figure*}


\label{lastpage}
\end{document}